\documentclass[11pt,a4paper]{article}
\usepackage{jheppub,xcolor}
\pdfoutput=1
\usepackage[utf8]{inputenc}
\graphicspath{{./figs/}}

\usepackage{amssymb}
\usepackage{dcolumn}	% Align table columns on decimal point
\usepackage{bm}			% bold math
\usepackage{bbm} 		% blacboard math for \mathbbm{1}
\usepackage{multirow}
\usepackage{slashed}
\usepackage{blkarray}
\usepackage{booktabs}
\usepackage{makecell}
\usepackage{tikz}
\usepackage[compat=1.1.0]{tikz-feynman}
\usepackage{longtable}
\usepackage{placeins}
\usepackage{listings}
\usepackage{amsmath}
\usepackage{xcolor}
\usepackage{pythonhighlight}
\usepackage{subcaption} 

\def\gsim{\mathrel{\rlap{\lower4pt\hbox{\hskip1pt$\sim$}}
    \raise1pt\hbox{$>$}}}         %greater than or approx. symbol
\def\lsim{\mathrel{\rlap{\lower4pt\hbox{\hskip1pt$\sim$}}
    \raise1pt\hbox{$<$}}}         %less than or approx. symbol

\definecolor{codegreen}{rgb}{0,0.6,0}
\definecolor{codegray}{rgb}{0.5,0.5,0.5}
\definecolor{codepurple}{rgb}{0.58,0,0.82}
\definecolor{backcolour}{rgb}{0.95,0.95,0.92}

\lstdefinestyle{mystyle}{
    backgroundcolor=\color{backcolour},   
    commentstyle=\color{codegreen},
    keywordstyle=\color{magenta},
    numberstyle=\tiny\color{codegray},
    stringstyle=\color{codepurple},
    basicstyle=\ttfamily\footnotesize,
    breakatwhitespace=false,         
    breaklines=true,                 
    captionpos=b,                    
    keepspaces=true,                 
    numbers=left,                    
    numbersep=5pt,                  
    showspaces=false,                
    showstringspaces=false,
    showtabs=false,                  
    tabsize=2
}
\definecolor{Red}{rgb}{1.,0.,0.}

\newcommand{\mcO}{\mathcal{O}}

\usepackage[normalem]{ulem}
\definecolor{ao}{rgb}{0.0, 0.5, 0.0}
\definecolor{mt}{RGB}{108, 210, 92}

\newcommand{\lp}{\left(}
\newcommand{\rp}{\right)}

\newcommand{\bc}{\begin{center}}
\newcommand{\ec}{\end{center}}
\newcommand{\ba}{\begin{array}}
\newcommand{\ea}{\end{array}}

\newcommand{\be}{\begin{equation}}
\newcommand{\ee}{\end{equation}}
\newcommand{\bea}{\begin{align}}
\newcommand{\eea}{\end{align}}

\usepackage{tabularx}

 \newcolumntype{C}[1]{>{\centering\arraybackslash}p{#1}}

\usepackage[normalem]{ulem}
\definecolor{ao}{rgb}{0.0, 0.5, 0.0}
\definecolor{mt}{RGB}{108, 210, 92}

\newcommand{\smefit}{{\sc \small SMEFiT}}

\numberwithin{equation}{section}
\numberwithin{figure}{section}
\numberwithin{table}{section}

\title{Connecting Scales: RGE Effects in the SMEFT at the LHC and Future Colliders}

\author[a]{Jaco~ter~Hoeve,}
\author[b]{Luca~Mantani,}
\author[c,d]{Juan~Rojo,}
\author[e]{Alejo~N.~Rossia,}
\author[f]{and Eleni~Vryonidou}
\affiliation[a]{The Higgs Centre for Theoretical Physics, University of Edinburgh,\\[0.1cm]
JCMB, KB, Mayfield Rd, Edinburgh EH9 3FD, Scotland}
\affiliation[b]{Instituto de F\'isica Corpuscular (IFIC), Universidad de Valencia-CSIC, E-46980 Valencia, Spain}
\affiliation[c]{Nikhef Theory Group, Science Park 105, 1098 XG Amsterdam, The Netherlands}
\affiliation[d]{Department of Physics and Astronomy, Vrije Universiteit Amsterdam, NL-1081 HV Amsterdam, The Netherlands}
\affiliation[e]{Dipartimento di Fisica e Astronomia ``G. Galilei'', Universit\`a di Padova, and Istituto Nazionale di Fisica Nucleare, Sezione di Padova, Via F. Marzolo 8, I-35131, Padova, Italy.}
\affiliation[f]{Department of Physics and Astronomy, University of Manchester, Oxford Road, Manchester M13 9PL, United Kingdom}

\emailAdd{jaco.ter.hoeve@ed.ac.uk}
\emailAdd{luca.mantani@uv.es}
\emailAdd{j.rojo@vu.nl}
\emailAdd{alejonahuel.rossia@unipd.it}
\emailAdd{eleni.vryonidou@manchester.ac.uk}

\date{\today}
\abstract{
Global interpretations of particle physics data within the framework of the Standard Model Effective Field Theory (SMEFT), including their matching to UV-complete models, involve energy scales potentially spanning several orders of magnitude. 
Relating these measurements among them in terms of a common energy scale is enabled by the Renormalisation Group Equations (RGEs).
Here we present a systematic assessment of the impact of RGEs, accounting for QCD, electroweak, and Yukawa corrections, in a global SMEFT fit of LEP and LHC data where individual cross-sections are assigned a characteristic energy scale.
We also quantify the impact of the RGE effects in projected global fits at the HL-LHC and the FCC-ee.
Finally, we assess the role that RGEs play on the sensitivity at HL-LHC and FCC-ee to representative one-particle UV models matched onto SMEFT either at tree and  one-loop level.
Our study emphasizes the importance of a consistent treatment of energy scales to achieve the best precision and accuracy in indirect searches for heavy new physics through precision measurements. 
}


\begin{document}
\begin{flushright}
\end{flushright}

\maketitle

\newpage
\section{Introduction}
\label{sec:intro}

The interpretation of particle physics data in the framework of effective field theories (EFTs) such as the SMEFT~\cite{Brivio:2017vri,Isidori:2023pyp} often involves combining experimental observables which are sensitive to very disparate energy transfers.
For instance, a factor of 30 arises between the characteristic energy scales of electroweak precision  observables (EWPOs) at LEP, $Q\sim m_Z$, and of the high-energy tails of top quark pair production at the LHC, reaching up to $m_{t\bar{t}}\sim 3$ TeV.
These scale differences become yet more marked in analyses including flavour data ($Q\sim m_b$) and low energy measurements, for example, information on the muon magnetic moment at $Q\sim m_\mu$.
The importance of treating consistently energy scales is also apparent in the case of sensitivity projections for future facilities such as the Future Circular Collider~\cite{FCC:2018byv,FCC:2018evy,FCC:2018vvp} operating in the electron-positron mode (FCC-ee).
Indeed, as highlighted in recent studies~\cite{Celada:2024mcf,Allwicher:2024sso,Gargalionis:2024jaw,Maura:2024zxz}, precision electroweak and Higgs measurements at the FCC-ee provide constraints on UV-complete models reaching mass scales up to several tens of TeV. 

These considerations suggest that, given the scope and precision of current and future experimental data, assuming all measurements entering a global SMEFT fit correspond to the same underlying energy scale is in general
not justified.
A theoretically consistent interpretation of data encompassing very different energy scales is possible via the Renormalisation Group Equations (RGEs),
 which relate Wilson coefficients associated to operators at a given reference scale $\mu_0$ with those at any other scale, $\mu\ne \mu_0$. 
 The RGE evolution is described by the anomalous dimension matrix which has been computed at one loop for the SMEFT~\cite{Jenkins:2013wua,Jenkins:2013zja,Alonso:2013hga} and implemented in several codes \cite{Aebischer:2018bkb,Celis:2017hod,DiNoi:2022ejg}.
Furthermore, partial results on the two-loop anomalous dimensions have become available recently \cite{Bern:2019wie, Bern:2020ikv,Jenkins:2023rtg,Jenkins:2023bls,Fuentes-Martin:2023ljp,Fuentes-Martin:2022vvu,Aebischer:2022anv,Aebischer:2024xnf,DiNoi:2024ajj,Born:2024mgz,Naterop:2024ydo,deVries:2019nsu,Jin:2020pwh,Fuentes-Martin:2024agf}. 
Crucially, RGE effects do not only induce operator running (where the value of the Wilson coefficient becomes scale dependent) but also operator mixing, where in particular non-zero coefficients at  $\mu\ne \mu_0$ can be generated even in scenarios where they vanish at the reference scale $\mu_0$. In the context of global EFT analyses, mixing can lead to enhanced sensitivities for certain Wilson coefficients. 

With this motivation, several groups have recently studied the impact that RGEs have in the context of SMEFT analyses of particle physics data~\cite{Battaglia:2021nys,Aoude:2022aro,DiNoi:2023onw,Bartocci:2024fmm,Allwicher:2023shc,Allwicher:2024sso,Maltoni:2024dpn}.
Building up on and complementing these studies, the goal of the present work is to extend the {\sc\small SMEFiT} global analysis of~\cite{Celada:2024mcf} with a systematic assessment of the impact of RGEs at the level of both Wilson coefficients and at the level of the masses and couplings of representative UV-complete models, when matching is carried out either at tree-level or at one-loop~\cite{terHoeve:2023pvs}.
In addition, we quantify the role that RGE effects play in the sensitivity projections at future high-energy colliders, in particular concerning the FCC-ee.
There, a careful inclusion of RGE effects is of outmost importance, given the potentially large energy gap between the scales at which data is collected ($Q \sim 91-360$ GeV) and the  mass scale of the UV scenarios being probed through quantum effects (up to $Q\sim$ tens of TeV).
We also emphasize how, in general, one cannot predict the qualitative consequences of neglecting RGE effects.
Indeed, ignoring RGE effects may either lead to an overestimate of the bounds, when operator mixing dilutes the constraints from the data, or to an underestimate thereof, in cases for which the same mixing introduces sensitivity to hitherto uncovered directions in the parameter space.

The starting point of our analysis is the global {\sc\small SMEFiT3.0} fit of LEP and LHC data of~\cite{Celada:2024mcf}, complemented by HL-LHC and FCC-ee projections, and based on the  {\sc\small SMEFiT} framework~\cite{Ethier:2021ydt,Ethier:2021bye,Hartland:2019bjb,Giani:2023gfq}.  
By interfacing {\sc\small SMEFiT} with the RGE solver {\sc\small Wilson} \cite{Aebischer:2018bkb}, we account for RGE effects in the observables entering the fit in an automated manner.
We have validated this automated implementation with existing calculations of RGE effects in the SMEFT, finding good agreement.
Our implementation of RGE effects in {\sc\small SMEFiT} is seamlessly incorporated in our pipeline for UV models~\cite{terHoeve:2023pvs} based on the  {\sc\small match2fit} interface to {\sc\small MatchMakerEFT}~\cite{Carmona:2021xtq}. 

Our results illustrate how the overall impact of RGEs in the global SMEFT fit depends heavily on the input dataset, whether the fit is linear or quadratic in the EFT expansion, and which operators are being considered.
For some operators, RGE effects lead to much more stringent bounds due to the additional sensitivity induced by the mixing.
For other operators, instead, the bounds become looser, showing that neglecting RGE effects may lead to a too optimistic estimate of the reach in the parameter space. 
We demonstrate the crucial role played by RGE effects to constrain heavy new physics via precision measurements at the FCC-ee, both in terms of precision and accuracy.

The outline of this paper is as follows.
First of all, in Sect.~\ref{sec:settings} we summarise the theoretical formalism underpinning the implementation of RGEs in {\sc\small SMEFiT} and the choice of analysis settings.
A global SMEFT fit with RGE effects included is presented in Sect.~\ref{sec:results}, where we focus on those aspects of the analysis for which RGEs have the largest impact.
Sect.~\ref{sec:uv_fcc} considers sensitivity projections for the reach on UV physics from precision measurements at the FCC-ee in the presence of RGE corrections. 
Finally, Sect.~\ref{sec:summary} summarises our main findings and outlines some possible directions for future developments.

Technical details are provided in the appendices.
The validation of RGEs in {\sc\small SMEFiT} is discussed in App.~\ref{app:benchmarking}.
App.~\ref{app:implementation} provides information on the RGE implementation and its performance in the fit.
App.~\ref{app:uv_models} summarizes the main features of the UV models considered. App.~\ref{app:additional_FCC_results} provides additional results on the impact of different FCC-ee runs on the same UV models considered in Sect.~\ref{sec:uv_fcc}. Finally, App.~\ref{app:numerical_bounds} contains the numerical values of the bounds on the Wilson coefficients from the linear and quadratic fits, both with and without RGE effects, presented in Sect.~\ref{sec:results}.

\section{Analysis settings}
\label{sec:settings}

We begin by outlining the framework adopted in this work to incorporate RGE effects into the global SMEFT analysis. 
First, we describe an update in our operator basis as compared to the analysis in~\cite{Celada:2024mcf}.  
Second, we review the main aspects of the RGEs relevant to dimension-six SMEFT operators, providing the theoretical foundation for their inclusion.  
Finally, we detail the renormalisation scale choice in {\sc\small SMEFiT}. 

\paragraph{Operator basis.}
Unless otherwise specified, the input dataset, theoretical calculations, and methodological settings follow those of the {\sc\small SMEFiT3.0} analysis of~\cite{Celada:2024mcf}.
In particular, we adopt the same SMEFT operator basis~\cite{Celada:2024mcf} with one modification: the inclusion of the purely Higgs dimension-six operator
\be
\label{eq:operator_purelyhiggs}
\mathcal{O}_\varphi = \lp \varphi^\dagger \varphi - \frac{v^2}{2}\rp^3  \, ,
\ee
modifying the Higgs self-coupling interactions and where the associated Wilson coefficient is denoted by $c_\varphi$. 
With the addition of $\mathcal{O}_\varphi$, the \smefit~operator basis now comprises $n_{\rm op}=51$ linearly-independent dimension-six operators. 

The Wilson coefficient $c_\varphi$ associated to Eq.~(\ref{eq:operator_purelyhiggs}) can be constrained at hadron colliders primarily through double Higgs production. Furthermore, additional constraints at the LHC arise from NLO electroweak corrections, which affect single Higgs production \cite{Gorbahn:2016uoy, Degrassi:2016wml, Maltoni:2017ims,Bizon:2016wgr}.
Similar considerations apply 
for lepton colliders.
At circular colliders such as the FCC-ee
and CEPC, the one-loop electroweak contribution to the $ZH$ production process offers an especially sensitive probe to $c_{\varphi}$ \cite{McCullough:2013rea}. 
In order to take profit of this effect, here we incorporate the full one-loop electroweak corrections in the SMEFT to the $Zh$ production process at the FCC-ee recently computed in~\cite{Asteriadis:2024xts}.
At linear $e^+e^-$ colliders such as ILC or CLIC, the higher energy
configurations can access Higgs pair production and thus directly probe $c_\varphi$ at tree level. 

We point out that a companion paper of this work, focusing on the Higgs trilinear coupling, is in preparation~\cite{smefit:trilinear}.
Furthermore, an updated version of the future collider SMEFT projections of~\cite{Celada:2024mcf}, which also includes linear leptonic and circular muon colliders, is being finalized~\cite{esppu2026}.
The latter will account for the direct constraints on $c_\varphi$ from Higgs pair production for those colliders in which this process is kinematically accessible.

\paragraph{RGE evolution.}
The leading-order (LO) RGEs governing the scale evolution of dimension-six SMEFT operators have been available for some time~\cite{Jenkins:2013wua,Jenkins:2013zja,Alonso:2013hga}, with numerical solutions implemented in public codes~\cite{Aebischer:2018bkb,Celis:2017hod,DiNoi:2022ejg}.
These RGEs capture the running of SMEFT operators mediated at the one loop  level by any of the SM interactions: QCD, electroweak, and Yukawa interactions. 
The RGEs for the dimension-six operators studied in this work can be expressed as
\be
\label{eq:rge1}
\frac{dc_i(\mu)}{d \ln \mu} = \sum_{j=1}^{n_{\rm op}}\gamma^{(6)}_{ij}\lp \bar{g}\rp c_j(\mu) \, ,\qquad i=1,\ldots,n_{\rm op} \, ,
\ee
where $\mu$ is the renormalisation scale and the sum runs over all the operators in the considered basis.

The anomalous dimension matrix $\gamma^{(6)}_{ij}(\bar{g})$ associated to dimension-six operators depends generically on the dimension-four Lagrangian parameters, 
\be
\bar{g} = \{g_s, g, g^\prime, y_u, y_d, y_e, \mu_h^2, \lambda\} \, ,
\ee
with $g_s,g,g'$ being the strong and electroweak couplings, $y_u,y_d,y_e$ being the Yukawa couplings, and $\mu_h,\lambda$ being the Higgs potential parameters. 
These parameters of the dimension-four Lagrangian are also subject to their own scale evolution described by the
corresponding RGEs:
\begin{equation}
\label{eq:rge2}
    \frac{d\bar{g}_i(\mu)}{d \ln \mu} = \sum_{j=1}^{n_{\rm op}}\gamma^{(4)}_{ij}\lp \bar{g}, \boldsymbol{c} \rp \bar{g}_j(\mu) \, ,
\end{equation}
where the relevant anomalous dimensions $\gamma_{ij}^{(4)}$ depend on both the SM parameters $\bar{g}$ and of the Wilson coefficients $\boldsymbol{c}$ associated 
to the dimension-six SMEFT operators.

In general, the system of coupled 
differential equations composed of Eqns.~(\ref{eq:rge1})--(\ref{eq:rge2})
cannot be solved analytically and demands  numerical solutions. 
Here we employ instead an approximation often called the Evolution Matrix Approximation (EMA) to linearize the problem~\cite{Proceedings:2019rnh,Fuentes-Martin:2020zaz}.
This approximation involves two steps: 
\begin{itemize}

    \item Dropping terms proportional to the dimension-six parameters (the Wilson coefficients $\boldsymbol{c}$) in the anomalous dimensions $\gamma_{ij}^{(4)}$
    associated to the SM dimension-four parameters, which decouples Eq.~(\ref{eq:rge1}) from Eq.~(\ref{eq:rge2}) and allows the SM couplings to evolve independently with the scale.
    
    \item Neglecting the non-linear contributions to $\gamma_{ij}^{(6)}$ arising from redefinitions of the SM parameters due to input measurements involving dimension-six Wilson coefficients.
    
\end{itemize}
Formally, all terms neglected in this approximation correspond to higher order corrections in the EFT expansion of the RGE evolution of the Wilson coefficients, i.e. $\mathcal{O}(\Lambda^{-4})$.
These neglected corrections are therefore expected to be negligible unless the Wilson coefficients are excessively large, which may occur for some loosely constrained coefficients.

Under the EMA assumptions, the system of differential equations Eq.~(\ref{eq:rge1}) can be solved exactly, yielding the scale dependence of the Wilson coefficients as:
\be
\label{eq:rge_solution}
c_i(\mu) = \sum_{j=1}^{n_{\rm op}}\Gamma_{ij}(\mu,\mu_0;\bar{g}) c_j(\mu_0) \, ,\qquad i=1,\ldots,n_{\rm op} \, ,
\ee
where $\mu_0$ is a reference energy scale, and $\Gamma_{ij}$ is the evolution operator matrix
which determines the scale evolution between $\mu_0$ and $\mu$ as a function of the values of the SM couplings $\bar{g}$.
This evolution matrix operator is given schematically by:
\begin{equation}
    \Gamma_{ij}(\mu,\mu_0;\bar{g}) = \exp\lp {\int^{\mu}_{\mu_0} d\log(\mu^\prime) \gamma_{ij}^{(6)}}(\bar{g}(\mu^\prime)) \rp \, ,
\end{equation}
which accounts also for the scale dependence of the SM couplings $\bar{g}(\mu)$.

Let us illustrate how RGE effects
modify the theoretical predictions of physical observables in the presence of dimension-six SMEFT operators.
If RGE effects are neglected, the Wilson coefficients are scale-independent and the theoretical predictions in the SMEFT, $T_{\rm EFT}$, are expressed as:
\be
\label{eq:theory_EFT_1}
T_{\rm EFT}({\boldsymbol{c}}/\Lambda^2)= T_{\rm SM} + \sum_{i=1}^{n_{\rm op}} \kappa_i\frac{c_i}{\Lambda^2} + \sum_{i,j=1}^{n_{\rm op}} \widetilde{\kappa}_{ij}\frac{c_ic_j}{\Lambda^4} \, ,
\ee
where $\Lambda$ is the EFT cutoff scale
and $T_{\rm SM}$ is the SM prediction for this observable. 
When RGE effects are included, the Wilson coefficients acquire scale dependence, and the expression becomes:
\be
\label{eq:theory_EFT_2}
T_{\rm EFT}({\boldsymbol{c}(\mu)}/\Lambda^2)= T_{\rm SM} + \sum_{i=1}^{n_{\rm op}} \kappa_i\frac{c_i(\mu)}{\Lambda^2} + \sum_{i,j=1}^{n_{\rm op}} \widetilde{\kappa}_{ij}\frac{c_i(\mu)c_j(\mu)}{\Lambda^4} \, ,
\ee
with cross-sections evaluated at the characteristic energy scale of the process $\mu$, e.g.\ the momentum transfer in the hard scattering reaction.
Employing now the EMA as a solution of the RGEs, we can express all theory predictions as a function of the Wilson coefficients at the same reference energy scale $\mu_0$:
\begin{eqnarray}
T_{\rm EFT}({\boldsymbol{c}(\mu_0)}/\Lambda^2)&=& T_{\rm SM} + \sum_{i,j=1}^{n_{\rm op}} \kappa_i\Gamma_{ij}\frac{c_j(\mu_0)}{\Lambda^2} + \sum_{i,j,k,\ell=1}^{n_{\rm op}} \widetilde{\kappa}_{ij}\Gamma_{ik}\Gamma_{j\ell}\frac{c_k(\mu_0)c_\ell(\mu_0)}{\Lambda^4} \,, \nonumber \\
&=& T_{\rm SM} + \sum_{j=1}^{n_{\rm op}} \kappa_j'\frac{c_j(\mu_0)}{\Lambda^2} + \sum_{k,\ell=1}^{n_{\rm op}} \widetilde{\kappa}_{k\ell}'\frac{c_k(\mu_0)c_\ell(\mu_0)}{\Lambda^4} \,,
\label{eq:theory_EFT_3}
\end{eqnarray}
where we see that, effectively, in this approximation the linear and quadratic EFT cross-section are multiplied with the evolution matrix operator, namely
\be
\kappa_j' = \sum_{i=1}^{n_{\rm op}} \kappa_i\Gamma_{ij} \, ,\qquad \widetilde{\kappa}_{k\ell}'=\sum_{i,j=1}^{n_{\rm op}} \widetilde{\kappa}_{ij}\Gamma_{ik}\Gamma_{j\ell} \, .
\ee
Eq.~(\ref{eq:theory_EFT_3}) hence adopts a common reference scale $\mu_0$ for all coefficients entering the theoretical predictions.

In this work, the solution of the RGEs, Eq.~(\ref{eq:rge_solution}), is implemented by means of the {\sc\small wilson} package~\cite{Aebischer:2018bkb} integrated in \smefit. 
Technical details of this implementation are presented in App.~\ref{app:implementation}.
Furthermore, we note that we adopt a scheme where the only non-zero Yukawa coupling is that of the top quark. Consequently, the anomalous matrix $\gamma^{(6)}_{ij}$ in \eqref{eq:rge1} will depend only on the SM parameters $\{g_s, g, g^\prime, y_t, \mu_h^2, \lambda\}$. 
With this restriction, we have verified that RGE effects do not violate our flavour symmetry choice, i.e.\ the RG evolution closes in our chosen operator basis.

%%%%%%%%%%%%%%%%%%%%%%%%%%%%%%%%%%%%%%%
\begin{table}[t!]
    \centering
    \footnotesize % Reduce font size
    \renewcommand{\arraystretch}{1.9} % Reduce vertical row spacing
    \setlength{\tabcolsep}{4pt} % Reduce horizontal column spacing
    \begin{tabular}{l|l||l|l}
        \toprule
        \textbf{Process} & \textbf{Scale Choice $\mu$} & \textbf{Process} & \textbf{Scale Choice $\mu$} \\ \midrule
        Higgs (ggF) & $\sqrt{m_H^2 + (p_T^H)^2}$ & $t\bar{t}b\bar{b}$ & $2m_t$ \\ \hline
        Higgs (VBF) & $\sqrt{m_H^2 + (p_T^H)^2}$ & $t\bar{t}V$ & $\sqrt{(2m_t + m_V)^2 + (p_T^V)^2}$ \\ \hline
        $VH$ & $\sqrt{(m_V + m_H)^2 + (p_T^V)^2}$ & $tV$ & $m_t + m_V$ or $\sqrt{(m_t + m_V)^2 + (p_T^t)^2}$ \\ \hline
        $t\bar{t}H$ & $\sqrt{(2m_t + m_H)^2 + (p_T^H)^2}$ & $W$-helicities & $m_t$ \\ \hline
        $tH$ & $m_t + m_H$ & $WZ$ & $m_T^{WZ}$ or $\sqrt{(m_Z + m_W)^2 + (p_T^Z)^2}$ \\ \hline
        $t\bar{t}$ & $m_{tt}$ & $WW$ & $m_{e\mu}$ \\ \hline
        Single-$t$ & $m_t$ & $V$ pole (incl. EWPOs) & $m_V$ \\ \hline
        $t\bar{t}\gamma$ & $2m_t$ & Bhabha scattering & $\sqrt{s}$ \\ \hline
        $t\bar{t}t\bar{t}$ & $4m_t$ & $e^+e^- \to WW\,/\,t\bar{t}\,/\,f\bar{f}$ & $\sqrt{s}$ \\ \hline
        $HH$ & $2m_H$ & $e^+e^- \to ZH$ & $\sqrt{s}$        
        \\ \bottomrule
    \end{tabular}
    \caption{The definition of the central scale $\mu$ for each of the processes
    considered in the present analysis.
    For differential distributions,
    the central scale is defined at the level
    of individual bins.
    For some observables, such as $WZ$ production, the scale choice depends on the specific differential distribution measured.
    }
    \label{tab:scale_definitions}
\end{table}
%%%%%%%%%%%%%%%%%%%%%%%%%%%%%%%%%%

\paragraph{Renormalisation scale choice.}
In the \smefit~dataset, the fitted observables span more than an order of magnitude in energy, ranging from $\mu\sim m_Z$ in  weak boson pole measurements at LEP  to $\mu\sim {\rm multi\,TeV}$ in differential distributions for top-quark pair production at the LHC.
Here we adopt a quasi-dynamic scale setting approach, where each individual data bin in the fit is assigned a characteristic energy scale $\mu$.
For most processes, the choice of scale $\mu$ typically aligns with the optimal renormalisation and factorisation scales $\mu\sim \mu_F=\mu_R$, which are determined based on perturbative convergence arguments and are closely related to the momentum transfer in the underlying hard scattering process.

\begin{figure}[t]
\centering
\includegraphics[width=0.99\textwidth]{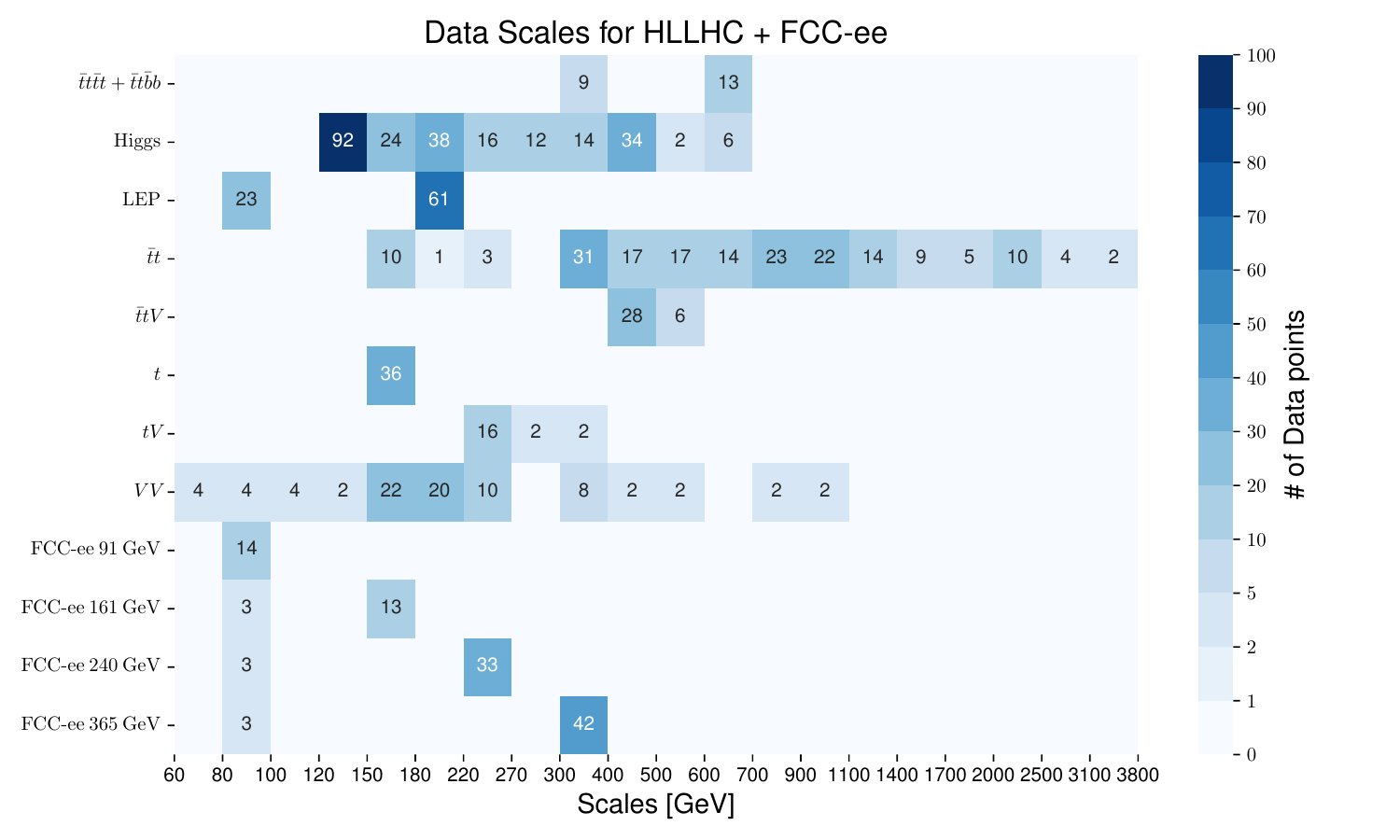}
\caption{Distribution of energy scales $\mu$ over the cross-sections entering the present analysis.
Each entry in the heat map indicates how many
data points associated to a given energy scale bin correspond to each of the considered datasets.
}
\label{fig:scales}
\end{figure}

The general prescription to set the scale $\mu$ assigned to each cross-section in the fit is  as follows.
For inclusive cross-sections, we adopt the threshold production energy required to produce the final state.
For instance,  we take $\mu = 2m_t$ for top-quark pair production total cross-sections. 
For differential distributions and simplified template cross-sections (STXS) instead, the energy scale is determined based on the kinematics of the corresponding bin. 
Table~\ref{tab:scale_definitions} summarizes the scale choices for the observables included in our analysis. Fig.~\ref{fig:scales} shows the distribution of energy scales for the data or projections used in the SMEFT analysis conducted in this work. 

The probed energy scales span nearly two orders of magnitude, ranging from the EWPOs centred at $\mu \sim m_Z$ to $t\bar{t}$ production reaching up to $\mu \sim 3$ TeV in the invariant mass differential distributions $m_{t\bar{t}}$. 
While the choice of the associated scale is not unique, moderate variations around the central value are not expected to significantly affect the overall impact of RGE effects. 
This expectation is verified in Sect.~\ref{sec:results}.

%--------------------------
%--------------------------
\section{RGE effects in the global SMEFT fit of LEP and LHC data}
\label{sec:results}

Here we quantify the impact of the RGE in a global SMEFT fit which takes into account experimental data collected at LEP and at Runs 1 and 2 of the LHC~\cite{Celada:2024mcf}.
This SMEFT analysis provides the baseline scenario to which we will compare the impact of the  projected HL-LHC and FCC-ee measurements in Sect.~\ref{sec:uv_fcc}.
We assess the stability of our results with respect to the choice of energy scale $\mu$ entering the observable calculation, Eq.~(\ref{eq:theory_EFT_1}), by varying its reference value specified in Table~\ref{tab:scale_definitions} within a given range.

As compared to~\cite{Celada:2024mcf}, the only difference in terms of the input dataset is that now we also account for the constraints on Higgs boson pair production at $\sqrt{s}=13$ TeV provided by the ATLAS collaboration~\cite{ATLAS:2024ish}. 
Probing the Higgs trilinear self-coupling $\lambda$, this dataset is sensitive to the purely-Higgs operator $\mathcal{O}_{\varphi}$. 
We use {\sc\small SMEFT@NLO} \cite{Degrande:2020evl} interfaced to {\sc mg5\_aMC@NLO}~\cite{Alwall:2014hca} to obtain the corresponding LO theory predictions. 
Otherwise, theoretical calculations are the same as in~\cite{Celada:2024mcf}, and in particular we  adopt the same electroweak input scheme, namely $\{\hat{m}_W, \hat{m}_Z, \hat{G}_F\}$.
Higher-statistics runs have been used to reduce the Monte Carlo uncertainties affecting some high-mass bins of top quark pair differential distributions, see App.~\ref{app:implementation}.

\subsection{Baseline results}
\label{subsec:results}

\begin{figure}[htbp]
    \centering
\includegraphics[width=0.838\linewidth]{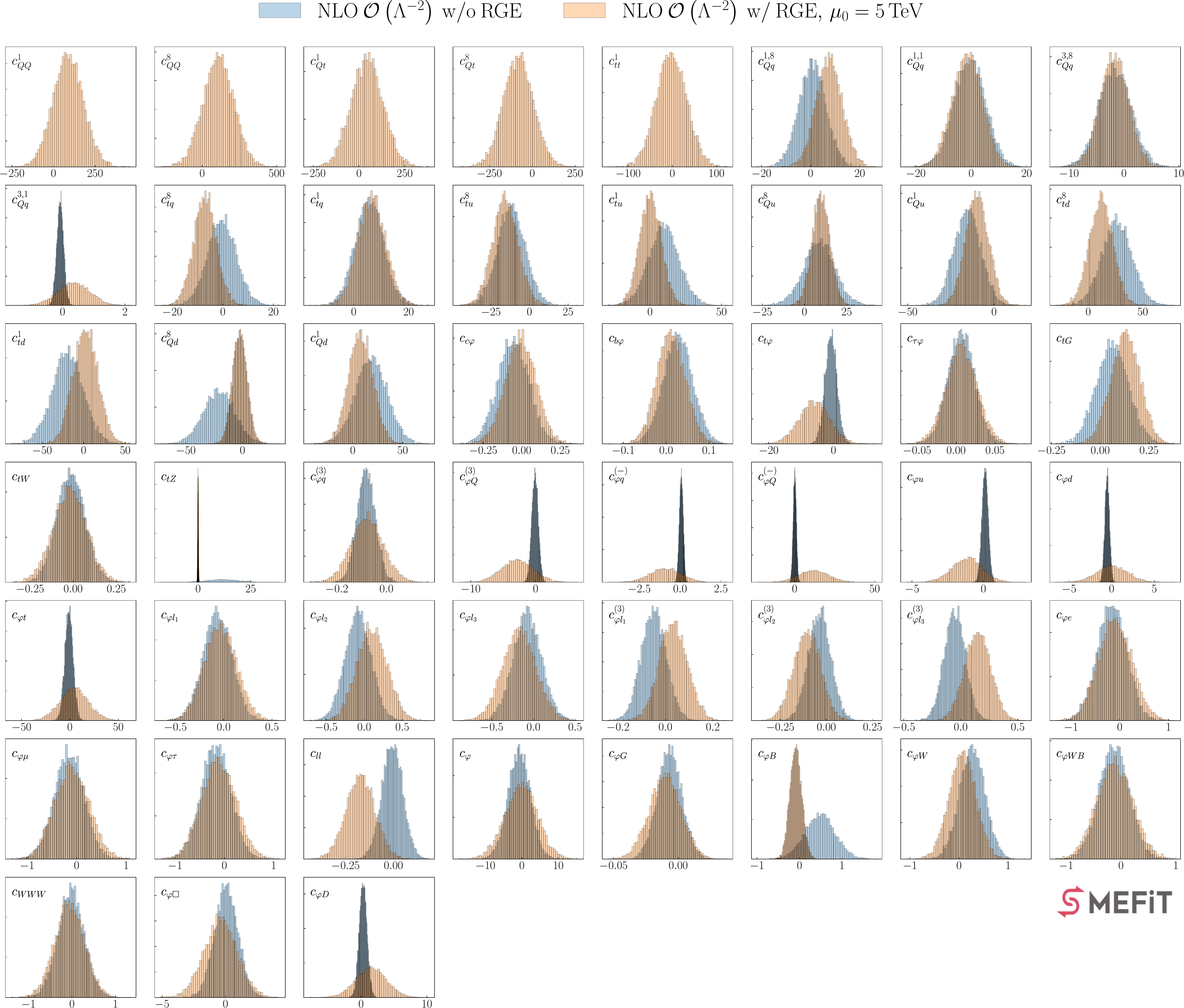}
    \\[0.3cm]
\includegraphics[width=0.838\linewidth]{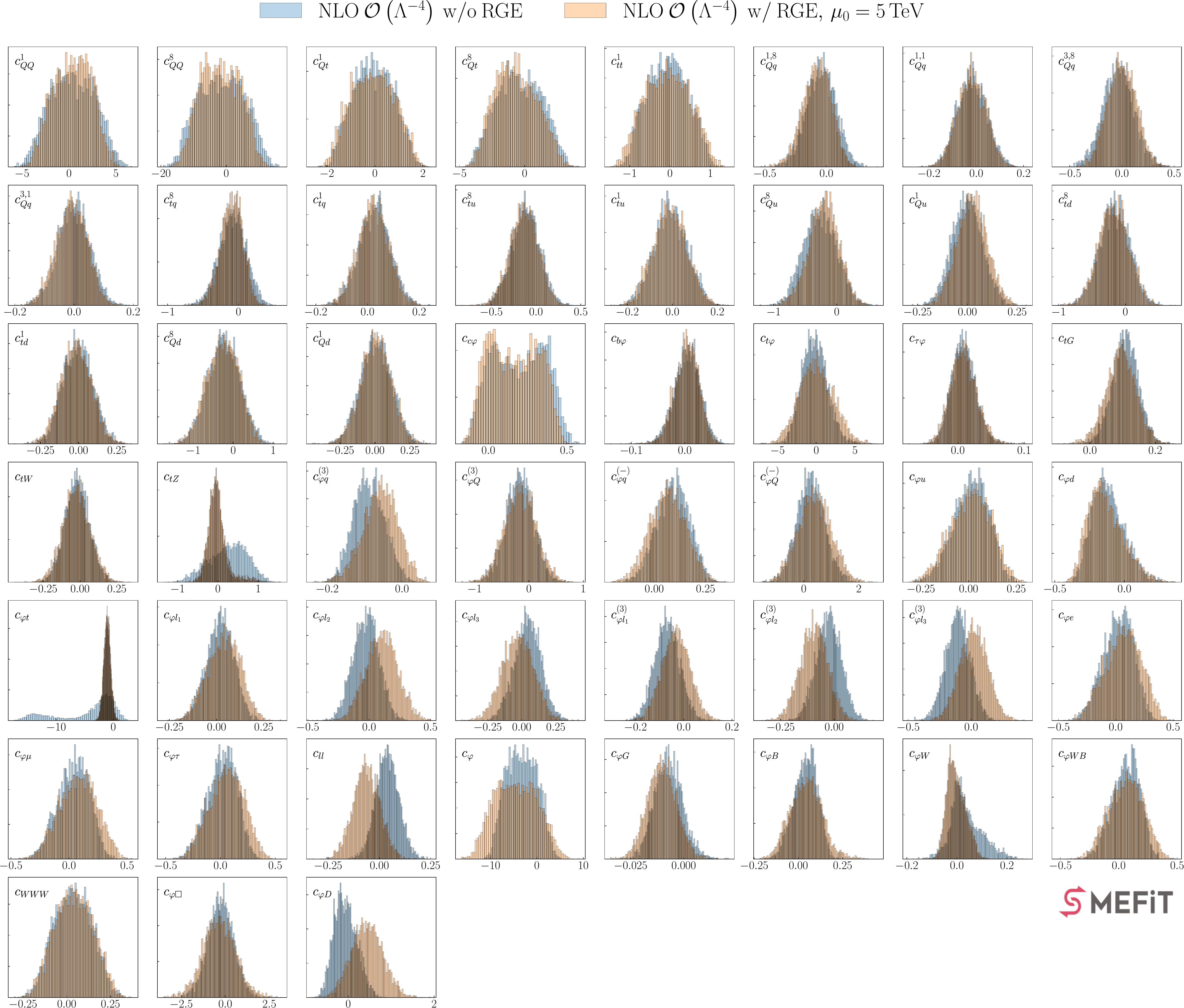}
    \caption{Posterior distributions at $\mu_0=5$ TeV for the $n_{\rm eft} = 51$ Wilson coefficients
    in the linear (upper) and quadratic (lower) EFT fits, comparing the results with and without RGE effects accounted for.
    }    \label{fig:posterior_rg_vs_no_rg_linear}
\end{figure}

The upper (lower) panel of Fig.~\ref{fig:posterior_rg_vs_no_rg_linear} displays the posterior distributions for linear (quadratic) fits, quantifying the impact of RGE effects in our global SMEFT fit to top, Higgs, diboson and EWPO data.
We present constraints for the Wilson coefficients $c({\mu_0})$ evaluated at ${\mu_0}=5$~TeV.
 We consider this choice of scale $\mu_0$ to be appropriate given that our dataset has the highest bins probing energies of at most 3 TeV.

%%%%%%%%%%%%%%%%%%%%%%%%%%%%%%%%%%%%%%%%%%%%%%%%%%%%
%%%%%%%%%%%%%%%%%%%%%%%%%%%%%%%%%%%%%%%%%%%%%%%%%%%%%

\paragraph{Linear EFT results.}
Discussing first the results of the linear analysis, we observe how the effects of the RGEs depends significantly on the specific operator being considered. 
In general, the impact of RGEs in the global fit can lead, depending on the operator, either to a tightening of the bounds, a loosening thereof, or to stable results.
Furthermore, for many coefficients also the most likely values of the distributions change, showcasing that without RGEs the fit central values may be biased.

RGE effects help in some cases to constrain new directions in the parameter space that were previously unconstrained in a fit without them. 
The most striking example is found in the four-heavy-quark operators.  
In a linear EFT fit without the inclusion of RGEs, these operators remain undetermined due to flat directions in the theory predictions for the $t\bar{t}t\bar{t}$ and $t\bar{t}b\bar{b}$ cross-section measurements.  
However, once RGEs are taken into account, these four-heavy-quark operators mix into operators that modify the EWPOs at the $Z$-pole, thereby lifting flat directions.  
Nonetheless, the resulting constraints are not particularly stringent, as strong correlations persist in the global fit, leading to bounds that extend beyond the region considered reliable in terms of perturbativity.

Another Wilson coefficient for which the inclusion of the RGEs leads to greatly improved bounds is $c_{tZ}$.
This operator modifies the coupling of the $Z$ boson to the top quark and contributes to both the associated production of top quarks with $Z$ bosons and the loop-induced decay channels $H\rightarrow Z\gamma$ and $H\rightarrow \gamma\gamma$.  
In the absence of RGE effects, the constraints on this operator are primarily driven by the loop-induced Higgs decay channels.  
However, once a global analysis is performed, a strong correlation emerges with operators such as $\mathcal{O}_{\varphi B}$ in Higgs decays, shifting the dominant source of information on the operator to $t\bar{t}Z$ production.
The situation changes drastically once RGE effects are taken into account, leading to a significant improvement in the bound on this operator.  

\begin{figure}[tp]
    \centering
    \includegraphics[width=0.35\linewidth]{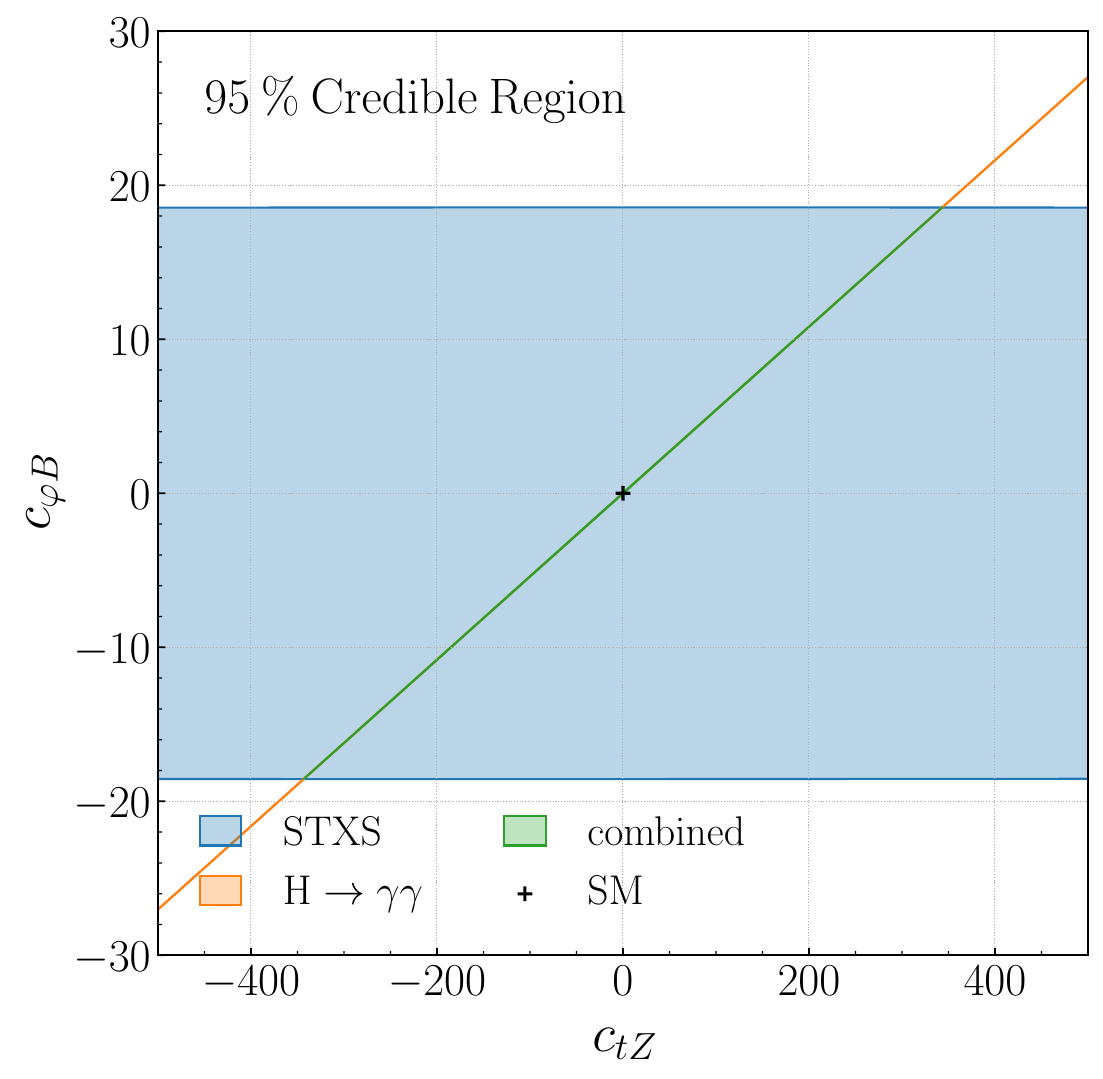}
    \includegraphics[width=0.35\linewidth]{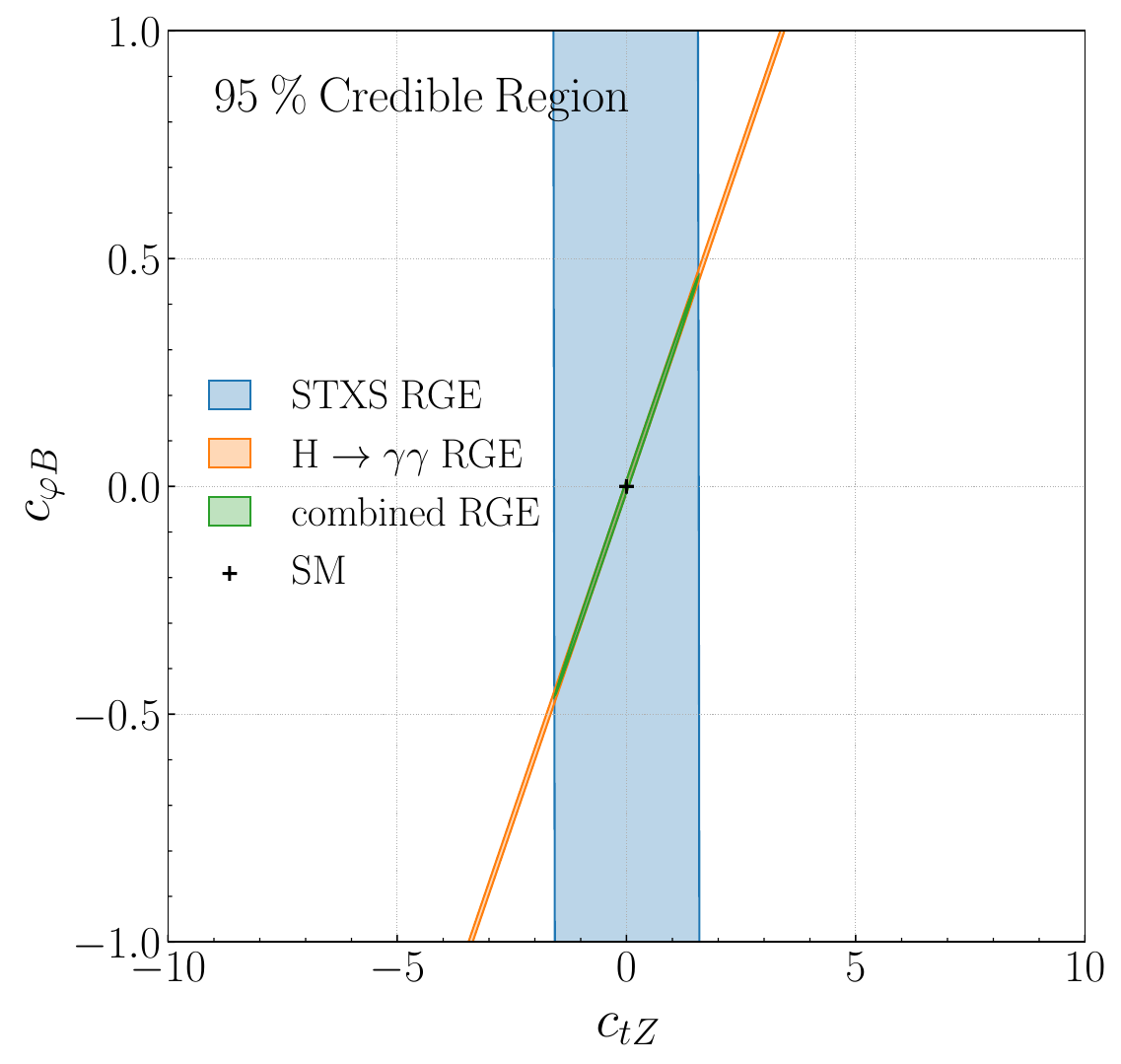}\\
    \includegraphics[width=0.35\linewidth]{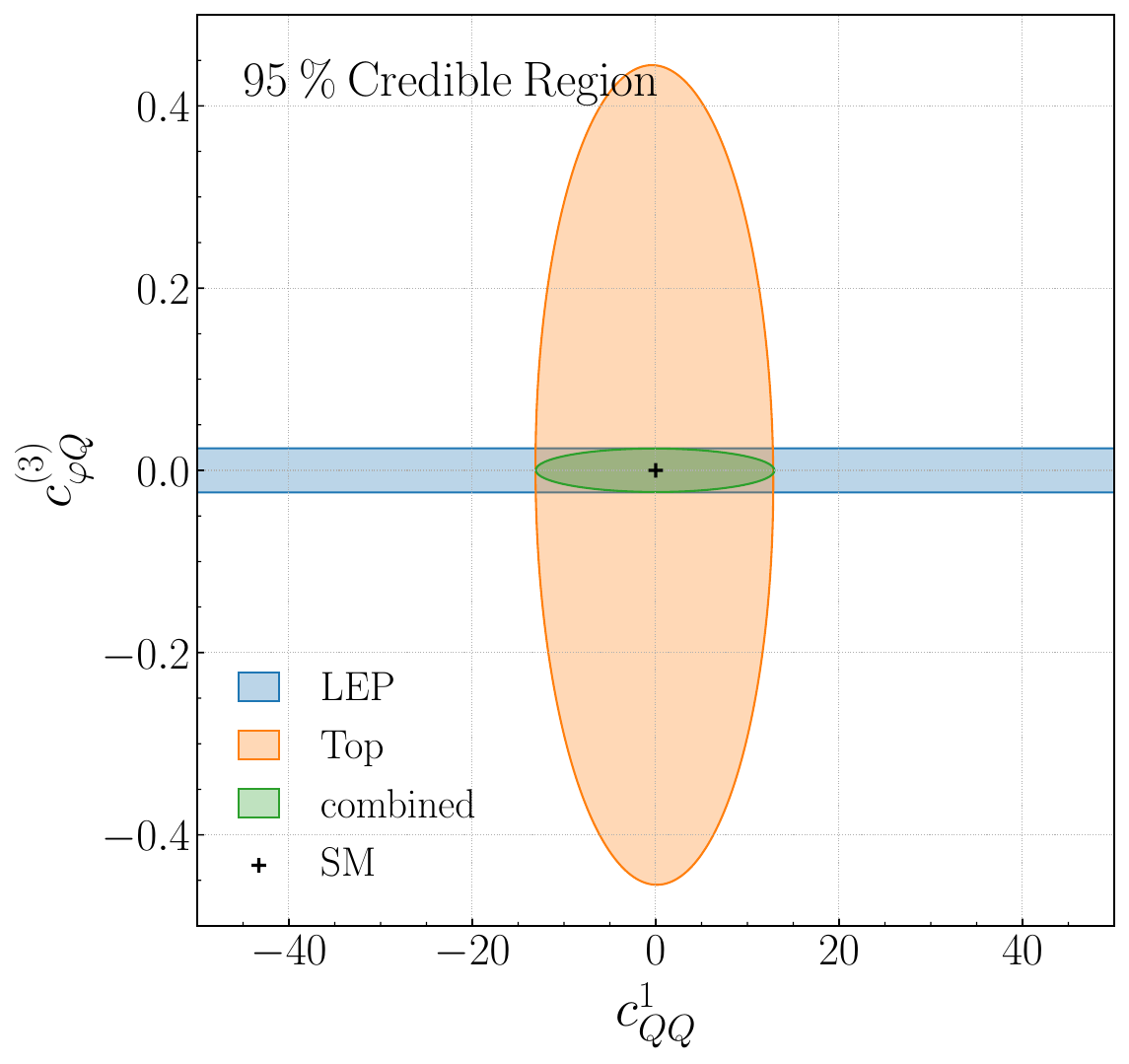}
    \includegraphics[width=0.35\linewidth]{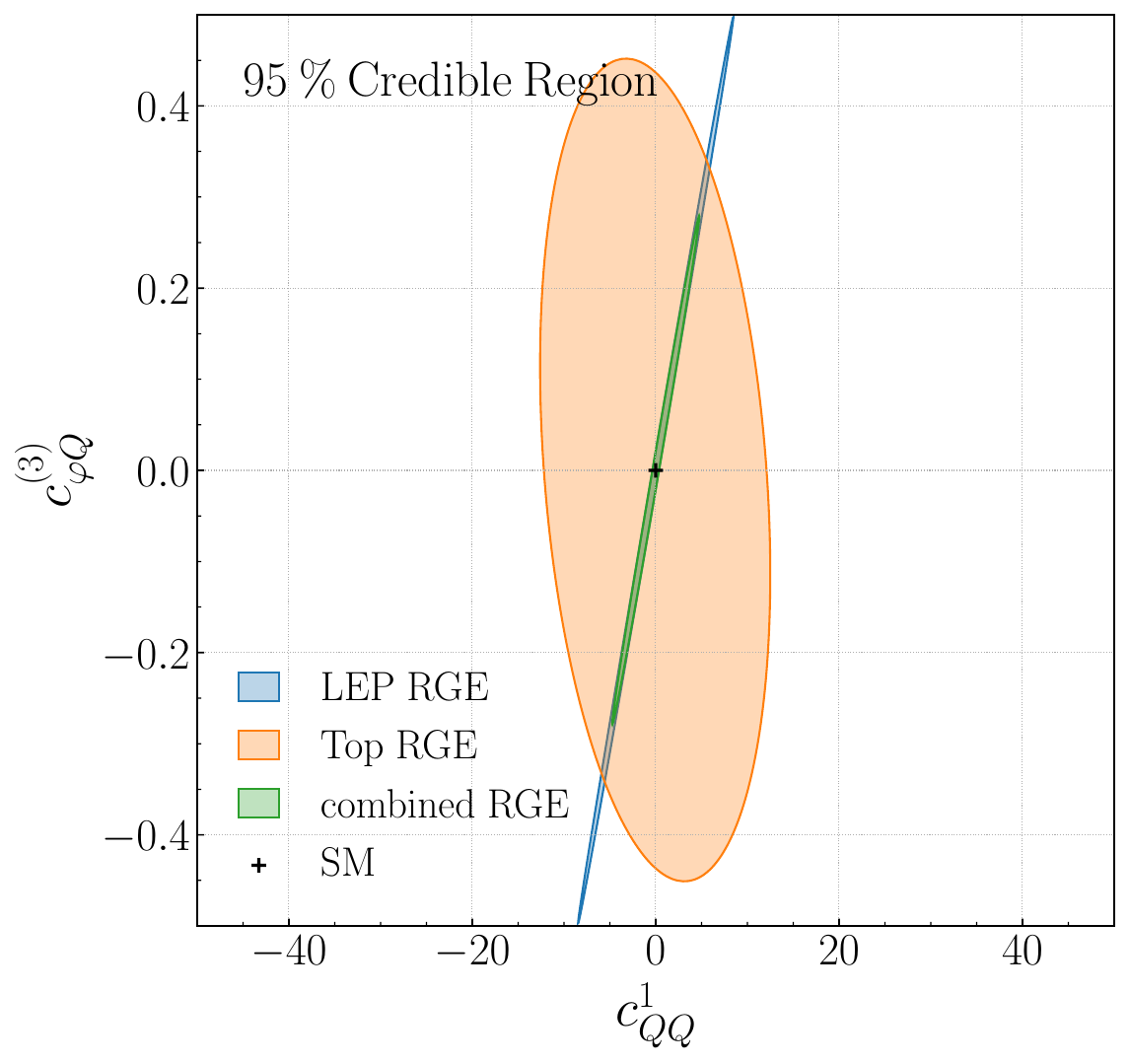}\\
    \caption{95\% C.~I. in two-parameter fits to different datasets
    illustrating the impact of RGE. 
The upper panels display the results for $\lp c_{\varphi B},c_{tZ}\rp$ without (left) and with (right panel) RGE in the fit,
comparing the separate impact of the ATLAS STXS and the $H\to \gamma\gamma$ signal strengths with that of their combination.
The lower panels show the corresponding results for $( c_{\varphi Q}^{(3)},c_{QQ}^{1})$
using separately the LEP EWPOs and the LHC top quark data, as well as their combination. 
Both cases correspond to linear EFT fits.
}
    \label{fig:2D_rge_fits}
\end{figure}

To further pinpoint the origin of this improvement, Fig.~\ref{fig:2D_rge_fits} (first row) presents two-dimensional linear fits of $c_{tZ}$ and $c_{\varphi B}$.  
Specifically, we perform separate fits to the ATLAS STXS Higgs production data~\cite{ATLAS:2022vkf}, which does not include Higgs decays and is therefore sensitive to $c_{\varphi B}$ through $VH$ and VBF production, and to a CMS gluon-fusion dataset where the Higgs decays into $\gamma \gamma$~\cite{CMS:2019xnv}.
The left panel displays a fit performed without RGE effects.  
One observes that, despite the high sensitivity of the Higgs decay channel $H\to \gamma\gamma$, no meaningful constraints arise due to the presence of a flat direction induced by the interplay with $c_{\varphi B}$.  
This degeneracy is partially lifted when including constraints on $c_{\varphi B}$ from STXS production modes; however, the resulting bound remains weak, as the probed directions are nearly parallel.  
As a consequence, Higgs data does not significantly constrain $c_{tZ}$, which instead remains primarily determined by top-quark measurements in the global fit.
However, the inclusion of RGE effects drastically alters this picture.  
The interplay between the two Higgs measurements is fully unlocked: the probed directions in parameter space are optimally rotated, and the sensitivity from STXS production modes is significantly enhanced.  
This improvement arises because $\mathcal{O}_{tZ}$ runs into $\mathcal{O}_{\varphi B}$ and $\mathcal{O}_{\varphi W B}$ under RGE evolution, both of which contribute to Higgs observables at tree level.
This study identifies how the significant improvement on the  $c_{tZ}$ bound arises in the linear fit. 

As discussed above, incorporating RGEs does not necessarily lead to more stringent bounds for all operators.  
In some cases, it can instead result in weaker constraints, particularly when the information from a given measurement becomes spread across a higher-dimensional parameter space due to RGE-induced mixing.  
This effect is evident for certain operators contributing to the EWPOs, such as $c_{\varphi q}^{(-)}$, $c_{\varphi D}$, and $c_{\varphi Q}^{(3)}$.  
For these cases, we observe a significant deterioration of the bounds upon including RGE effects, highlighting how a global fit that neglects these effects can severely overestimate its sensitivity to the underlying Wilson coefficients.
To better understand this effect, in the bottom panels of Fig.~\ref{fig:2D_rge_fits}, we present two-dimensional fits of the Wilson coefficients $c_{\varphi Q}^{(3)}$ and $c_{QQ}^{1}$, considering the constraints from LEP and top data separately.  
In the absence of RGE effects, the constraint on $c_{\varphi Q}^{(3)}$ is dominated by LEP measurements, specifically from $Z \to b\bar{b}$ decays, while the four-heavy-quark operator $c_{QQ}^{1}$ is constrained solely by top data, particularly four-heavy-quark production.  
However, once RGE effects are included, the running of the four-heavy-quark operator into $Z \to b\bar{b}$ induces a rotation of the flat direction in the parameter space.  
This results in a loss of constraining power on $c_{\varphi Q}^{(3)}$, which then becomes constrained only by the combined information from LEP and top data.

A similar degradation of the sensitivity is obtained for the top Yukawa coupling $c_{t\varphi}$ and for some of two-heavy-two-light four-quark operators. 
All in all, it is clear that including RGEs in the global fit will improve its accuracy but not necessarily its precision, since the latter is overestimated for many operators in the fits without RGE effects.
%

%%%%%%%%%%%%%%%%%%%%%%%%%%%%%%%%%%%%%%%%%%%%%%%%%%%%%%%
%%%%%%%%%%%%%%%%%%%%%%%%%%%%%%%%%%%%%%%%%%%%%%%%%%%%%%

\paragraph{Quadratic EFT results.}

Moving on to the results of the quadratic EFT fits (lower part of Fig.~\ref{fig:posterior_rg_vs_no_rg_linear}), we observe that the impact of RGEs is significantly less pronounced compared to the linear EFT fits, both in terms of central values and on the impact on the magnitude of the 95\% Credible Interval (C.I.) bounds.  
For instance, the posterior distributions for the two-light-two-heavy operators remain stable upon the inclusion of RGEs.  
Quadratic corrections can substantially improve constraints, as they often lead to a significant reduction of correlations between coefficients.  
Since RGE effects primarily introduce correlations between coefficients, their impact is less pronounced in the quadratic fit, where the coefficients are generally more constrained.  

Interestingly, and in contrast to the linear fits, we do not observe a major degradation of the constraining power for any coefficient in the quadratic fits upon including RGE effects.  
Nevertheless, for a small subset of operators, we still find notable RGE corrections, either in the form of reduced uncertainties or shifts in the central values.  
Two examples of the former are $c_{\varphi t}$, whose double-peak structure disappears as RGE effects disfavour the non-SM solution, and $c_{tZ}$, whose bounds improve for the same reason as in the linear case, albeit in a less dramatic manner. Secondary minima typically disappear when new sensitivities are incorporated into the EFT analysis, as degeneracies far from the SM are observable-dependent and become disfavoured by additional observables with different parametric dependencies. RGE-improved predictions effectively add such sensitivities; for example, $c_{\varphi t}$ becomes sensitive to EWPOs. Consequently, the second minimum found in $ttV$ data flows into regions excluded by LEP constraints.
Furthermore, for operators primarily constrained by the EWPOs, such as $c_{\ell\ell}$, $c_{\varphi D}$, and $c_{\varphi q}^{(3)}$, the main effect of RGEs is now a moderate shift in their central values, while uncertainties remain stable.

\begin{figure}[t]
    \centering
\includegraphics[width=0.95\linewidth]{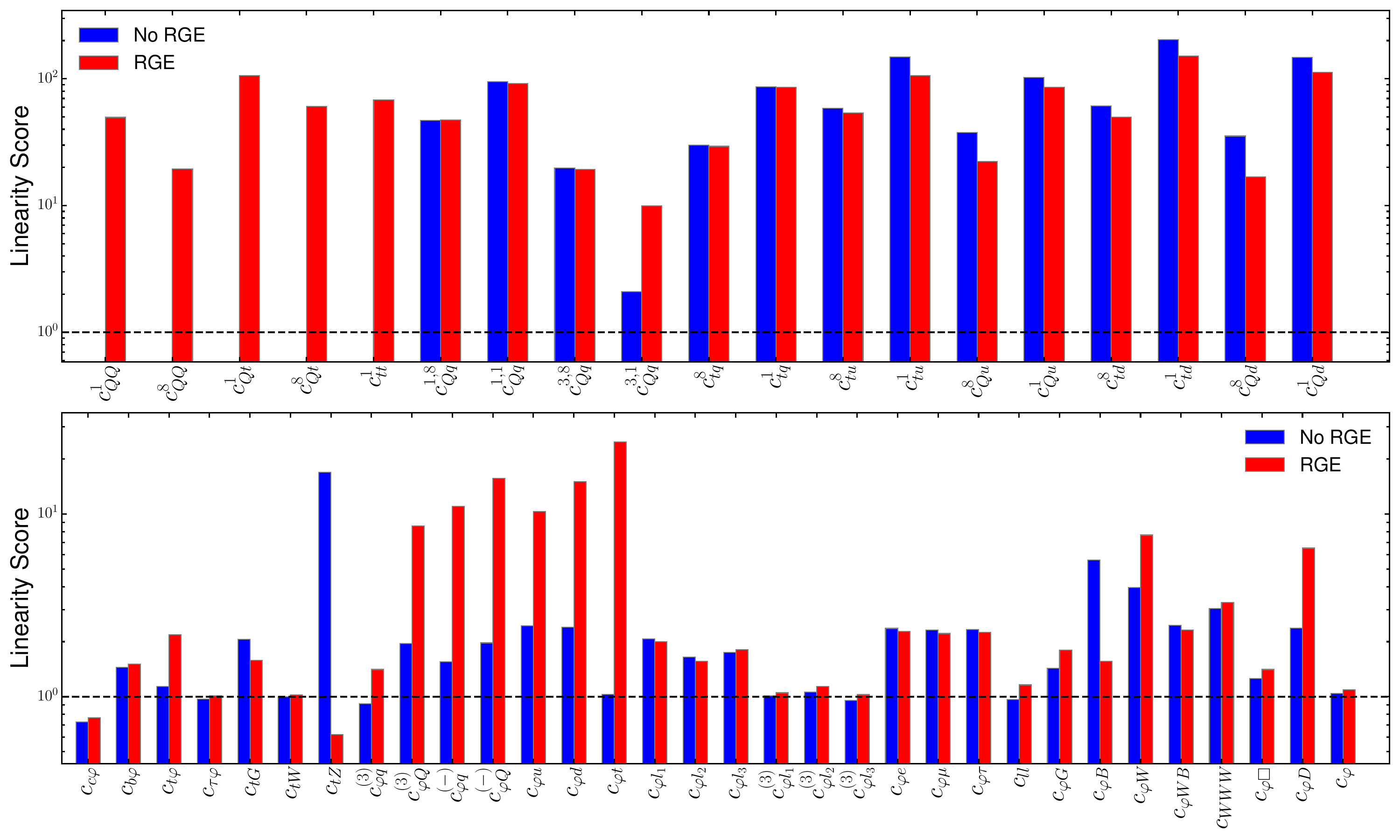}
    \caption{Linearity scores, defined as the ratio of the 95\% C.I.  marginalised bound on a given Wilson coefficient between the linear fits and the quadratic ones. 
    A value of $\mathcal{O}(1)$ indicates that for this operator the linear EFT contributions dominate.
    The horizontal dashed line indicates a linearity score of 1.
    }
    \label{fig:linear_vs_quad_scores}
\end{figure}

\paragraph{Linearity scores.}
Fig.~\ref{fig:linear_vs_quad_scores} displays the linearity scores, defined as the ratio of widths of the 95\% C.I. marginalised intervals of a given Wilson coefficient between the linear fits and the quadratic ones, comparing fits both with and without RGE effects.  
A linearity score of $\mathcal{O}(1)$ indicates that the linear EFT contributions dominate for that operator.\footnote{We note that a linearity score which is significantly smaller than 1 is a sign of a double minima structure and occurs in cases where the data cannot distinguish between the two solutions. 
One such example is the charm Yukawa operator, which is explained from the fact that our observables are insensitive to the sign of the charm Yukawa coupling. }

The figure illustrates the impact of including RGE effects on this metric, assessing whether the enhanced theoretical precision brings us closer to or farther from the regime where quadratic corrections can be neglected.  
For four-heavy operators and two-light-two-heavy four-fermion operators, we find that the constraints remain dominated by quadratic corrections.  
However, there are two notable cases where RGEs shift the behavior of the fits.  

Firstly, the coefficient $c_{tZ}$, which without RGE effects is entirely within the quadratic regime, moves towards the linear regime once RGEs are included.  
In fact, we even observe that the quadratic bound is slightly larger than the linear one, though this is merely an artifact of a residual double solution that slightly inflates the quadratic bound.  
On the other hand, a striking effect occurs for the current operators involving quarks, both right-handed and left-handed.  
While these operators were initially in the linear regime, the inclusion of RGEs shifts them towards a quadratic-dominated regime.  
This can be attributed to the loss of precision in the linear fits, where the interplay with operators running to the $Z$ pole, such as the four-heavy ones, dilutes the overall sensitivity.
   
%%%%%%%%%%%%%%%%%%%%%%%%%%%%%%%%%%%%%%%%%%%%%%%%%%%
%%%%%%%%%%%%%%%%%%%%%%%%%%%%%%%%%%%%%%%%%%%%%%%%%%%%%

\paragraph{Fisher information analysis.}
As is customary in SMEFT analyses, employing the Fisher information framework~\cite{Ellis:2020unq,Ethier:2021bye}, evaluated at either the linear or quadratic level in the EFT expansion, provides valuable insight into the sensitivity of different Wilson coefficients to the datasets included in the fit.
In our specific analysis, the Fisher information matrix is computed at the reference scale $\mu_0$ using Eq.~(\ref{eq:theory_EFT_3}). This allows us to assess how the inclusion of RGE effects impacts sensitivity within the fit.

\begin{figure}[htbp] % Allow placement at the top of the page or on a float page
    \centering
    \begin{subfigure}[t!]{0.43\textwidth}
        \centering
        \includegraphics[width=0.8\textwidth]{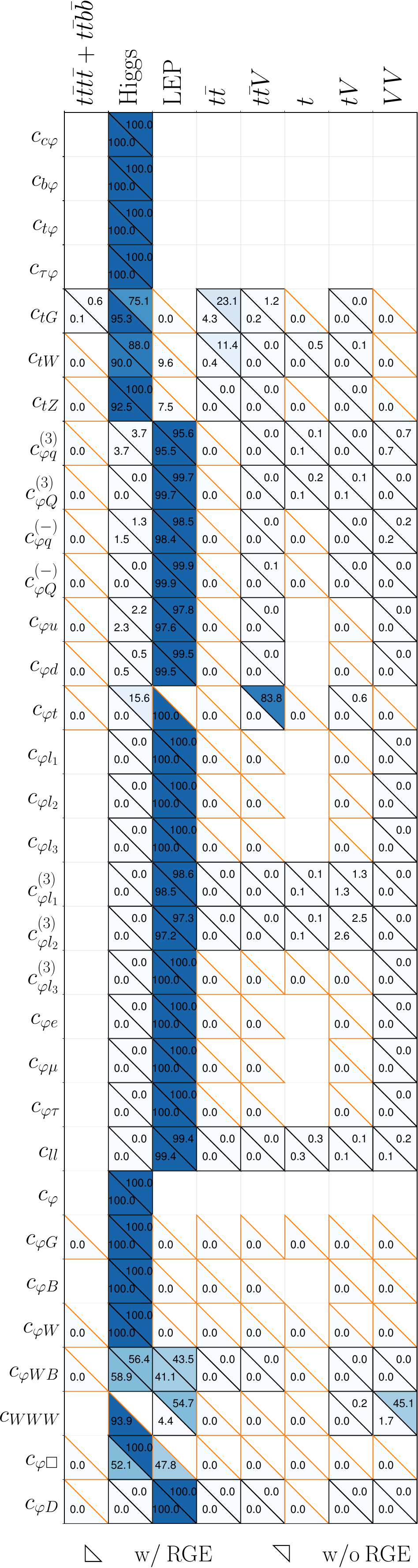}
        % \caption{Description of the left figure.}
        \label{fig:subfig1}
    \end{subfigure}
    \hfill
    \begin{subfigure}[t!]{0.43\textwidth}
        \centering
        \includegraphics[width=\textwidth]{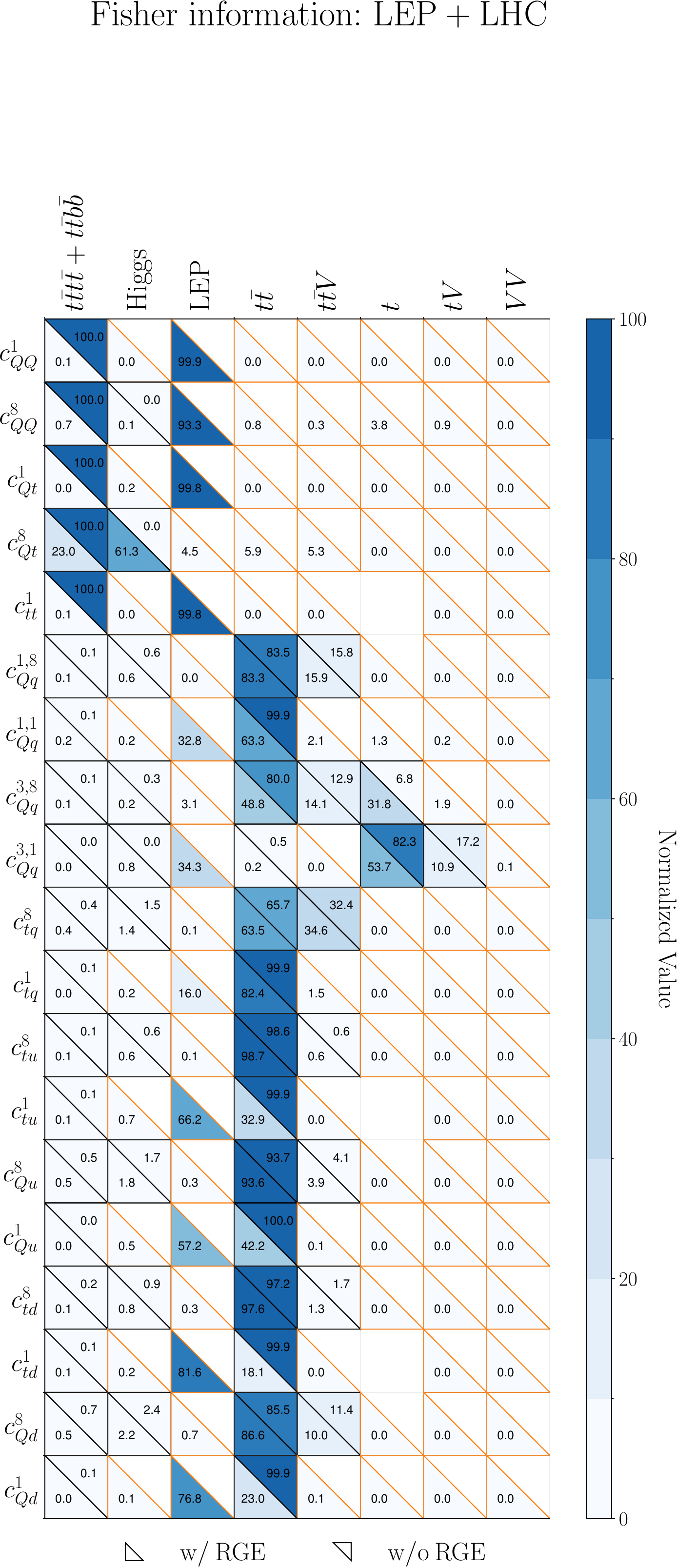}
        % \caption{Description of the right figure.}
        \label{fig:subfig2}
    \end{subfigure}
    \caption{
    Left: diagonal entries of the linear Fisher information matrix normalised (to 100) by row for the 2-fermion, 4-lepton, and purely bosonic dimension-six  SMEFT operators evaluated at $\mu_0 = 5$ TeV. 
    For each coefficient, the higher the value of an entry, the higher its sensitivity to the corresponding dataset. 
    For each entry, the bottom left (upper right) triangles denote the value of the Fisher information with (without) RG effects accounted for in the theory calculations. Right: same, now for the 4-heavy and 2-light-2-heavy operators. 
    The coloured entries label coefficients for which the sensitivity enters exclusively due to RGE effects.
    }
    \label{fig:fisher_rge}
\end{figure}

To illustrate this, Fig.~\ref{fig:fisher_rge} presents the diagonal elements of the Fisher information matrix, evaluated at linear order in the EFT expansion, both with and without RGE effects.
For each Wilson coefficient, we indicate its relative sensitivity to the different process categories in our fit, normalizing the values by row to a maximum of 100.
The lower-left (upper-right) triangle corresponds to the fits where RGE effects are (are not) included in the calculation of physical observables.
The left panel displays the two-fermion, four-lepton, and purely bosonic operators, while the right panel shows the four-heavy and two-light-two-heavy operators.
Wilson coefficients for which sensitivity to a given dataset arises entirely from RGE effects are highlighted in orange.

The Fisher information analysis in Fig.~\ref{fig:fisher_rge} provides a quantitative basis for validating several key findings of the linear EFT fit.
First, we confirm that once RGEs are included, the four-heavy operators are primarily constrained by LEP and Higgs measurements.
Their impact on various processes removes flat directions, allowing them to be constrained, albeit still weakly, at the linear level.
Similarly, in the absence of RGEs, the two-light-two-heavy operators derive the majority of their sensitivity from $t\bar{t}$ production. However, once RGE effects are included, a subset of these operators, such as $c_{td}^{1}$ and $c_{Qd}^1$, gains up to 80\% of their Fisher information from LEP’s EWPOs due to their contributions to weak boson decays. These colour singlet operators are poorly constrained at the linear level as they do not interfere with the leading contributions to the $t\bar{t}$ process.  

Additional notable features from the Fisher analysis include:
(i) an increased role of Higgs measurements in constraining $c_{tG}$ in the presence of RGEs,
(ii) a shift in the dominant constraint on $c_{\varphi t}$ from Higgs and $t\bar{t}V$ data (in the non-RGE case) to LEP’s EWPOs via its running into $c_{\varphi D}$, and
(iii) a change in the primary sensitivity to the triple-gauge operator $c_{WWW}$, which transitions from diboson production to Higgs measurements when RGEs are included.

\begin{figure}[t!]
    \centering
    \includegraphics[width=0.35\linewidth]{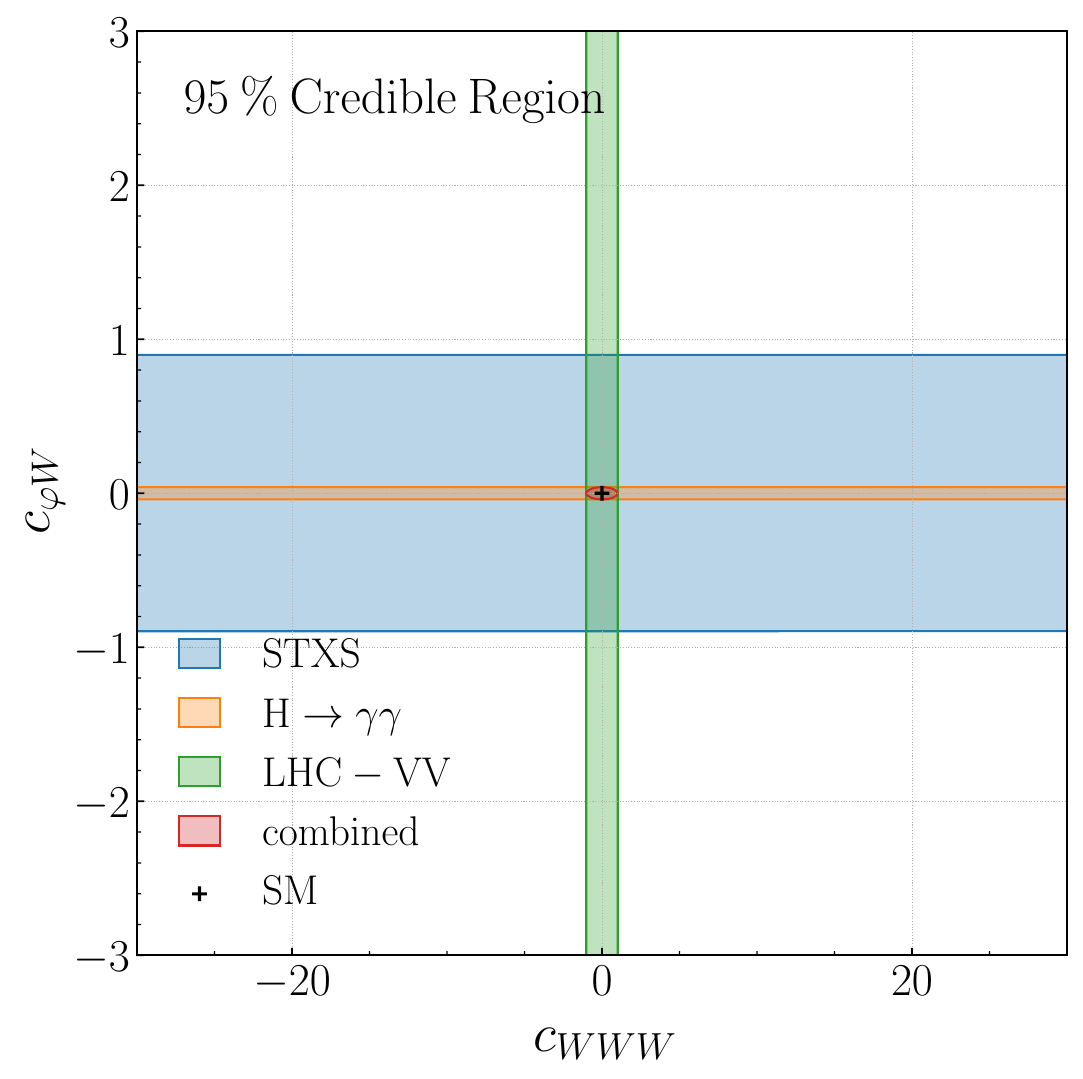}
    \includegraphics[width=0.35\linewidth]{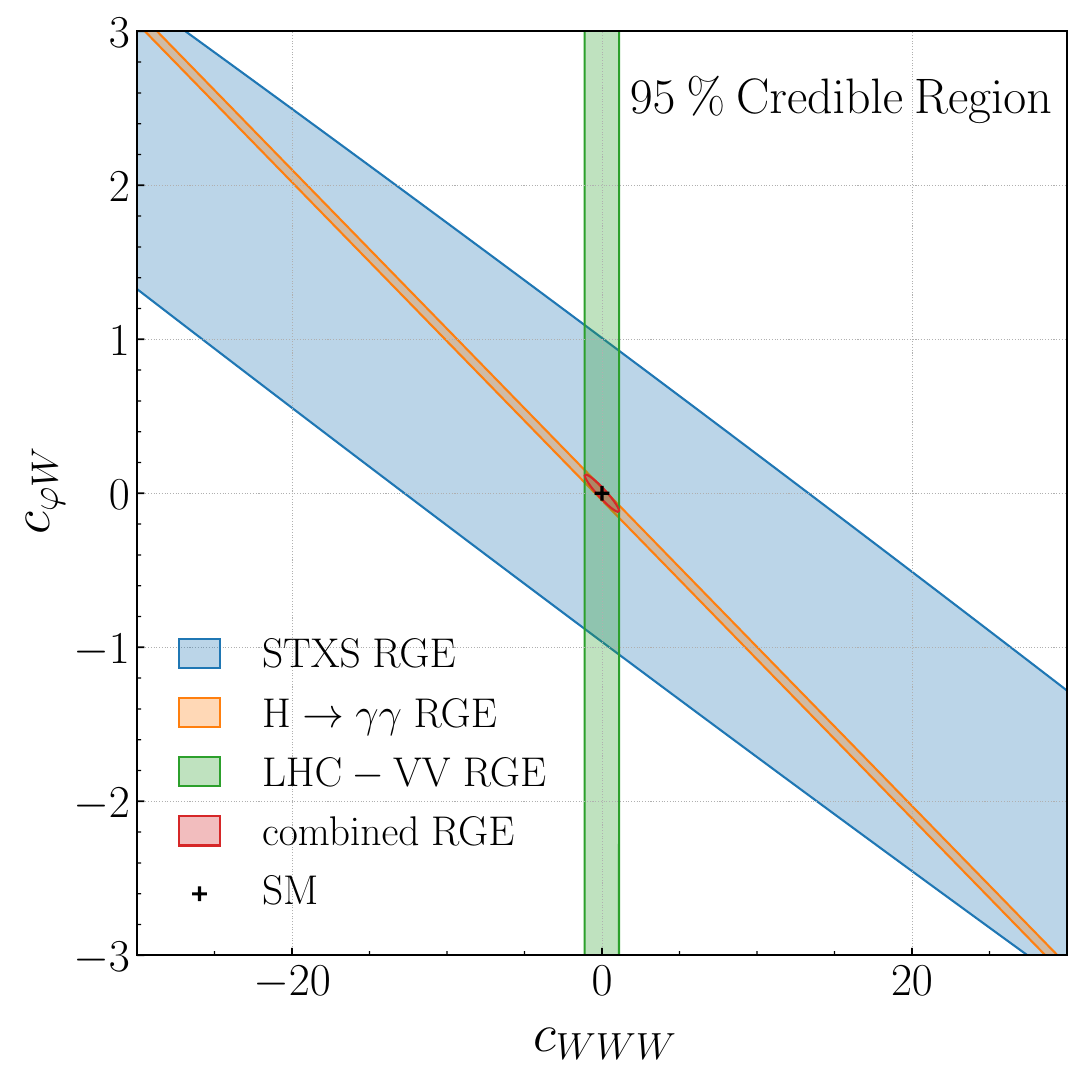}\\
    \caption{Same as Fig.~\ref{fig:2D_rge_fits}, now for $\lp c_{WWW},c_{\varphi W}\rp$,
    comparing separately the impact of ATLAS STXS, $H\to \gamma\gamma$, and LHC diboson datasets with that of their combination.
    }
    \label{fig:cWWW-2D}
\end{figure}

This last point is particularly intriguing. To understand the underlying mechanism, we present in Fig.~\ref{fig:cWWW-2D} a two-dimensional fit of $c_{WWW}$ and $c_{\varphi W}$, both with and without RGE effects, while breaking down contributions from different datasets.
Without RGE effects, the $c_{WWW}$ coefficient is entirely constrained by diboson production at the LHC, with no sensitivity from Higgs data.
However, when RGEs are included, the triple-gauge operator runs into $\mathcal{O}_{\varphi W}$ and $\mathcal{O}_{\varphi W B}$, making it sensitive to Higgs decays into $\gamma \gamma$ and STXS measurements.
Despite this, in a simultaneous fit, the strong correlation between these operators prevents Higgs data from significantly impacting the marginalized bound, explaining why the posterior distributions in Fig.~\ref{fig:posterior_rg_vs_no_rg_linear} remain largely unchanged when including RGEs.
Nevertheless, in an individual fit, the $c_{WWW}$ coefficient would be primarily constrained by Higgs data, which explains why the diagonal elements of the Fisher information matrix in Fig.~\ref{fig:fisher_rge} indicate that the dominant sensitivity to $c_{WWW}$ arises from Higgs measurements.

%%%%%%%%%%%%%%%%%%%%%%%%%%%%%%
%%%%%%%%%%%%%%%%%%%%%%%%%%%%%%%%

%%%%%%%%%%%%%%%%%%%%%%%%%%%%%%%%%%%%%%%%%%%%%%%%%%%%%%%%%%%%%%%%%
%%%%%%%%%%%%%%%%%%%%%%%%%%%%%%%%%%%%%%%%%%%%%%%%%%%%%%%%%%%%%%%%%%%

One possible limitation of the Fisher information matrix as displayed in Fig.~\ref{fig:fisher_rge} is that it ignores the correlations between different Wilson coefficients since it considers only the diagonal entries.
When provided in such a format, the information contained in the Fisher matrix indicates the sensitivity to a given operator in a linear individual one-parameter fit, which in general will be different to that obtained in the global fit upon marginalisation.
Ascertaining the role played by the theory-induced correlations between Wilson coefficients is especially important in the presence of RGEs, which connect among them many operators which were previously decoupled.

This consideration motivates the introduction of a modified Fisher information heat map where the data is integrated out, illustrating how new sensitivities on the Wilson coefficients emerge in the operator space at the observable scale, rather than in the data space. 
Neglecting the effects of RGEs and at the linear level in the EFT expansion, the Fisher information matrix is given by \cite{Ethier:2021bye}:
\begin{equation}
\label{eq:fisher_matrix_linear}
    F_{ii'} = \sum_{j,j'=1}^{n_{\rm dat}}\kappa_{i,j} \, \Sigma^{-1}_{jj^\prime} \, \kappa_{i^\prime, j^\prime} \, , \qquad i,i'=1,\ldots,n_{\rm op}\, ,
\end{equation}
where $\Sigma$ denotes the total covariance matrix including theory and experimental uncertainties, and $\kappa_{i,j}$ represents the EFT linear cross-section for the $i$-th Wilson coefficient and the  $j$-th cross section entering the fit, see Eq.~(\ref{eq:theory_EFT_1}) for its definition. 
The Fisher information matrix $F_{ii'}$, Eq.~(\ref{eq:fisher_matrix_linear}), can be understood as the inverse covariance matrix in the Wilson coefficient space. 
To see this, note that its diagonal elements provide the uncertainties associated with each Wilson coefficient, $\delta {\boldsymbol{c}}$,  as determined from the considered dataset in individual one-parameter fits:
\begin{equation}
    \text{diag}(F) = \left( \frac{1}{(\delta c_1)^2}, \dots, \frac{1}{(\delta c_{n_{\rm op}})^2} \right) \, .
    \label{eq:new_fisher_1}
\end{equation}

When the effects of RGEs are accounted for in the theoretical calculations entering the fit, Eqns.~(\ref{eq:theory_EFT_2})--(\ref{eq:theory_EFT_3}), and assuming a common scale $\mu$ for all the fitted observables, the Fisher information matrix of Eq.~(\ref{eq:fisher_matrix_linear}) is replaced by
\begin{equation}
    F^{\text{RGE}}_{kk'} = \sum_{i,i'=1}^{n_{\rm op}}\sum_{j,j'=1}^{n_{\rm dat}}\Gamma_{ki} \, \kappa_{i,j} \, \Sigma^{-1}_{jj^\prime} \, \kappa_{i^\prime, j^\prime} \, \Gamma_{k^\prime i^\prime} = \sum_{i,i'=1}^{n_{\rm op}} \Gamma_{ki} \, F_{ii^\prime} \, \Gamma_{k^\prime i^\prime} \, , \qquad k,k'=1,\ldots,n_{\rm op}\, ,
    \label{eq:rge_fisher_fixed_scale}
\end{equation}
where the second step explicitly demonstrates that the total covariance matrix can be integrated out. 
If we disregard the off-diagonal elements of the $F$ matrix (setting $i=i'$ in the sum), focusing exclusively on individual fits of the Wilson coefficients at the observable scale, and examine the diagonal elements of $F^{\text{RGE}}$, we find:
\begin{equation}
    \text{diag}(F^{\text{RGE}}) = \left( \frac{\Gamma_{11}^2}{(\delta c_1)^2} + \cdots + \frac{\Gamma_{1n_{\rm op}}^2}{(\delta c_{n_{\rm op}})^2}, \dots, \frac{\Gamma_{n_{\rm op}1}^2}{(\delta c_1)^2} + \cdots + \frac{\Gamma_{n_{\rm op}n_{\rm op}}^2}{(\delta c_{n_{\rm op}})^2} \right) \, ,
\end{equation}
which allows us to decompose, for each Wilson coefficient at the reference scale, the contribution of information originating from the various Wilson coefficients as these are probed at the observable scale.

%%%%%%%%%%%%%%%%%%%%%%%%%%%%
\begin{figure}[htbp]
    \centering
    \includegraphics[width=0.84\linewidth]{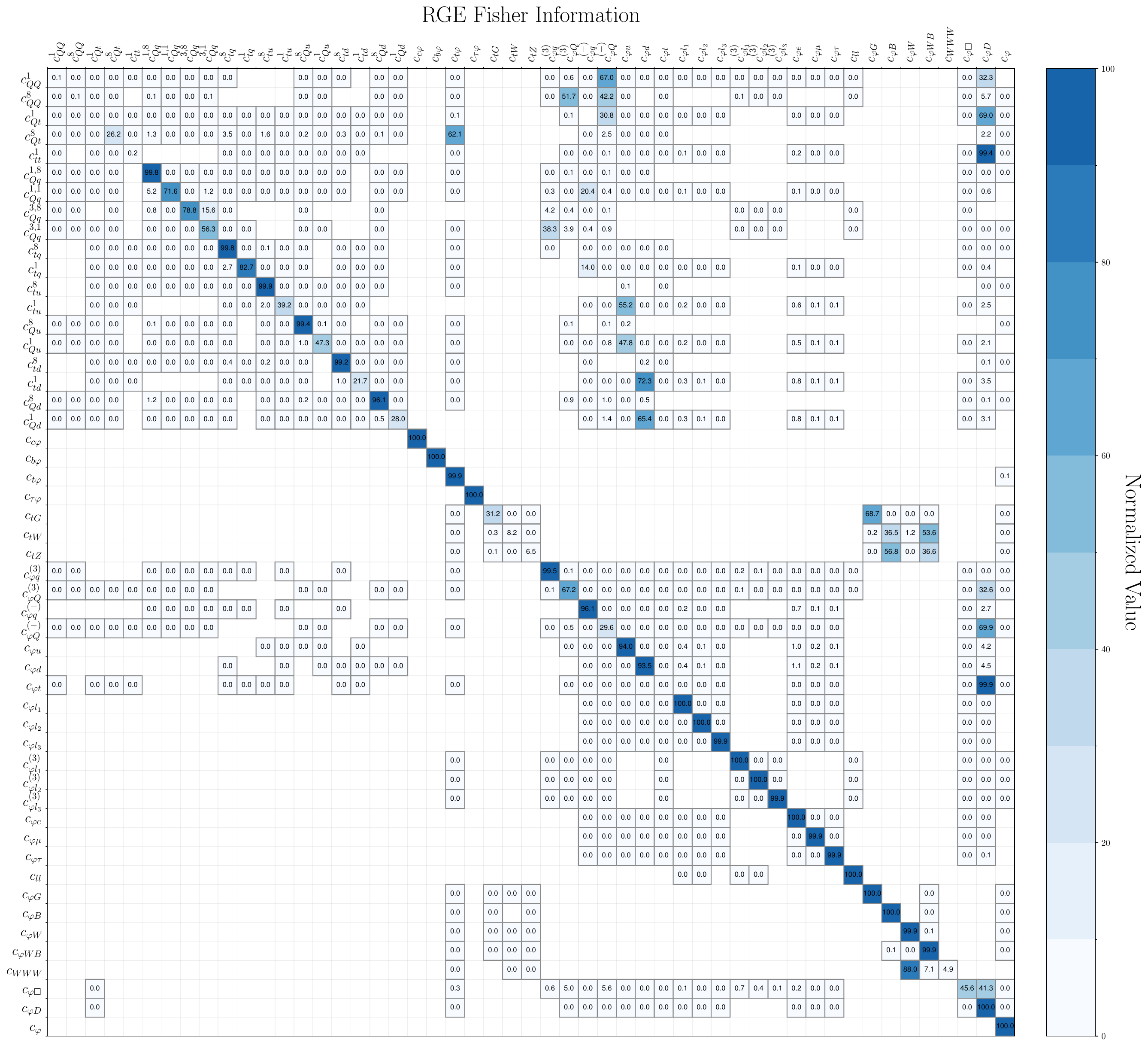}\hfill
    \caption{The Fisher information matrix, $F^{\rm RGE}_{k}$, 
    in the space of Wilson coefficients, Eq.~(\ref{eq:new_fisher_3}).
    Each of the rows is normalised to 100.
    The larger the value of a given entry $(k, i)$,
    the stronger the correlation between the $k$-th operator at the initial scale $\mu_0=5$ TeV and the $i$-th operator at the observable scale $\mu$.
    }
    \label{fig:RGE_fisher_lhc}
\end{figure}
%%%%%%%%%%%%%%%%%%%%%%%%%%%%

When considering the dynamic scale setup, namely the association of a different scale $\mu$ for each of the data bins included in the fit, the RGE matrix $\Gamma$ now depends on the data point.
In this case, the Fisher information matrix
Eq.~\eqref{eq:rge_fisher_fixed_scale} is replaced by:
\begin{equation}
     F^{\text{RGE}}_{kk'} = \sum_{i,i'=1}^{n_{\rm op}}\sum_{j,j'=1}^{n_{\rm dat}}\Gamma^j_{ki} \, \kappa_{i,j} \, \Sigma^{-1}_{jj^\prime} \, \kappa_{i^\prime, j^\prime} \, \Gamma^{j^\prime}_{k^\prime i^\prime} \, , \qquad k,k'=1,\ldots,n_{\rm op}\, ,
     \label{eq:new_fisher_2}
\end{equation}
and due to this dependence the summation over the data space cannot be carried out as an intermediate step. Consequently, the diagonal component of the Fisher information matrix (where off-diagonal elements at the observable scale are neglected) is expressed as
\begin{equation}
    F^{\text{RGE}}_{k} = \sum_{i=1}^{n_{\rm op}} \sum_{j,j'=1}^{n_{\rm dat}} \Gamma^j_{ki} \, \kappa_{i,j} \, \Sigma^{-1}_{jj^\prime} \, \kappa_{i,j^\prime} \, \Gamma^{j^\prime}_{ki} \equiv \sum_{i=1}^{n_{\rm op}} f_k^i \, ,\qquad k=1,\ldots,n_{\rm op}\, ,
    \label{eq:new_fisher_3}
\end{equation}
where $F^{\text{RGE}}_{k}$ quantifies the Fisher information associated with the $k$-th Wilson coefficient at the reference scale. 
This formulation enables a detailed breakdown of the contributing information $f_k^i$ associated with the $i$-th Wilson coefficient at the observable scale.

Fig.~\ref{fig:RGE_fisher_lhc} displays the breakdown of the reduced diagonal part of Fisher information matrix, $F^{\rm RGE}_{k}$,  in the space of Wilson coefficients given by Eq.~(\ref{eq:new_fisher_3}).
Each of the rows displaying the $f_k^i$ is normalised to 100.
As discussed above, for a given matrix entry $(k,i)$, a larger value indicates a stronger information transfer from the  i-th operator at the observable scale $\mu$ to the k-th operator at the reference scale $\mu_0$.
Those operators whose diagonal entries are less than 100 correspond to operators for which effects due to running and mixing from the reference scale $\mu_0$ down to the observable scale $\mu$ are sizeable and hence induce a decrease in the auto-correlation.
Since in the absence of RGEs one would have a trivial identity matrix, deviations of $F^{\rm RGE}$ from $\mathbb{I}$ quantify the impact of RGEs in terms of connections between the Wilson coefficients at different scales.

The information provided by Fig.~\ref{fig:RGE_fisher_lhc} is complementary, and well aligned, with that of  Fig.~\ref{fig:fisher_rge}.
For instance, we observe that information on the four-heavy operators originates from the RGE induced coefficients $c_{\varphi D}$ and $c_{\varphi Q}^{(-)}$. These two operators are probed by EWPOs which explains also the patterns seen in Fig. \ref{fig:fisher_rge}. 

Likewise, some of the two-light-two-heavy operators, such as $c_{Qq}^{3,1}$, are constrained through their running into some of the EWPO operators, such as $c_{\varphi q}^{(3)}$.
Other operator pairs connected in a significant manner through the RGEs in Fig.~\ref{fig:RGE_fisher_lhc} are $(c_{tG},c_{\varphi G})$, with $\mathcal{O}_{tG}$ running into $\mathcal{O}_{\varphi G}$  leading to an enhanced sensitivity to Higgs data in the presence of RGE effects, and $(c_{WWW},c_{\varphi W})$, consistent with the observation from Fig.~\ref{fig:fisher_rge} that the triple gauge operator in the presence of RGEs, instead of via diboson production, becomes mostly constrained from Higgs production observables in individual fits.

\subsection{Impact of scale choice}
\label{sec:impact_scale_choice}

As discussed in Sect.~\ref{sec:settings}, for each physical observable entering the fit, we associate a characteristic energy scale $\mu$ when evaluating theoretical predictions in the SMEFT, as given in Eq.~(\ref{eq:theory_EFT_2}).  
Our chosen scales for all processes included in the global analysis are summarized in Table~\ref{tab:scale_definitions}.  
However, this choice is not unique. Similar to the renormalization $\mu_R$ and factorization $\mu_F$ scales in QCD calculations, it is crucial to assess the stability of the fit results under variations of $\mu$.  
Such a check is facilitated by the {\sc\small SMEFiT} functionalities (see App.~\ref{app:implementation}), which allow for a constant rescaling of the central scales associated with physical observables, $\mu \to \tilde{\mu} = \kappa \mu$, where $\kappa$ is a positive real number.  
Motivated by this, we have repeated the linear and quadratic fits presented in Sect.~\ref{subsec:results} for different values of $\kappa$, specifically $\kappa = 0.5$ and $\kappa = 2$.
The effects of scale variations at the level of Wilson coefficients can be quantified by computing the envelope of half the length of the 95\% C.I., $[c_i^{\rm (sc,min)},c_i^{\rm (sc,max)}]$,  defined as
\be 
c_i^{\rm (sc, min)} = \min_{\kappa} {\rm C.I.}^{\rm (95, halfwidth)}(\kappa)\, ,~ c_i^{\rm (sc, max)} = \max_{\kappa} {\rm C.I.}^{\rm (95, halfwidth)}(\kappa), \qquad \kappa\in\{0.5, 1.0, 2.0\} \, .
\label{eq:envelope_scales}
\ee
Note\footnote{We display half the length of the 95\% C.I., as this corresponds to a 2$\sigma$ deviation in the case of Gaussian statistics.} that the envelope defined this way will not be necessarily symmetric with respect to the central fit with $\kappa=1$.

\begin{figure}[htbp!]
    \centering
\includegraphics[width=0.77\linewidth]{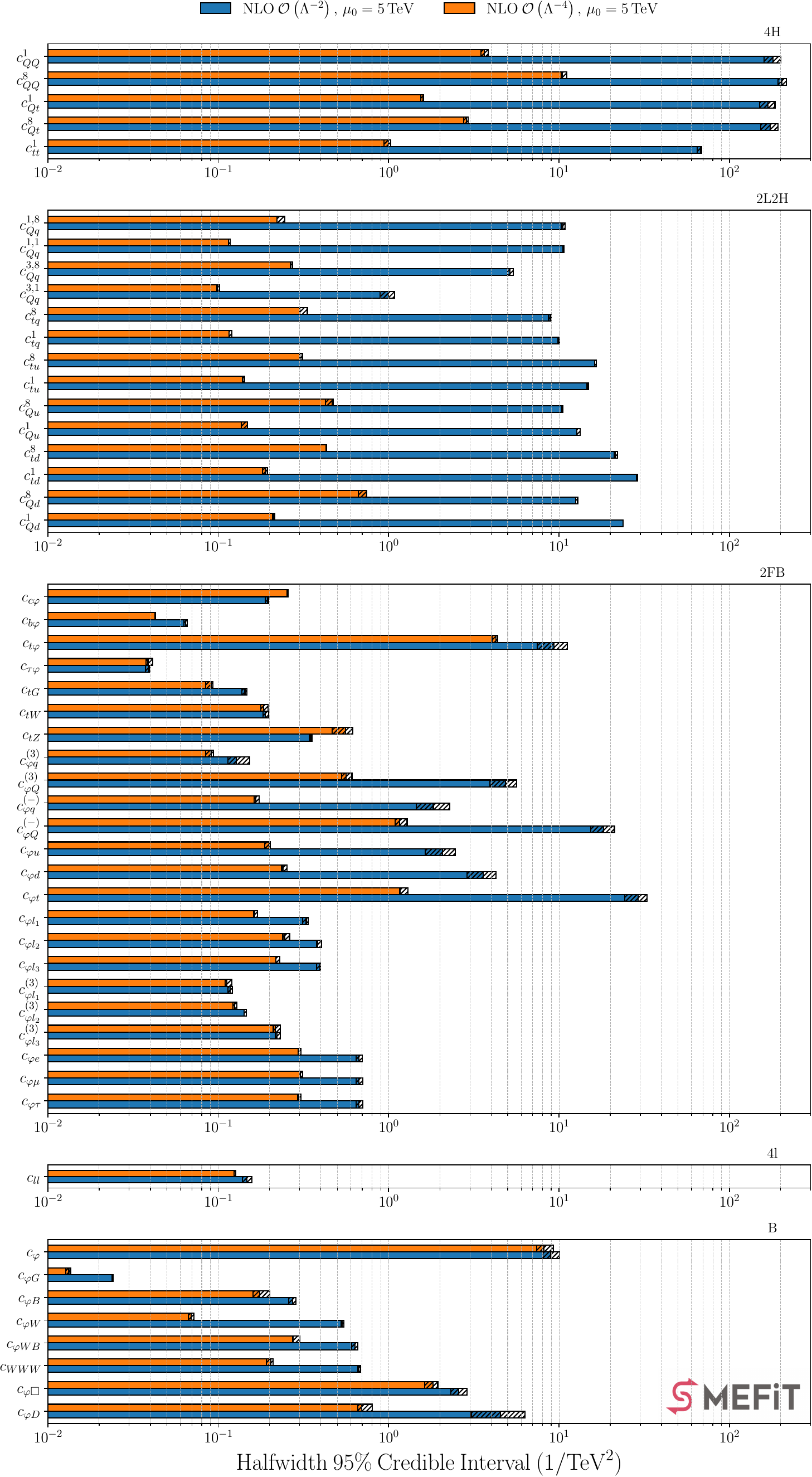}\hfill
    \caption{Half the width of the marginalised 95\% C.I on the same Wilson
    coefficients shown in Fig.~\ref{fig:posterior_rg_vs_no_rg_linear}
    for the linear and quadratic EFT fits.
    The hatched regions indicate the variation of the results whenever the central scale $\mu$ assigned to the fitted observables is varied by a factor two and defined by the envelope in Eq.~(\ref{eq:envelope_scales}).}
    \label{fig:scale_variation_rge}
\end{figure}

Fig.~\ref{fig:scale_variation_rge} presents the 95\% C.I. width on the same Wilson coefficients as in Fig.~\ref{fig:posterior_rg_vs_no_rg_linear}, comparing results from the linear and quadratic EFT fits. In addition, hatched regions indicate the impact of varying the central scale $\mu$ assigned to the fitted observables by a factor of two, as defined by the envelope in Eq.~(\ref{eq:envelope_scales}).
Overall, the effect of scale variations is mild, typically resulting in differences below the few-percent level for most operators. As expected, the impact is more pronounced for coefficients that are weakly constrained, making scale variations particularly relevant for linear fits.
Larger effects are observed for operators contributing to EWPOs at tree level in the linear fit, where scale variations can introduce uncertainties of up to 45\% in the case of $c_{\varphi q}^{(-)}$. However, this estimate is likely conservative, as the hard scale associated with EWPOs is generally well determined, especially for $Z$-pole measurements, making a smaller variation physically more justified.

Nonetheless, the analysis of Fig.~\ref{fig:scale_variation_rge}  underscores that scale uncertainties should ideally be accounted for. They represent a significant source of uncertainty in global EFT fits, particularly when interpreting potential deviations between experimental data and SM predictions, as well as when matching the SMEFT to specific UV-complete models.

%%%%%%%%%%%%%%%%%%%%%%%%%%%%%%%%
%%%%%%%%%%%%%%%%%%%%%%%%%%%%%%%%

\section{RGE effects in the SMEFT at future colliders}
\label{sec:uv_fcc}

As discussed in the introduction, accounting for RGE effects is known to play a particularly prominent role in the context of the SMEFT interpretation of Higgs, top quark, and electroweak measurements from future lepton colliders~\cite{Allwicher:2024sso,Gargalionis:2024jaw}, both at the level of Wilson coefficients and in terms of the parameter space of UV-complete models.
The reason is two-fold.
First, because of the large gap between the scale at which measurements are taken and the UV mass scale upon which these precise measurements are sensitive.
Second, as is generally true, SMEFT interpretations cannot directly constrain the absolute mass scale $\Lambda$ where the heavy UV particles live, but only the ratios ${\boldsymbol{c}}/\Lambda^{2}$.
Once RGEs are accounted for, we can extract $\boldsymbol{c}(\mu_0)/\Lambda^{2}$ and carry out a matching to UV models where all relevant mass scales are consistently defined. 

In this section, we quantify the role played by RGE effects when deriving constraints on the EFT coefficients, using projected data in the specific case of the FCC-ee. We analyse the resulting impact on the reach on masses and couplings of UV-complete models matched to the SMEFT, focusing on 
representative benchmark scenarios.
First, we revisit the SMEFT fits with FCC-ee projections presented in~\cite{Celada:2024mcf} in light of the new implementation of RGE effects in \smefit.
We then consider their impact on representative single-particle extensions of the SM probed by precision measurements at the FCC-ee, where we also study the role of theory errors and determine the relative sensitivity of the different FCC-ee runs.

Following the analysis of~\cite{Celada:2024mcf}, all observables included in the fits shown in this section, including the LHC and LEP ones, are generated from Level-0 closure test projections assuming the SM as the underlying law.
Unless said otherwise, we include current SM theory uncertainties for all observables where they are available (see App.~\ref{app:implementation} and~\cite{Celada:2024mcf} for details).
Hence we will not display results for central values of these fits, which coincide with the SM expectation by construction, and consider only the impact at the level of the uncertainties on the Wilson coefficients.

%%%%%%%%%%%%%%%%%

\subsection{RGE impact at the FCC-ee}
\label{eq:reg_futurecolliders}

First we revisit the results of~\cite{Celada:2024mcf} for the global SMEFT fit including both HL-LHC and FCC-ee (Level-0) projections now accounting for RGE effects.
The input observables and the associated theory calculations are the same as in~\cite{Celada:2024mcf} except for the following important updates.
We include dedicated HL-LHC projections on $t\bar{t}$-production differential in the invariant mass of the top pair, as well as the $t\bar{t}$ charge asymmetries, both taken from \cite{Durieux:2022cvf}. In addition, we add the inclusive signal strength projection on Higgs pair production at HL-LHC from \cite{Cepeda:2019klc}. Regarding projections for the FCC-ee, we include the complete NLO electroweak corrections to $e^+e^-\to Zh$ production in the SMEFT from dimension-six operators as presented in~\cite{Asteriadis:2024xts}. 
These EW loop effects in single Higgs production at the FCC-ee introduce a new sensitivity to the Higgs self-coupling operator $c_\varphi$ which is complementary to that arising from Higgs pair production measurements at the HL-LHC, FCC-hh, and in the high-energy runs of linear colliders. 

\begin{figure}[htp]
    \centering
    \includegraphics[width=0.76\linewidth]{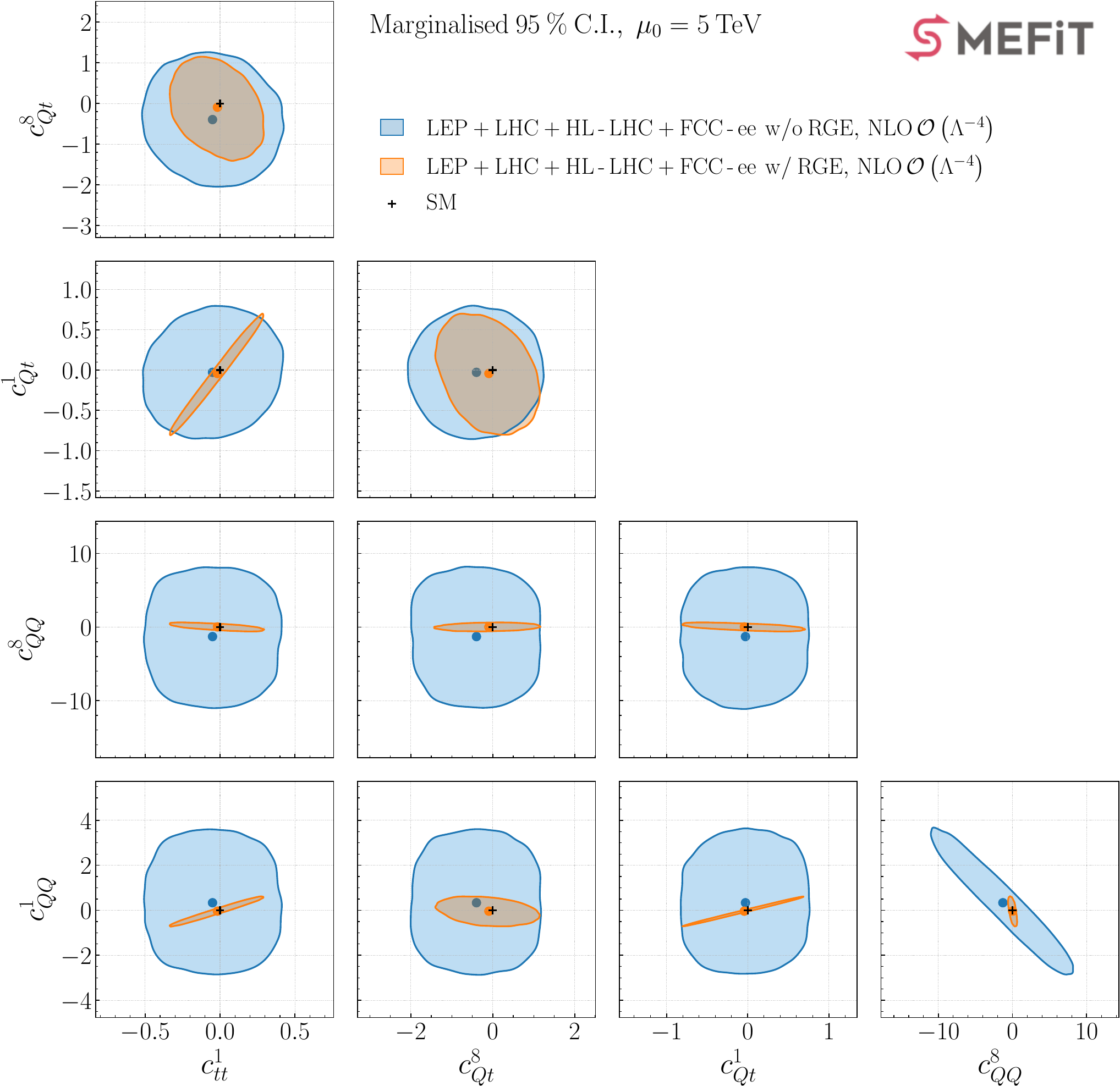}
    \caption{95\% C.I. evaluated at $\mu_{0}=5$ TeV in a quadratic SMEFT fit where the fitting basis is restricted to the four-heavy quark operators $c^1_{Qt}$, $c^8_{Qt}$,  $c^1_{QQ}$, $c^8_{QQ}$, and $c_{tt}^1$. 
     The input dataset is is the same LEP+LHC observables as in Sect.~\ref{sec:results} now complemented with the HL-LHC and FCC-ee projections.
     We compare the outcome of fits based on theory calculations with and without RGE effects. 
   }
    \label{fig:RGE_4H}
\end{figure}

\paragraph{Fits with a restricted operator basis.}
Before showing the global fit results, let us discuss for illustration the results of a fit including HL-LHC and FCC-ee projections in which only a subset of operators are allowed to float. 
This situation is most relevant in the context of the UV-matching results, to be discussed in Sect.~\ref{sec:uvmodels_fccee}, since typically each model activates only a subset of all the possible operators which constitute our basis.
Fig.~\ref{fig:RGE_4H} displays the impact of RGEs in a quadratic SMEFT fit to HL-LHC and FCC-ee projections (in addition to LEP and LHC observables) where the fitting basis is restricted to the four-heavy-quark operators: $c^1_{Qt}$, 
$c^8_{Qt}$,  $c^1_{QQ}$, $c^8_{QQ}$,
and $c_{tt}^1$, with all other coefficients 
set to their SM expectation.
Recall that in the absence of RGEs, these operators are only constrained by the $t\bar{t}t\bar{t}$
and $t\bar{t}b\bar{b}$ cross-sections at the (HL-)LHC.

%%%%%%%%%%%%%%%%%%%%%%%%%%%%%%%%
%%%%%%%%%%%%%%%%%%%%%%%%%%%%%%%%

From the results of Fig.~\ref{fig:RGE_4H}, one observes how the running and mixing enabled by RGEs lead to much more stringent constraints on the four-heavy-quark operators considered due to the breaking of degeneracies and flat directions. 
In this specific case, the sizeable improvement in sensitivity arises due to the running of the four-heavy operator coefficients into Higgs operators that can be constrained by the $ZH$ run of the FCC-ee, as also indicated by the Fisher information analysis displayed in Fig.~\ref{fig:fisher_rge_fcc} below.
This result illustrates the importance of the new sensitivity induced by RGE effects in the SMEFT fit: if new physics induces non-zero four-heavy coefficients, these can be probed through tree-level contributions to four-heavy quark production at the (HL-)LHC, but also at loop-level via the EWPOs and Higgs production at the FCC-ee. 

%%%%%%%%%%%%%%%%%%%%%%%%%%%%%%%%%%
\begin{figure}[t]
    \centering
     \includegraphics[width=0.49\linewidth]{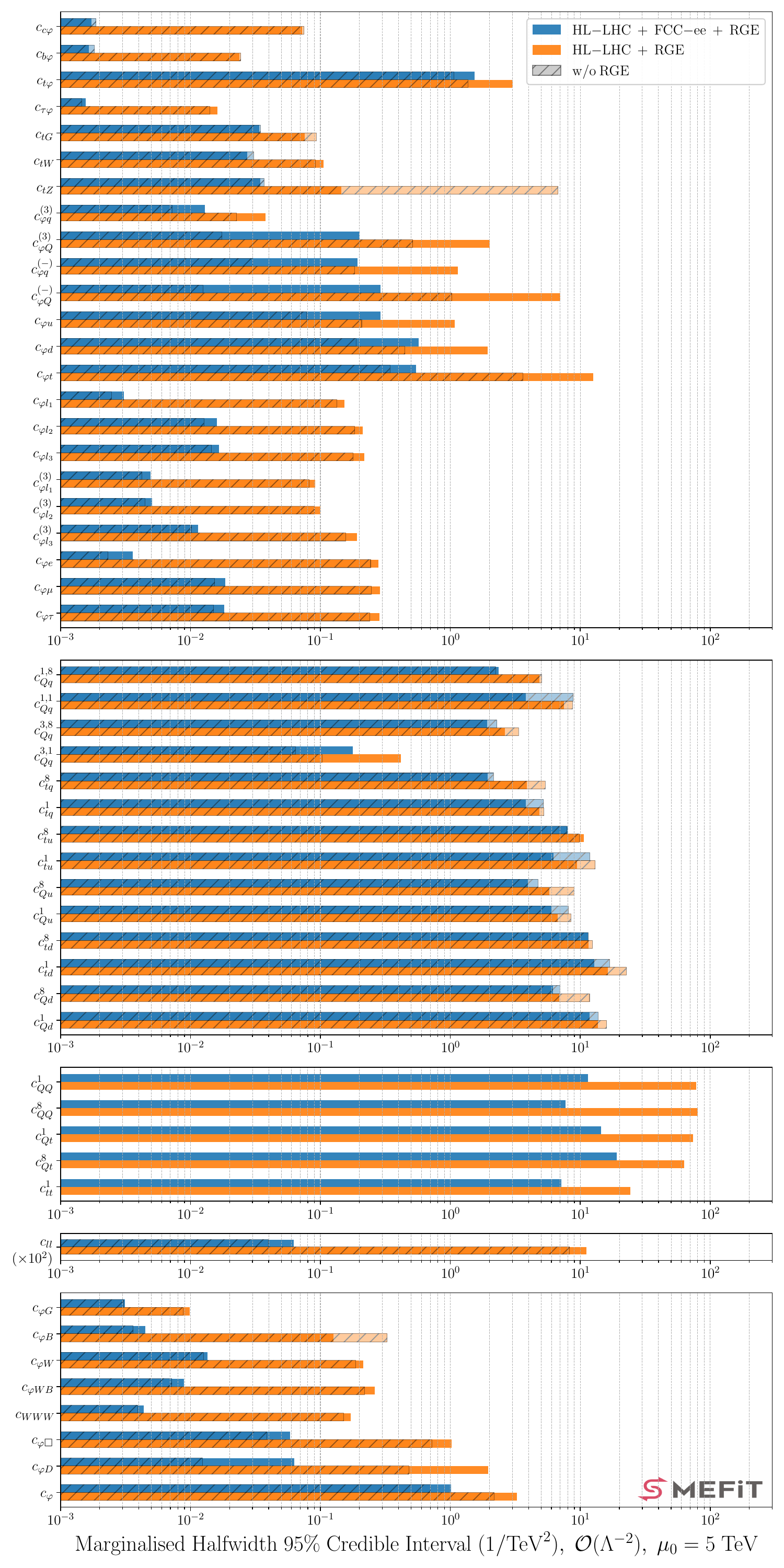}
    \includegraphics[width=0.49\linewidth]{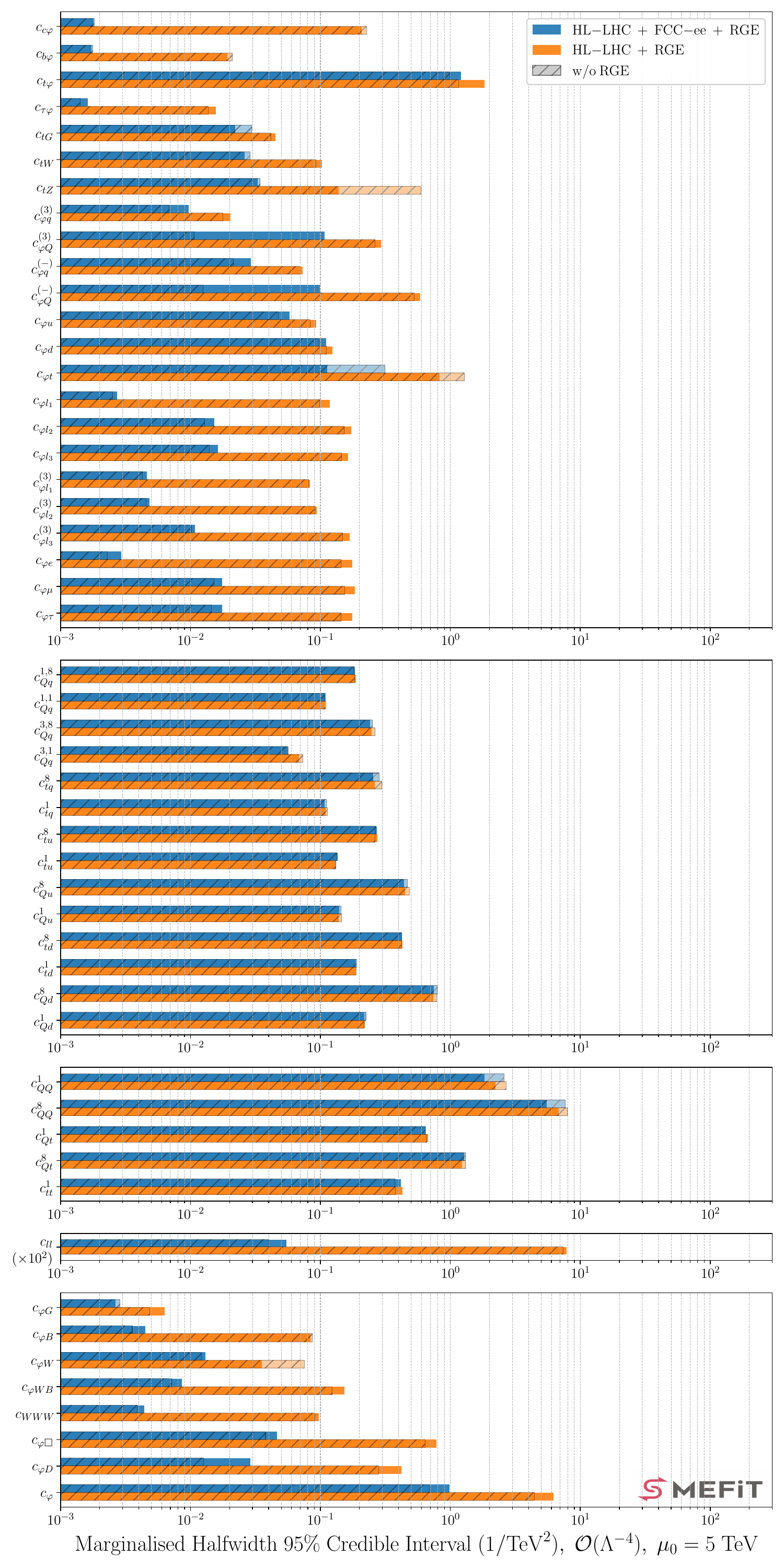}
    \caption{Half of the width of the projected $95\%$ C.I.  at HL-LHC (orange) and HL-LHC + FCC-ee (blue) with and without RGE effects (solid colour and lighter dashed bars respectively). 
    In both cases, the input dataset also includes LEP and LHC Run 2 data as baseline. 
In the left panel, we show the result of a linear fit and in the right panel, the corresponding results from a quadratic fit.
In all cases, we show the bounds for the Wilson coefficients at $\mu_0=5$~TeV.
}
\label{fig:spider_plot_fcc_hllhc_rge}
\end{figure}
%%%%%%%%%%%%%%%%%%%%%%%%%%%%%%%%%%%
\paragraph{Fits to the global dataset.}
We now move to present results of fits based on the global dataset, namely the same LEP+LHC dataset of Sect~\ref{sec:results} extended with the HL-LHC and FCC-ee projections.
Fig.~\ref{fig:spider_plot_fcc_hllhc_rge}
displays the results of linear (left) and quadratic (right panel) SMEFT fits to the full dataset.
We present the magnitude of the bounds, defined as half the total width of the $95\%$ confidence intervals. Note that a logarithmic scale is used for visualization.

In the linear case, we observe a moderate impact of the RGEs for most 4-fermion, 2-lepton and bosonic operators.
$c_{Qq}^{3,1}$, $c_{Qu}^{8}$, $c_{t\varphi}$ and $c_{t Z}$ are the only 4-quark, Yukawa and dipole Wilson coefficients affected noticeably by the RGEs, and the latter only in the case of HL-LHC projections.
Among the bosonic operators, $c_{\varphi B}$ and $c_{\varphi D}$ receive large corrections from the RGEs, although the former only at HL-LHC.
Finally, RGEs generate looser bounds for all 2-quark current operators at both HL-LHC and FCC-ee.
These highlighted operators show a similar behaviour under RGE effects in the LHC fit, as seen in Fig.~\ref{fig:posterior_rg_vs_no_rg_linear}.

In the quadratic fit case, shown in the right panel of Fig.~\ref{fig:spider_plot_fcc_hllhc_rge}, the impact of RGEs is milder as we saw in the LHC fit in Section \ref{sec:results}. 
The 2-quark current operators of the third generation, namely $c_{\varphi Q}^{(-)}$, $c_{\varphi Q}^{(3)}$, and $c_{\varphi t}$ still suffer a significant impact from the RGEs at FCC-ee: the bound is loosened by a factor $\sim 10$ for  $c_{\varphi Q}^{(-)}$ and $c_{\varphi Q}^{(3)}$, while it is tightened by a factor $\sim 3$ for $c_{\varphi t}$.
Two other Wilson coefficients that show large RGE effects are $c_{t Z}$ and $c_{\varphi D}$, which was also seen in the linear case.
The same behaviour is repeated in the LHC case except for $c_{\varphi Q}^{(-)}$ and $c_{\varphi Q}^{(3)}$.

\paragraph{Fisher information and correlation analysis.}
As in Sect.~\ref{sec:results}, we now present a Fisher information and correlation analysis that enables us to shed light on some of the fit results displayed in Fig.~\ref{fig:spider_plot_fcc_hllhc_rge}.
First, Fig.~\ref{fig:fisher_rge_fcc} presents the diagonal entries of the Fisher information matrix, in the same format as in Fig.~\ref{fig:fisher_rge}, now 
including also the contribution of the FCC-ee 
observables.
Comparing the two sets of plots, it is clear that FCC-ee cross-sections, in particular for the Tera-$Z$ run ($\sqrt{s}=91$ GeV) and the $ZH$ run ($\sqrt{s}=240$ GeV) dominate the  overall sensitivity for the majority of the operators included in the fit, especially once RGE effects are accounted for.

%%%%%%%%%%%%%%%%%%%%%%%%%%%%%%
\begin{figure}[htbp] % Allow placement at the top of the page or on a float page
    \centering
    \begin{subfigure}[t!]{0.46\textwidth}
        \centering
        \includegraphics[width=\textwidth]{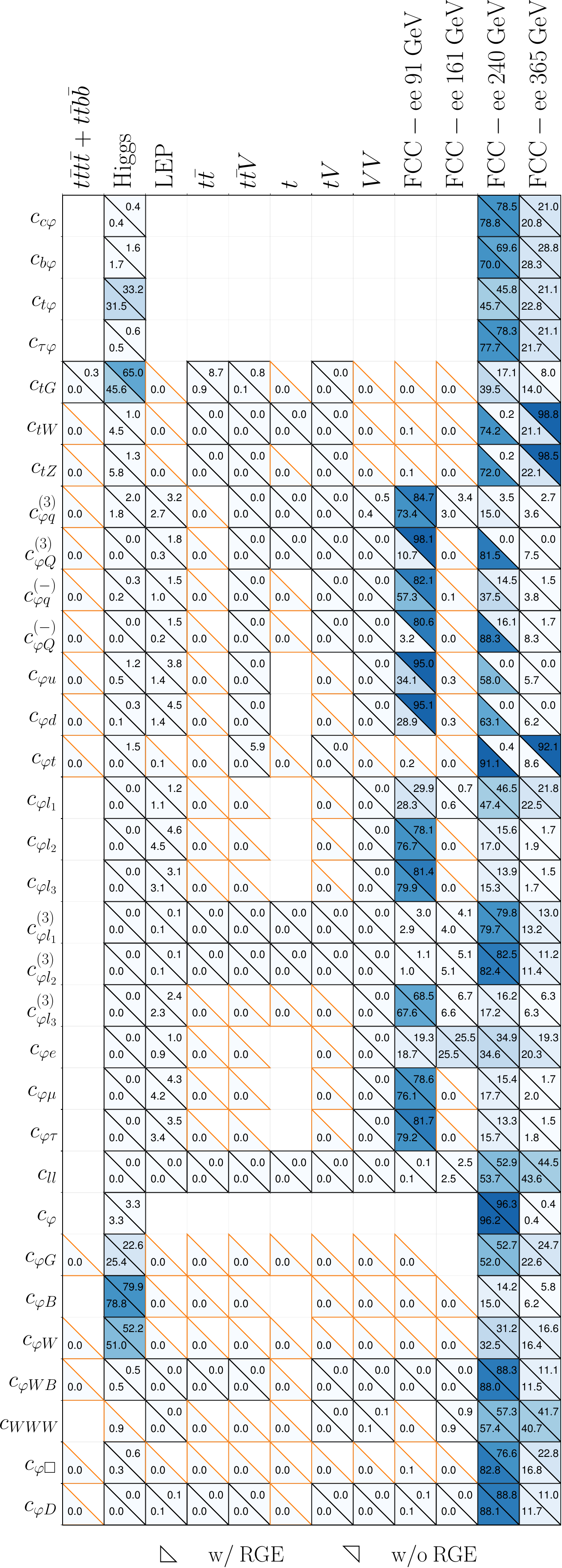}
        % \caption{Description of the left figure.}
        \label{fig:subfig1}
    \end{subfigure}
    \hfill
    \begin{subfigure}[t!]{0.46\textwidth}
        \centering
        \includegraphics[width=\textwidth]{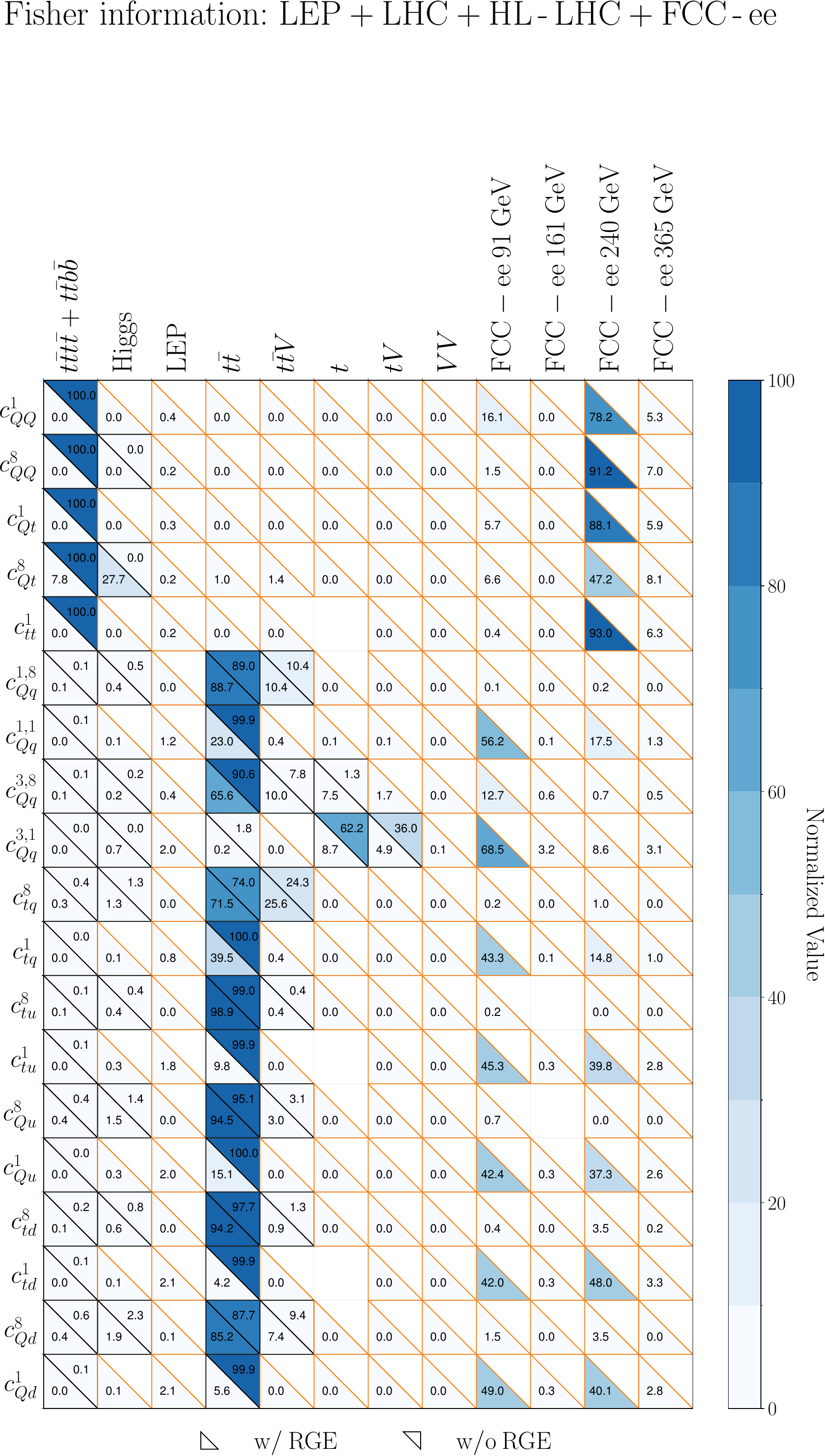}
        % \caption{Description of the right figure.}
        \label{fig:subfig2}
    \end{subfigure}
    \caption{
    Same as Fig.~\ref{fig:fisher_rge}
    now including the FCC-ee projections, which dominate the overall sensitivity for many of the operators included in the fit. 
    }
    \label{fig:fisher_rge_fcc}
\end{figure}
%%%%%%%%%%%%%%%%%%%%%%%%%%%%%%%%

In the case that RGEs are neglected, the Fisher analysis shows that FCC-ee observables dominate the sensitivity to all the purely bosonic or two lepton operators with the exception of $c_{\varphi B}$, and $c_{\varphi W}$, which are instead constrained by Higgs production measurements at the LHC.
Furthermore, without RGEs the only constraints on the four-heavy and two-light-two-heavy operators come from top quark production at the LHC + HL-LHC.
From Fig.~\ref{fig:fisher_rge_fcc} we also observe that the main effect of RGEs in this fit is to shift the sensitivity towards the FCC-ee observables, specifically for the $ZH$ run at $\sqrt{s}=240$ GeV.
Indeed, we see that for operators such as $c_{tW}$, $c_{tZ}$, and $c_{\varphi t}$ the dominant entry of the Fisher information matrix moves from the 365 GeV run to the 240 GeV one.
Likewise, many of the operators primarily constrained by the FCC EWPOs, such as $c_{\varphi u}$, $c_{\varphi d}$, $c_{\varphi Q}^{(3)}$,
$c_{\varphi Q}^{(-)}$, and $c_{\varphi q}^{(-)}$, are in the presence of RGEs dominated by the 240 GeV run observables, further highlighting how the RGE flows markedly enhances the constraining power of the $ZH$ run.
Also, in the presence of RGEs, the previously unconstrained four-heavy operators are now informed by the FCC-ee 240 GeV data. Finally, many of the two-light-two-heavy operators, in particular the singlet combinations, become dominated by the 240 GeV measurements as well, instead of by the top-quark pair production data which is the most relevant observable if RGEs are neglected. 

\begin{figure}[tp]
    \centering
    \includegraphics[width=0.84\linewidth]{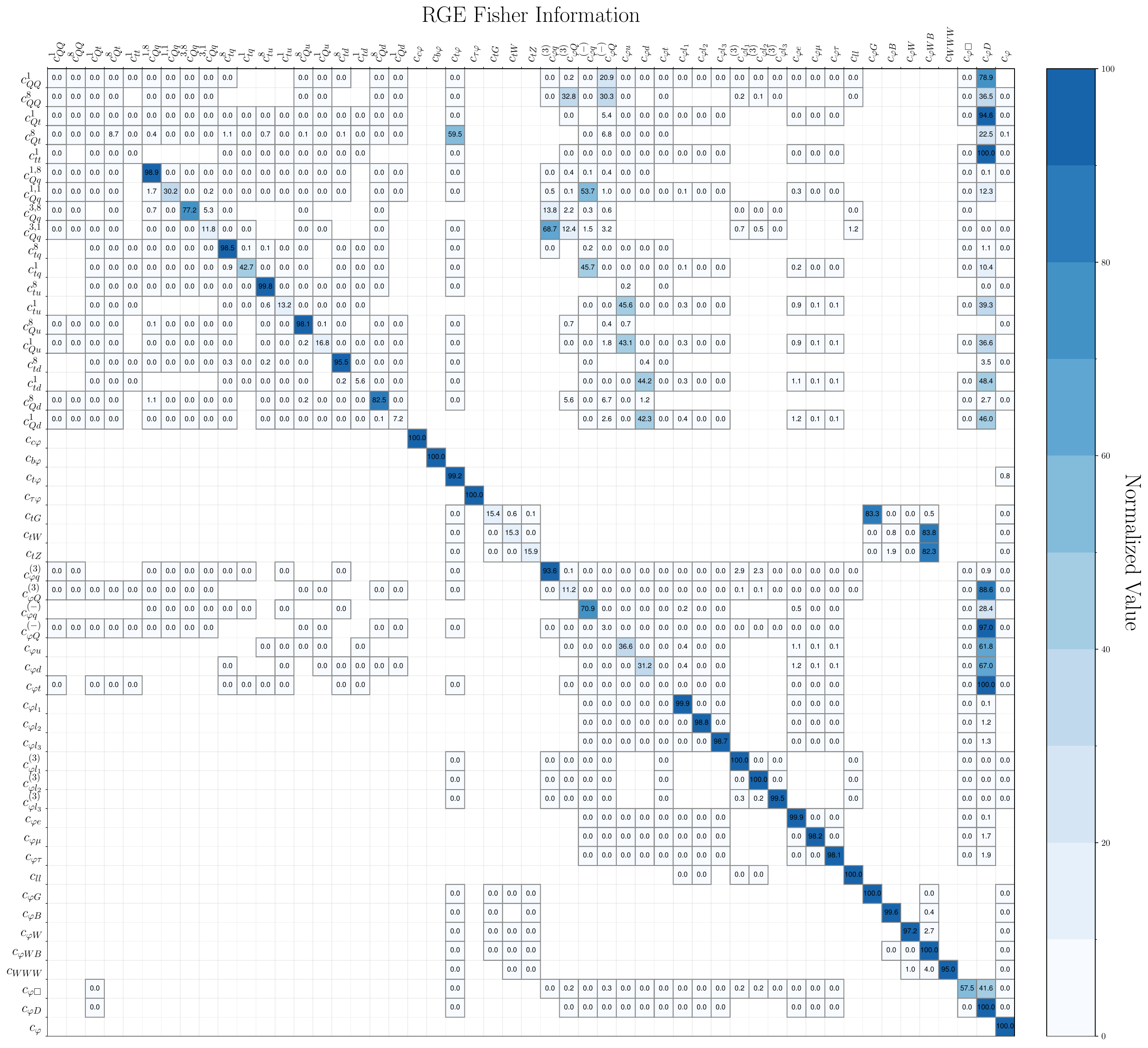}\hfill
    \caption{Same as Fig.~\ref{fig:RGE_fisher_lhc} for the fit including HL-LHC and FCC-ee observables.}
    \label{fig:RGE_fisher_fcc}
\end{figure}

A complementary handle to the Fisher information matrix of Fig.~\ref{fig:fisher_rge_fcc} in the understanding of the fit results is provided in Fig.~\ref{fig:RGE_fisher_fcc}, which shows the same Fisher operator correlation map as in Fig.~\ref{fig:RGE_fisher_lhc} now for the fit including the projected HL-LHC and FCC-ee observables.
As discussed in Sect.~\ref{sec:results}, the higher the value of an entry $(k,k')$ of this matrix, the stronger the (auto-)correlation between the $k$-th operator at the reference scale $\mu_0$ and the $k'$-th operator at the observable scale $\mu$.
In particular, diagonal entries much smaller than 100 indicate coefficients whose sensitivity in the fit is primarily induced by RGE running and mixing effects. 

%%%%%%%%%%%%%%%%%%%%%%%%%%%%%%%%%%
%%%%%%%%%%%%%%%%%%%%%%%%%%%%%%%

The map in Fig.~\ref{fig:RGE_fisher_fcc} illustrates the sensitivity flows between the Wilson coefficients as a consequence of activating RGE effects.
Some of the most noticeable features are the flow of the four-heavy operators such as $c_{Qq}^1$, fixed by four heavy quark production, onto $c_{\varphi D}$, constrained by Higgs  production and the EWPOs; the flow of two-light-two-heavy operators into operators mostly constrained at the FCC-ee, such as $c_{Q,q}^{3,1}$ into $c_{\varphi q}^{3}$ and $c_{tu}^{1}$ into $c_{\varphi u}$;
the flow of $c_{tG}$, constrained by top-quark pair production, into
$c_{\varphi G}$, fixed by Higgs production; the flow
of $c_{tW}, c_{tZ}$ (from single top) onto $c_{\varphi WB}$ (EWPOs),
and the flow of $c_{\varphi Q}^{(3)}$  onto $c_{\varphi D}$, relating EWPOs and $ZH$ processes at the FCC-ee.
Fig.~\ref{fig:fisher_rge_fcc} clearly highlights how the relative impact of the $\sqrt{s}=240$ GeV run of the FCC-ee is increased once RGEs are accounted for, with enhanced sensitivity which would otherwise come mostly from the 91 GeV run. 

\begin{figure}[tp]
    \centering    \includegraphics[width=0.49\linewidth]{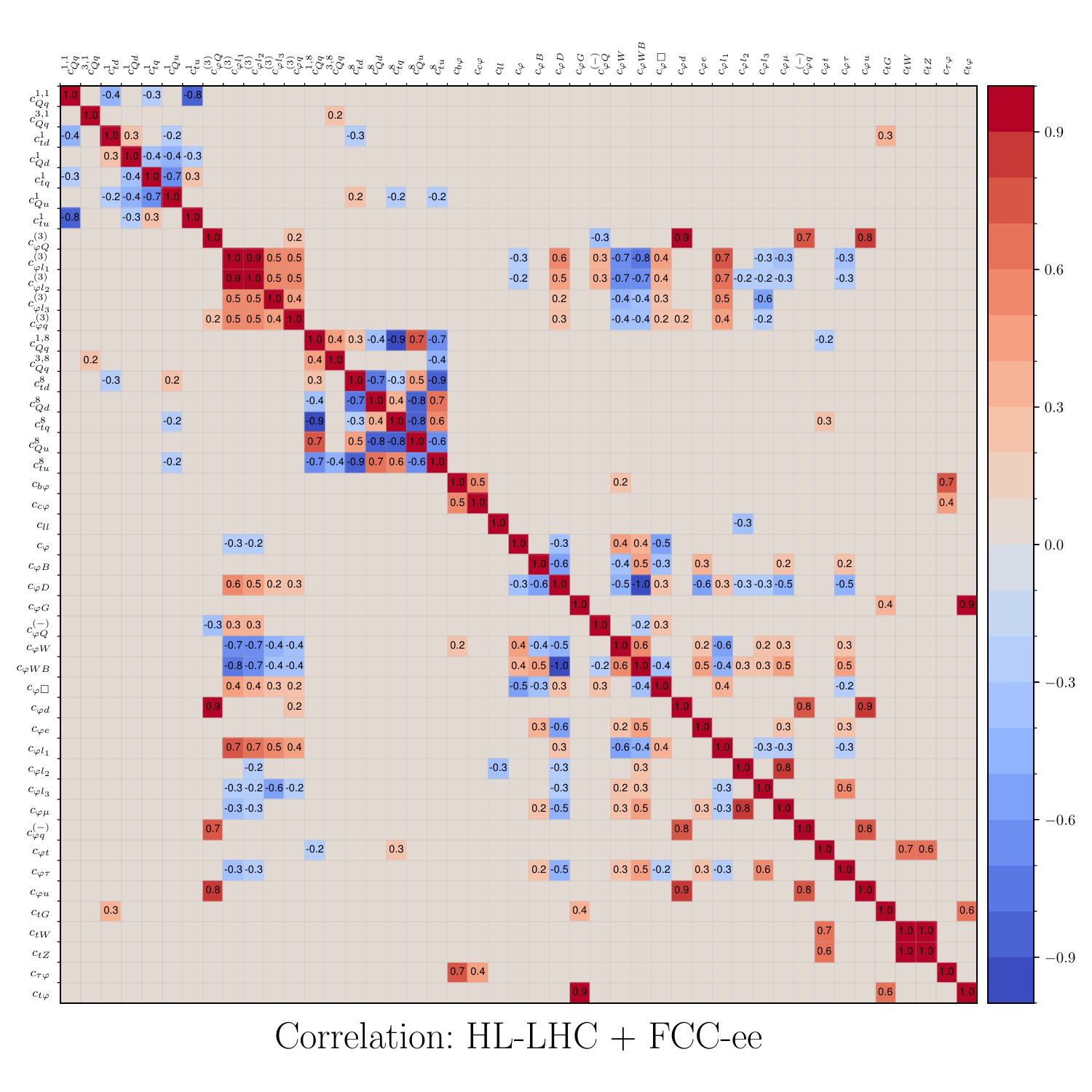}
\includegraphics[width=0.49\linewidth]{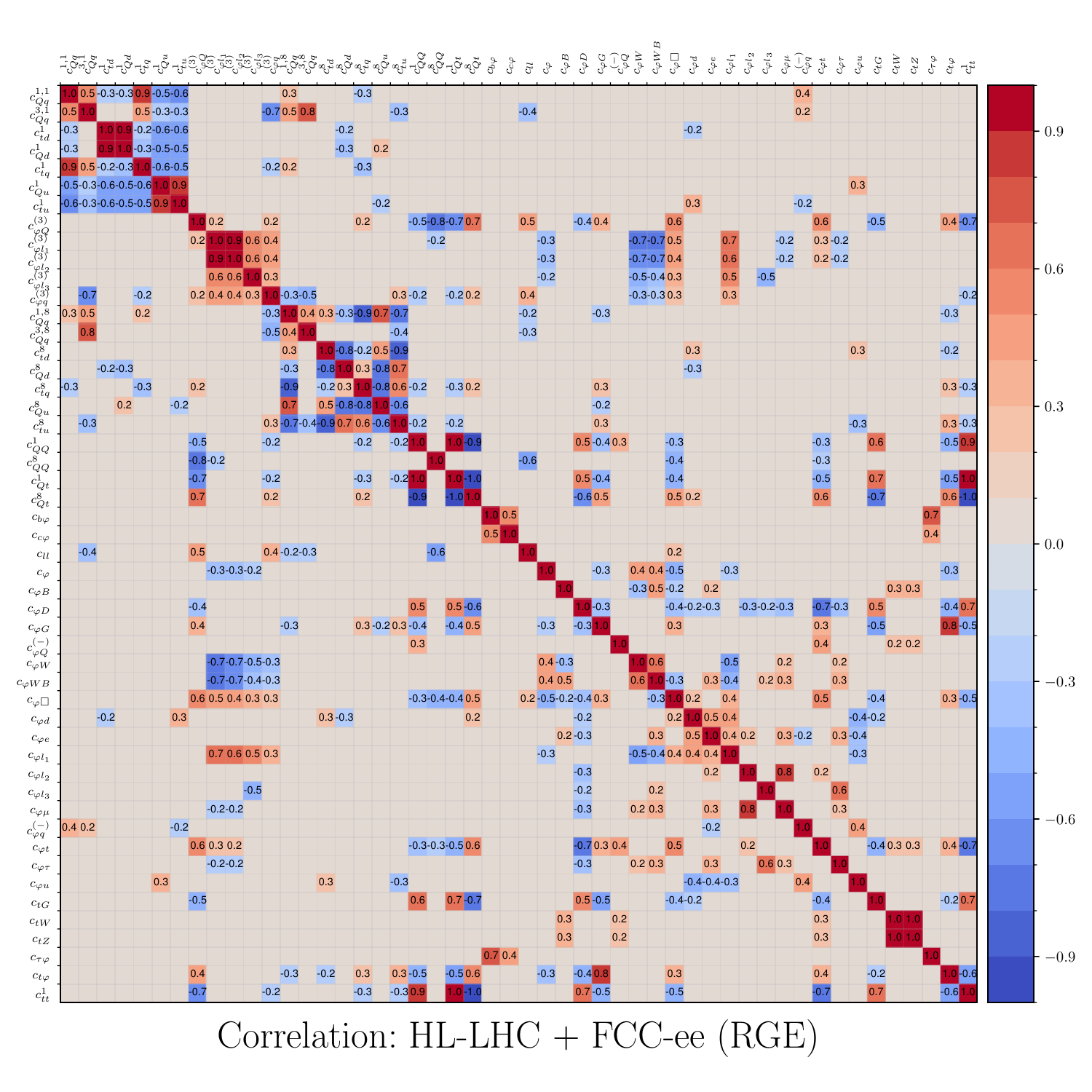}\\
    \caption{The correlation matrices between
    Wilson coefficients in the linear EFT fit to the dataset including HL-LHC and FCC-ee projections, comparing results without (left) and with (right panel) RGE effects. 
    Empty entries
    correspond to pairs of coefficients for
    which $|\rho_{ij}|\le 0.15$. 
    In the quadratic fits, the overall correlation pattern is mostly stable upon the inclusion of RGE.
    }
    \label{fig:corr_matrices}
\end{figure}

Finally, Fig.~\ref{fig:corr_matrices} shows the entries of the correlation matrices $\rho_{ij}$ evaluated for each pair of the Wilson coefficients entering the fit in either the linear or the quadratic fit.
These are fits based on the global dataset which includes both the HL-LHC and the FCC-ee projections.
Empty entries in the correlation maps
correspond to pairs of coefficients for
which $|\rho_{ij}|\le 0.15$.
%

%%%%%%%%%%%%%%%%%%%%%%%%%%%%%%%%%%%%%%%%%%%%%%
%%%%%%%%%%%%%%%%%%%%%%%%%%%%%%%%%%%%%%%%%%%%

Considering first the correlation maps in the case of the linear EFT fits, it is clear that RGE effects induce a large number of sizeable correlations between pairs of Wilson coefficients which were absent before.
This is particularly visible for the two-light-two-heavy operators and for the four-heavy ones, which are  absent in the non-RGE reference fit.
Another noteworthy example of this phenomenon is the top Yukawa operator $c_{t\varphi}$: without RGEs, the only appreciable correlations are with $c_{\varphi G}$ and $c_{tG}$, but once RGE effects are activated a much richer correlation pattern emerges.
In addition to inducing new correlations between the fit parameters, RGE effects can also break or reduce large correlations as a consequence of closing quasi-flat directions.
As an illustration, in the non-RGE baseline fit, the $c_{\varphi d}$ coefficient is highly correlated to both $c_{\varphi Q}^{(3)}$ and $c_{\varphi u}$, while these large correlations are partially washed out once RGE effects are activated.
Finally, we note that the purely Higgs operator $c_{\varphi}$ is very loosely correlated with the other parameters entering the fit, and that activating the RGEs does not change this picture. 

Similar considerations apply to the correlation maps in the case of the quadratic EFT fits. 
All in all, the picture which emerges from these correlation maps is consistent with that provided by the Fisher information analysis. 

%%%%%%%%%%%%%%%%%%%%%%%%%%%%%%%%%
%%%%%%%%%%%%%%%%%%%%%%%%%%%%%%%%%%

\subsection{Charting UV models at the FCC-ee}
\label{sec:uvmodels_fccee}

In this section, building upon the results presented in~\cite{terHoeve:2023pvs,Celada:2024mcf}, we quantify the impact of RGE effects in the projected sensitivity to representative UV extensions of the SM which can be constrained by FCC-ee measurements through their matching to the SMEFT operators.
Notably, it has been shown that FCC-ee observables can probe all renormalizable models that match onto the dimension-six SMEFT at tree level, and that this can be achieved with only the $Z$-pole and $WW$-threshold runs if one considers one-loop effects either via RGEs or matching~\cite{Allwicher:2024sso,Gargalionis:2024jaw}. 

Here we consider the impact of RGEs in single particle extensions of the SM, matched at either tree-level or one-loop, for models with heavy scalars, fermions, and vectors in different gauge representations.
We assume the matching scale to be the mass of the heavy particle, $M_{\text{UV}}$, in all cases.
The one-particle UV models considered here are the same as those studied in our previous analyses~\cite{terHoeve:2023pvs,Celada:2024mcf},
with the addition of two new scalar fields:  $\Theta_1\sim(1,4)_{1/2}$ and $\Theta_3\sim(1,4)_{3/2}$. 
The inclusion of the electroweak quadruplets $\Theta_{1}$ and $\Theta_{3}$ is motivated by the addition of the purely Higgs operator $\mcO_{\varphi}$ to our fit basis together with that of observables sensitive to the Higgs self-interaction either via tree-level or one-loop effects.
These EW quadruplets are matched up to one loop as motivated by their interesting phenomenology~\cite{Durieux:2022cvf}. 
We provide more details of all the models considered in App.~\ref{app:uv_models}.

\begin{figure}[t]
    \centering
    \includegraphics[width=0.5\linewidth]{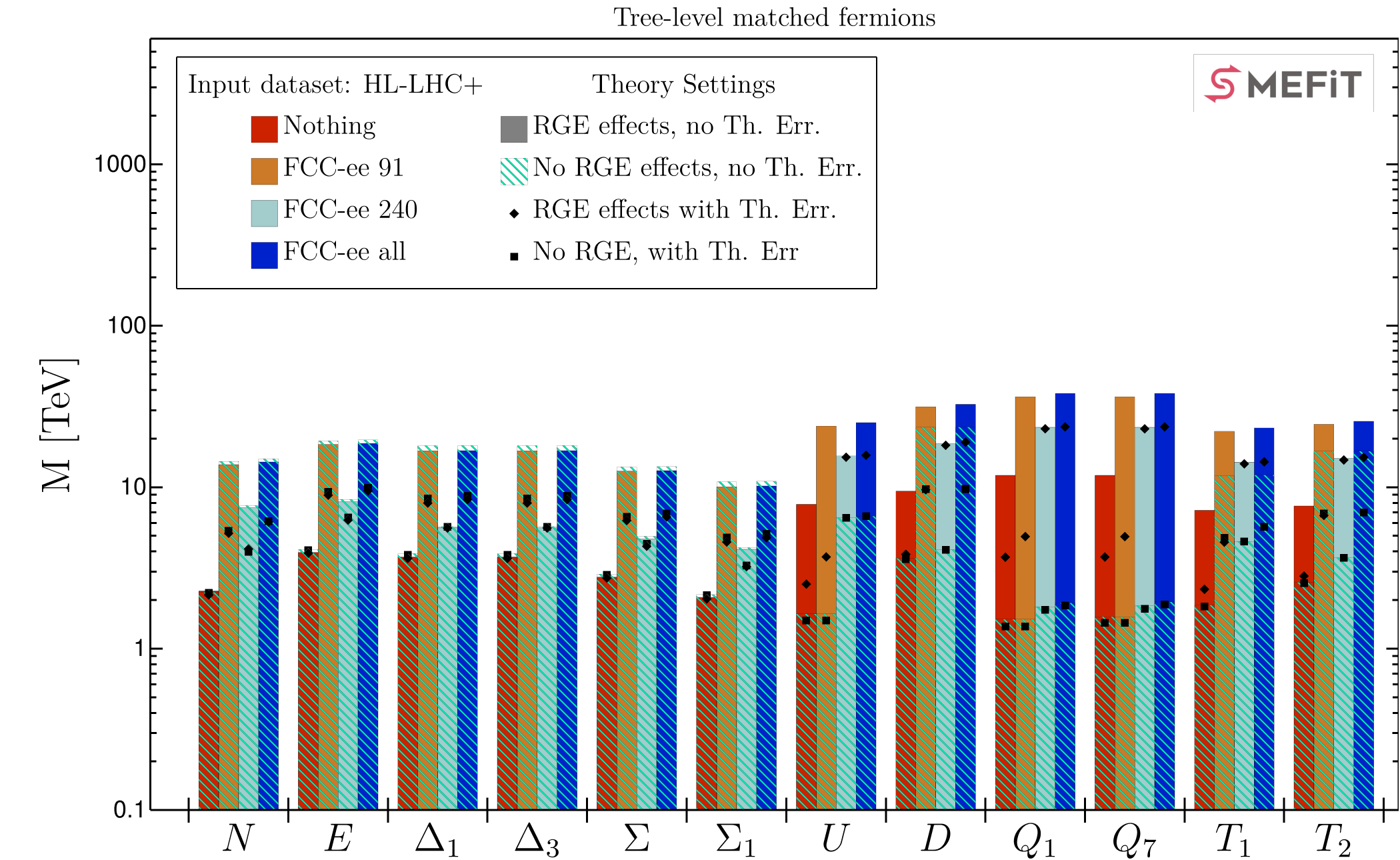}\hfill
    \includegraphics[width=0.5\linewidth]{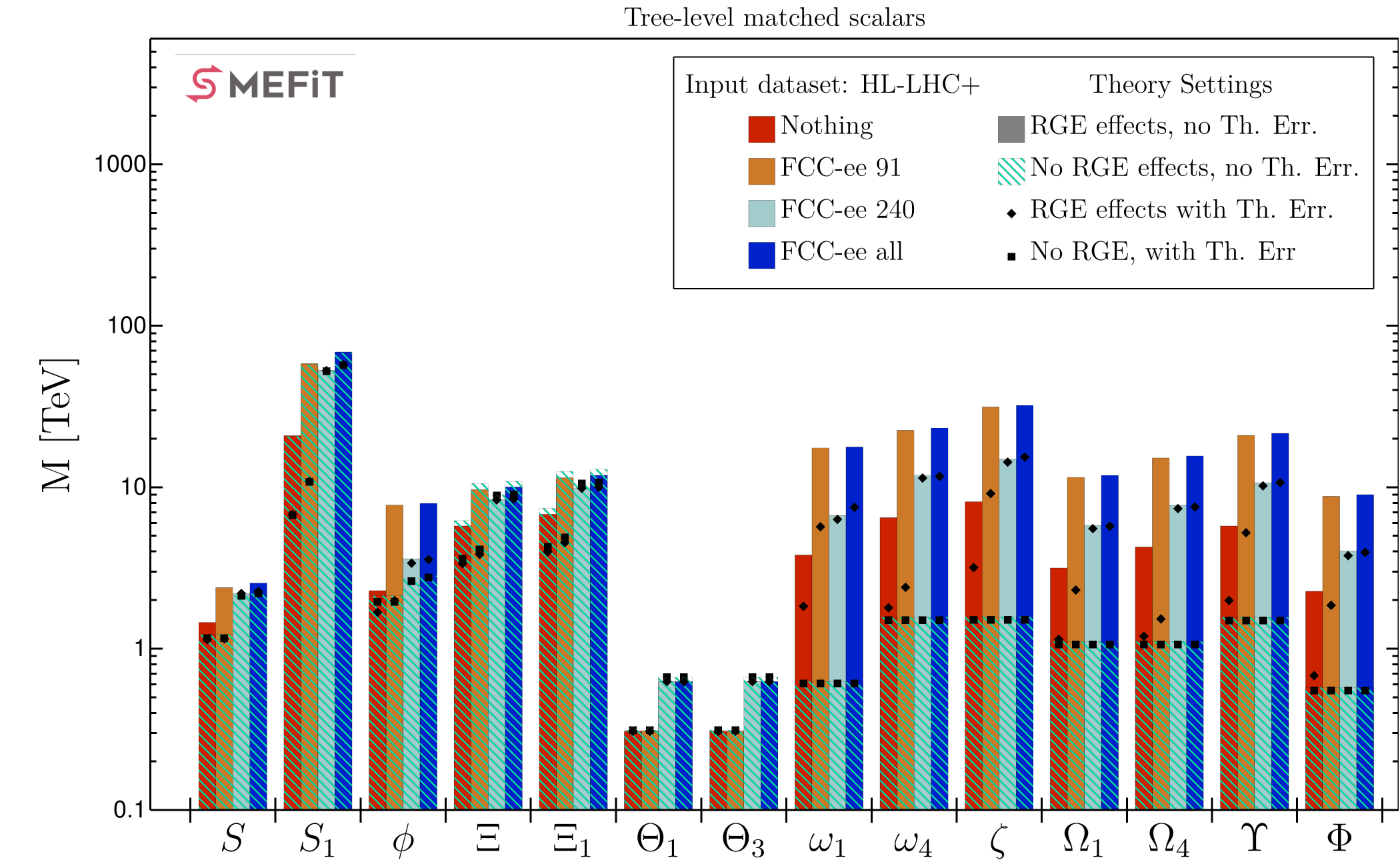}
    \includegraphics[width=0.5\linewidth]{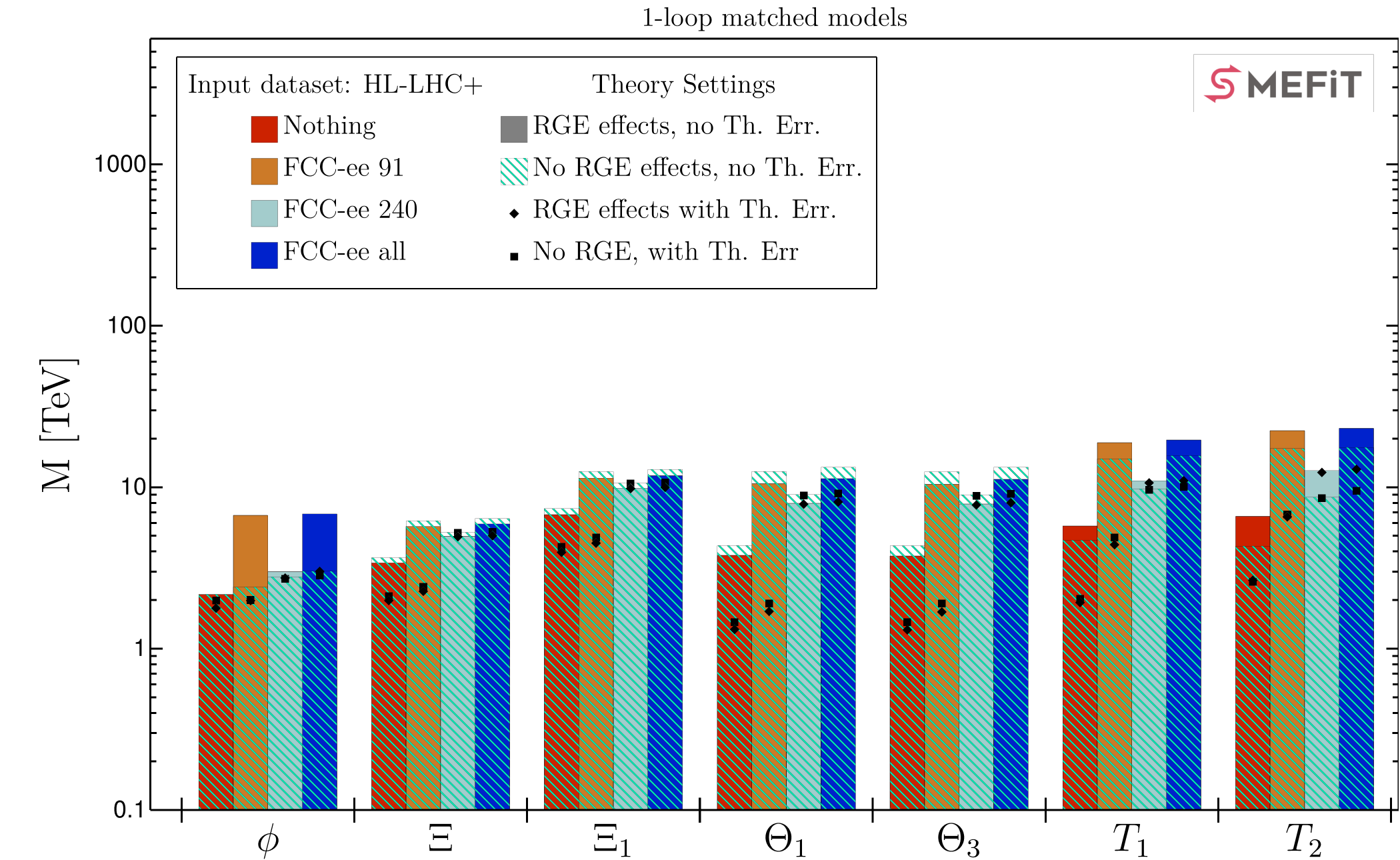}\hfill
    \includegraphics[width=0.5\linewidth]{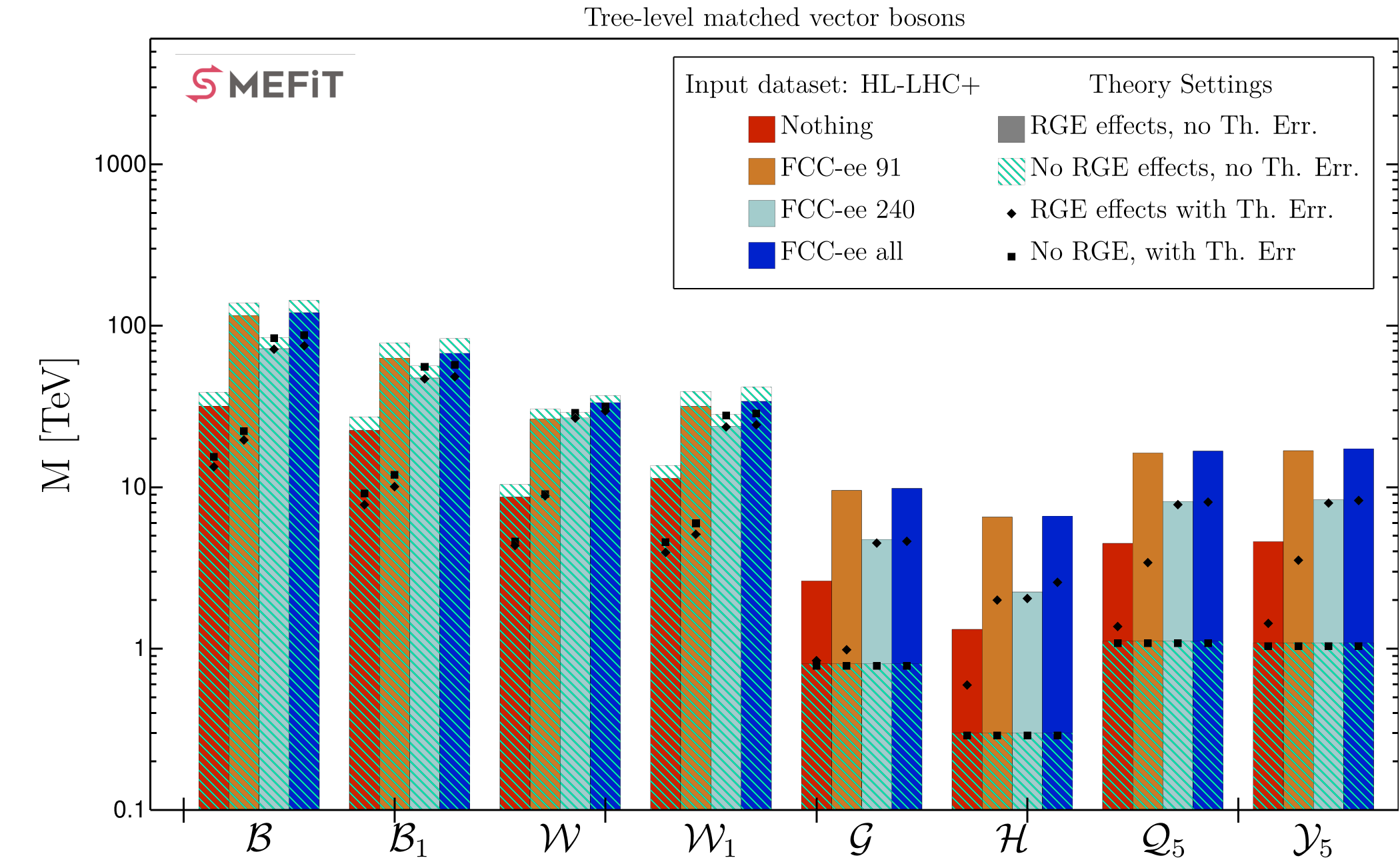}  
    \caption{The $95\%$ CL lower bounds on the heavy particle mass $M_{\rm UV}$ in representative single particle extensions of the SM  matched onto the SMEFT.
    This mass reach is determined by a $\chi^2$ scan by means of the condition $\chi^2(M_{\rm UV})=\chi^2_{\rm SM} + 3.84$.
    All UV couplings are set at the matching scale, $M_{\rm UV}$, to be either 1 or $1$~TeV, depending on their dimensionality. 
    We compare the baseline bounds, based on LEP + HL-LHC projections, with those associated to three different combinations of the FCC-ee dataset: the $91$~GeV or $240$~GeV runs on their own and the combination of all the FCC-ee runs, $91+161+240+365$~GeV. 
    Results are provided both with and without RGE effects accounted for, and we also compare the bounds with and without including theoretical uncertainties in the SM predictions for the EWPOs.
    The RGE evolution is performed from $M_{\rm UV}$ down to the scale of each dataset.   
    From the top left in a clockwise direction, we show the results for heavy vector-like fermions, heavy scalars, and heavy vector spin-1 bosons, all matched at tree level, as well as representative heavy scalar and fermion models matched at the one-loop level.
    }
    \label{fig:mass_reach_FCC}
\end{figure}

In Figure~\ref{fig:mass_reach_FCC}, we show the $95\%$ C.I. mass reach of HL-LHC and FCC-ee in several simple extensions of the SM matched onto the SMEFT. 
The HL-LHC dataset incorporates LEP observables, and all the FCC-ee runs considered use the HL-LHC dataset as a baseline.
We define the mass reach as the mass value at which $\Delta\chi^2 = 3.84$, obtained by scanning over masses and computing the $\chi^2$ at each point. The result is presented with dynamical RGE running, evolved down from the UV mass of each particle. We refer to App.~\ref{app:additional_FCC_results} for a more fine-grained version that includes additional energy breakdowns.
The bars show the bound obtained without considering current theoretical uncertainties on SM predictions, while the black square and rhomboidal markers indicate the result of including them. 
We consider these two extreme cases the most optimistic and pessimistic scenarios.
We consider a selection of UV couplings in each UV model, which are detailed in App.~\ref{app:uv_models}, and assume all dimensionless (dimensionful)  couplings between SM and heavy particles to be $g_{\rm UV}(M_{\rm{UV}})=1$ ($g_{\rm UV}(M_{\rm UV})=1$~TeV).

%%%%%%%%%%%%%%%%%%%%%%%%%%%%%%%%%%%%%%%%
%%%%%%%%%%%%%%%%%%%%%%%%%%%%%%%%%%%%%%%%

The results for heavy scalar bosons matched at tree level are shown in the upper-right panel of Fig.~\ref{fig:mass_reach_FCC}. These models can be grouped into two classes depending on whether the RGE effects significantly impact the mass reach. Models like $S$, $S_1$, $\Xi$ and $\Xi_1$ generate operators that enter in EWPOs at tree-level and they dominate the bound on the model, hence rendering RGE effects subleading. The EW quadruplets $\Theta_{1}$ and $\Theta_{3}$ only generate $\mcO_{\varphi}$ at tree-level, which explains both the poor bound and the absence of impact from the RGE running. The remaining scalar models see a large improvement in their bounds after including RGE effects, in particular at FCC-ee. Among them, $\phi$ stands out as the only colourless one and the only one where the HL-LHC bound is not heavily improved by RGE running since it is driven by its tree-level contribution to $c_{t\varphi}$, which shows no significant running. The rest of scalar models, namely $\omega_{1}$, $\omega_4$, $\xi$, $\Omega_1$, $\Omega_4$, $\Upsilon$ and $\Phi$, generate only 4-heavy-quark operators that run into FCC-ee observables, which explains the observed improvement in all considered datasets.

In the case of vector bosons, shown on the lower-right panel, those models that do not show a large improvement in the bound from RGE effects show instead a loosening of the bound driven by the self-running of the operators $c_{\varphi\Box}$ and $c_{\varphi D}$. As with the scalars, the larger RGE-induced improvements appear for coloured models. The case of vector-like fermions matched at tree-level, presented in the upper-left panel, is very similar to the scalar case. The colourless fermions show small decreases in the bounds induced by RGE effects. On the other hand, the coloured fermions show large improvements in their bounds after including RGE effects due to their tree-level contributions to $c_{\varphi Q}^{(-)}$, $c_{\varphi Q}^{(3)}$, and $c_{\varphi t}$.
Finally, we show a selection of scalar and vector-like fermion models matched at 1-loop in the bottom-right panel. 1-loop matching results do not remove the possibility of important RGE effects in certain models, such as $\phi$, $T_1$ and $T_2$. 

These different models also allow us to evaluate the physics potential of the FCC-ee program and the relevance of each of its energy runs. Overall, FCC-ee will improve the mass reach for each of these models by a factor of up to $\sim 6$. However, this improvement would be limited to very few models were not for the inclusion of RGE effects. These RGE effects are crucial to harness the power of FCC-ee to constrain coloured heavy particles. 

As found in the literature~\cite{Allwicher:2023shc}, the RGE running is the key to extracting all possible information from the Z-pole run of FCC-ee, which becomes in those conditions the leading probe of the entire FCC program for most models. The Higgs and top-pair runs contribute to the bound of several models such as $S_1$, $\phi$, $\Theta_{1,3}$ matched at 1 loop and $T_{1,2}$. We show the result of considering the 240 GeV run observables on their own to highlight that, in isolation, can provide bounds close to the ones from the Z-pole for very few models and how it benefits from the combination with the Z-pole run.

Moreover, we can assess the impact of current theory errors on LHC observables and EWPOs on the reach of FCC-ee. Current theory errors on the SM predictions for the EWPOs~\cite{Awramik:2006uz,Dubovyk:2019szj,Freitas:2014hra,Awramik:2003rn} reduce substantially the sensitivity of the $Z$-pole run of FCC-ee in all models that are constrained by it.  In several of them, e.g. $U$, $Q_{1}$, $Q_{7}$, $\omega_{4}$, the mass reach increases by an order of magnitude when the theory errors are neglected. For some models, e.g.\ $\mathcal{W}$, $\mathcal{W}_1$, $\Xi$, $T_1$, $Q_7$, the addition of higher-energy FCC-ee runs can compensate, and only partially,  the sensitivity lost due to theory errors in the EWPOs. Hence, the reduction of theory uncertainties on EWPO predictions by the time of FCC-ee is one of the most pressing issues of theoretical particle physics.  

%%%%%%%%%%%%%%%%%%%%%%%%%%%%%%%%
\begin{figure}[t!]
    \centering
    \includegraphics[width=0.60\linewidth]{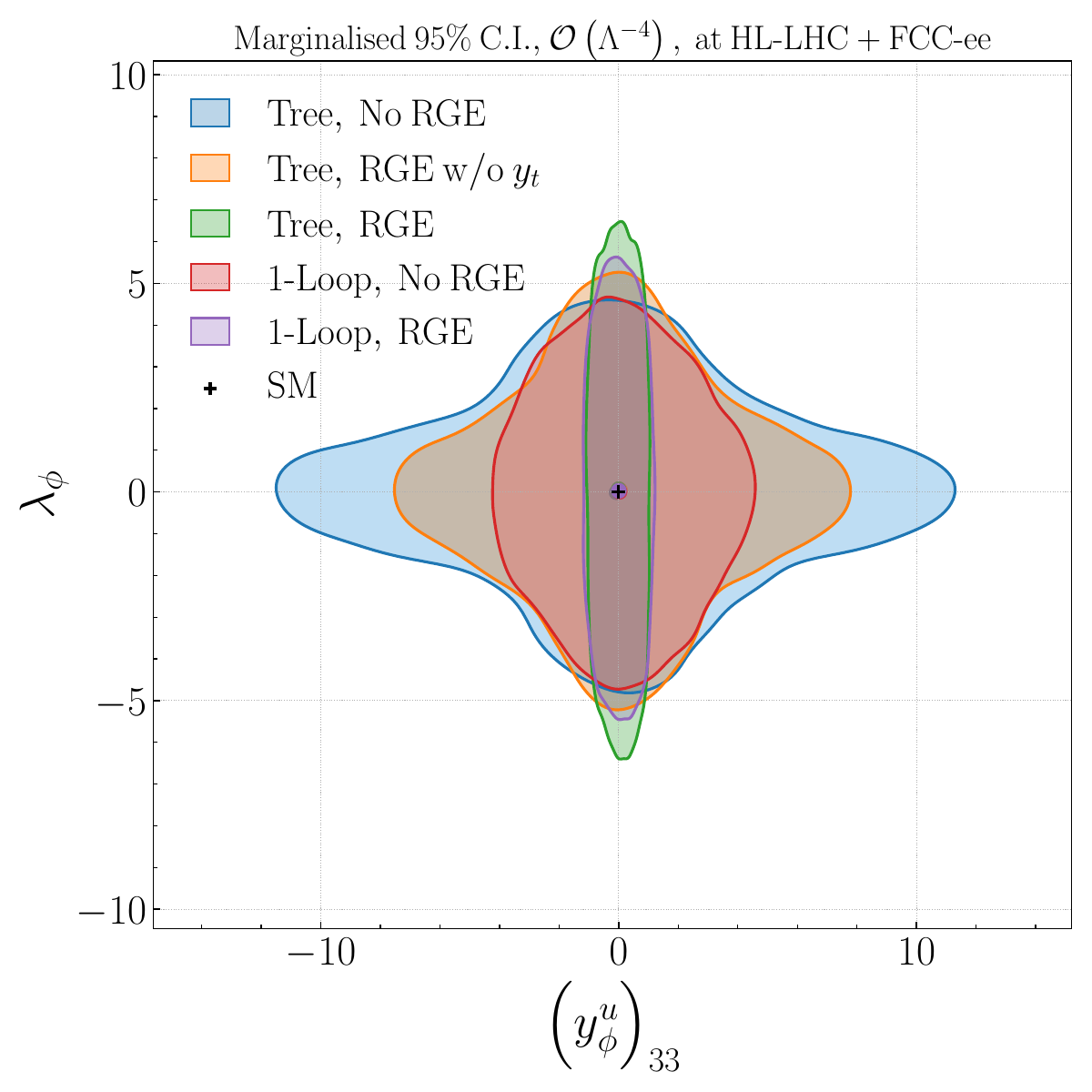}    \caption{The marginalized $95\%$ C.I. bounds obtained from the fit to the global dataset, including HL-LHC and FCC-ee projections, of the heavy scalar doublet $
    \phi$ model.
    The heavy scalar mass is set to $M_{\phi}=8$~TeV, which also coincides with the matching scale. 
    We compare results based on tree-level and one-loop matching, in each case with and without accounting for RGE effects.
    For the tree-level matched model, we also show the effect of removing the SM top Yukawa, $y_t$, from the RGE evolution.}
    \label{fig:heavy_doublet_fcc_rge}
\end{figure}
%%%%%%%%%%%%%%%%%%%%%%%%%%%%%%%%

To finalize this section, we want to highlight
the subtle interplay between RGE running and one-loop matching effects by considering the heavy scalar doublet ($\phi$) one-particle extension of the SM~\cite{terHoeve:2023pvs}.
In this model the only relevant UV couplings are  $\lambda_\phi$ (self-coupling interaction) and $(y^{u}_{\phi})_{33}$ (Yukawa-type interaction).
 The heavy scalar mass is assumed to be $M_{\phi}=8$~TeV, which also coincides with the matching scale. 
 We can then derive bounds on the two UV couplings of this model using either tree-level or one-loop matching, in each case with and without accounting for RGE effects, and  study in this manner the interplay between matching and RGEs.

We display in Fig.~\ref{fig:heavy_doublet_fcc_rge} the $95\%$ C.I. bounds on the two UV couplings of this heavy scalar doublet model from a fit which includes both HL-LHC and FCC-ee projections. 
In our previous studies~\cite{terHoeve:2023pvs,Celada:2024mcf}, this model had a flat direction along $\lambda_\phi$ when considering tree-level matching.
In the present analysis, this flat direction is closed since we consider the constraints on Higgs pair production at the (HL-)LHC as well as the information contained in the NLO EW corrections to $ZH$ production at the FCC-ee, both of which
constrain $\lambda_\phi$ via its tree-level contribution to $\mcO_\varphi$.

As compared to the baseline tree-level matching, no RGE results, one observes how accounting for RGE effects leads to a dramatic improvement in the sensitivity to $(y^{u}_{\phi})_{33}$ by up to an order of magnitude, while the constraints on the self-interaction $\lambda_\phi$ become looser.
Around half of the improvement on the sensitivity to $(y^{u}_{\phi})_{33}$ comes from the running proportional to the SM top Yukawa coupling.
The information provided by one-loop matching improves the sensitivity to the $\left(y^{u}_{\phi}\right)_{33}$ coupling only in the case in which RGEs are neglected, and once RGEs are activated, the bounds on this coupling are the same irrespective of whether one uses tree-level or one-loop matching.
Nevertheless, 1-loop matching effects improve slightly the bound along the $\lambda_\phi$ direction.

The relative importance of 1-loop RGE and 1-loop matching effects is highly model-dependent.
A clear counteract to the $\phi$ model studied before is provided by the scalar EW quadruplets $\Theta_{1}$ and $\Theta_3$, for which RGE effects have little relevance when matching them at either tree or 1-loop level.
However, as can be seen by comparing the upper right and lower left panels of Fig.~\ref{fig:mass_reach_FCC}, including the finite 1-loop contributions via matching increases the mass reach by a factor $\sim 10$ both at HL-LHC and FCC-ee.
These EW quadruplets have a rich phenomenology at future colliders~\cite{Durieux:2022hbu} which we explore in depth in a companion paper~\cite{smefit:trilinear}.
\section{Summary and outlook}
\label{sec:summary}

In this work we have presented a systematic assessment of the role that RGEs have in the context of global SMEFT analyses connecting measurements taken at different energy scales.
Considering first a global analysis of LEP and LHC data, subsequently complemented by HL-LHC and FCC-ee projections, we have demonstrated that the inclusion of RGE effects is instrumental to push forward the precision and accuracy frontiers of SMEFT fits.

We find that RGE-improved predictions can significantly alter the interplay between different datasets, as clearly demonstrated in our fit to LEP and LHC data.
First, neglecting RGE effects can result in a substantial loss of sensitivity. This is particularly evident in the case of four-heavy-quark operators, which remain poorly constrained due to degeneracies in a global linear fit without RGEs. 
However, incorporating RGE effects breaks these degeneracies, thereby enhancing sensitivity.
A similar, more pronounced, improvement is observed for the bounds on the four-heavy-quark operators at the FCC-ee.
Nevertheless, we also showed that RGE effects might also result in less stringent bounds, and identified cases where the bounds on the Wilson coefficients became looser because of information dilution to a broader parameter space. 
The distinct impact of RGEs in linear and quadratic fits was scrutinised via the linearity score analysis, which demonstrated how RGEs can shift a bound dominated by quadratic effects into the linear regime and vice versa.

Perhaps the most striking result of this work is the spectacular impact that accounting for RGEs has on the mass reach of new heavy particles at the FCC-ee.
As illustrated by Fig.~\ref{fig:mass_reach_FCC}, we identified several BSM one-particle extensions for which, once RGE effects are considered, the mass reach of the FCC-ee measurements is improved by more than an order of magnitude.
Moreover, we showed how current theoretical uncertainties on $Z$-pole observables would reduce the overall impact of the 91 GeV FCC-ee run and increase the relevance of the higher-energy runs, while the $Z$-pole run would dominate in an ideal scenario.
This result highlights the importance of accounting for the most precise theory calculations available in the SM and the SMEFT to make the most of the exquisitely precise electroweak, Higgs, and top quark measurements that would be provided by the FCC-ee.

Another finding of this study is the complementarity of the different types of one-loop effects (related in particular to either RGEs or to matching) entering the analysis  of UV models such as those considered in Fig.~\ref{fig:heavy_doublet_fcc_rge}.
This complementarity illustrates how the combination of one-loop corrections increases the sensitivity to specific UV couplings.
It is, however, important to emphasize that, in general, all different one-loop effects should be included consistently and that the dominance of one specific one-loop correction over another is model- and process-dependent.

The results of this paper can be extended in several complementary directions. 
First of all, while here we focused on the FCC-ee projections, the same analysis methodology could be deployed for any other of the proposed future particle colliders, from CLIC and ILC to the CEPC, LHeC, and the muon collider, as well as to the FCC-hh.
Work in progress~\cite{esppu2026} motivated by the ongoing European Strategy for Particle Physics Update will tackle these future collider projects and present a comprehensive comparative SMEFT interpretation now fully including RGE effects.
Second, as a spin-off of our analysis, we are quantifying~\cite{smefit:trilinear} the potential of the FCC-ee to indirectly constrain the Higgs self-coupling via measurements of single Higgs production in the $ZH$ production channel, complementing the direct bounds at the HL-LHC from Higgs pair production.
Third, the availability of RGE evolution meets one of the main requirements to include flavour and low-energy constraints into \smefit~ as it enables the running and mixing of the Wilson coefficients to the $Q\sim m_b$ scale where they can be matched to the usually adopted Low Energy EFT (LEFT), which is relevant to interpret flavour data and to connect with one of the available flavour and low-energy likelihoods. 

\subsection*{Acknowledgments}
A.~R. and E.~V. are supported by the European Research Council (ERC) under the European
Union’s Horizon 2020 research and innovation programme (Grant agreement No. 949451) and by a Royal
Society University Research Fellowship through grant URF/R1/201553.
The work of A.~R. is also supported by the University of Padua under the 2023 STARS Grants@Unipd programme (Acronym and title of the project: HiggsPairs – Precise Theoretical Predictions for Higgs pair production at the LHC). 
A. R. acknowledges support from the COMETA COST Action CA22130.
A.~R. would like to acknowledge the Mainz Institute for Theoretical Physics (MITP) of the Cluster of Excellence PRISMA+ (Project ID 390831469) for enabling him to complete a portion of this work.
L.~M. acknowledges support from the European Union under the MSCA fellowship (Grant agreement N. 101149078) Advancing global SMEFT fits in the LHC precision era (EFT4ward).
The work of J.~t.~H. is 
supported by the UK Science and Technology Facility Council (STFC) consolidated grant ST/X000494/1. 
The work of J.~R. is supported by the Dutch Research
Council (NWO) and by the Netherlands eScience Center.

The authors are grateful to L. Allwicher, G. Durieux, C. Grojean, and M. McCullough for stimulating conversations. 
A.~R. is grateful to A. Biekoetter, J. Fuentes-Mart\'in and V. Miralles for useful discussions.
L.~M. thanks P. Stangl and K. Mimasu for useful conversations.

\appendix
\section{Validation and benchmarking}
\label{app:benchmarking}

In this appendix, we present two benchmark studies demonstrating the validity of our RGE implementation in the global fit: first,  a self-consistency test showing that the fit results close under RGE evolution, and second, a benchmark comparison with the results reported in~\cite{Aoude:2022aro} for representative top quark sector operators. 

\begin{figure}[t]
\centering
\includegraphics[width=\textwidth]{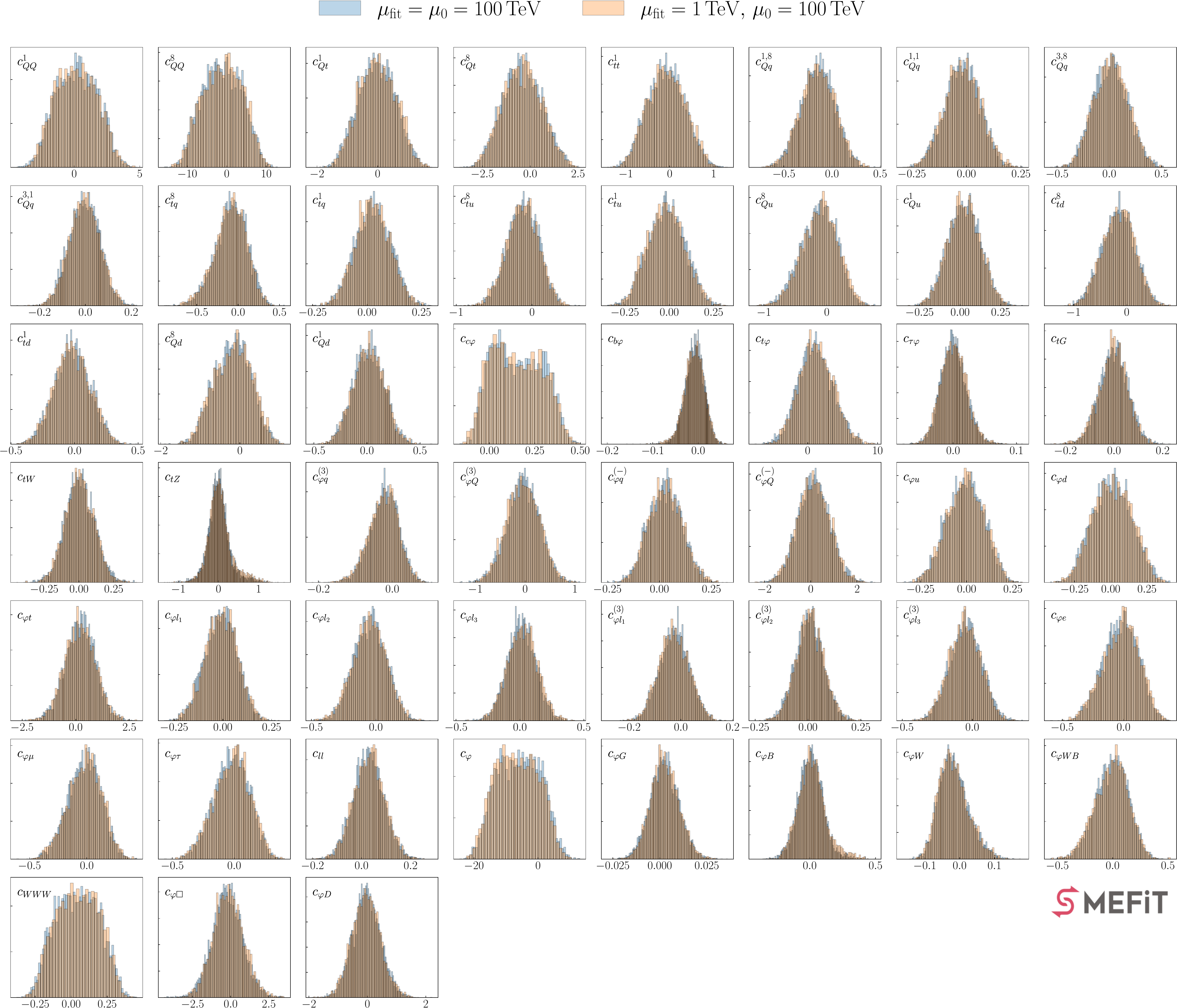}
\caption{Posterior distributions of the Wilson coefficients from the SMEFT fit of Sect.~\ref{sec:results}.
We compare the results of a fit carried out
using a fitting scale $\mu_{\rm fit}=100$ TeV
 with those of a fit carried out using $\mu_{\rm fit}=1$ TeV followed by evolving the obtained posteriors up to $\mu_0=$ 100 TeV.
The agreement found confirms that our fit closes under RG evolution.}
\label{fig:rg_closure}
\end{figure}

First, we demonstrate that fit results
are independent of the choice of scale $\mu_{\rm fit}$ at which the fit is performed, which is ultimately arbitrary and unphysical, used for the RGE evolution. 
To this end, we compare in Fig.~\ref{fig:rg_closure} the results of a fit carried out
using a value of $\mu_{\rm fit}=100$ TeV as the fitting scale
with those of a fit carried out instead using $\mu_{\rm fit}=1$ TeV, where the resulting posterior distributions are then
evolved up to $\mu_0=$ 100TeV.
This fit is done at NLO QCD in the QCD perturbative expansion and includes EFT corrections up to $\mathcal{O}\lp \Lambda^{-4}\rp$ in the EFT expansion.
The excellent agreement between the two sets of posterior distributions confirms that fit results are strictly independent of the choice of initial scale $\mu_{\rm fit}$, as expected, and therefore that the fit closes satisfactorily upon $\mu_{\rm fit}$ variations.

%--------------------------
%--------------------------

\begin{figure}[t]
\centering
\includegraphics[width=0.75\textwidth]{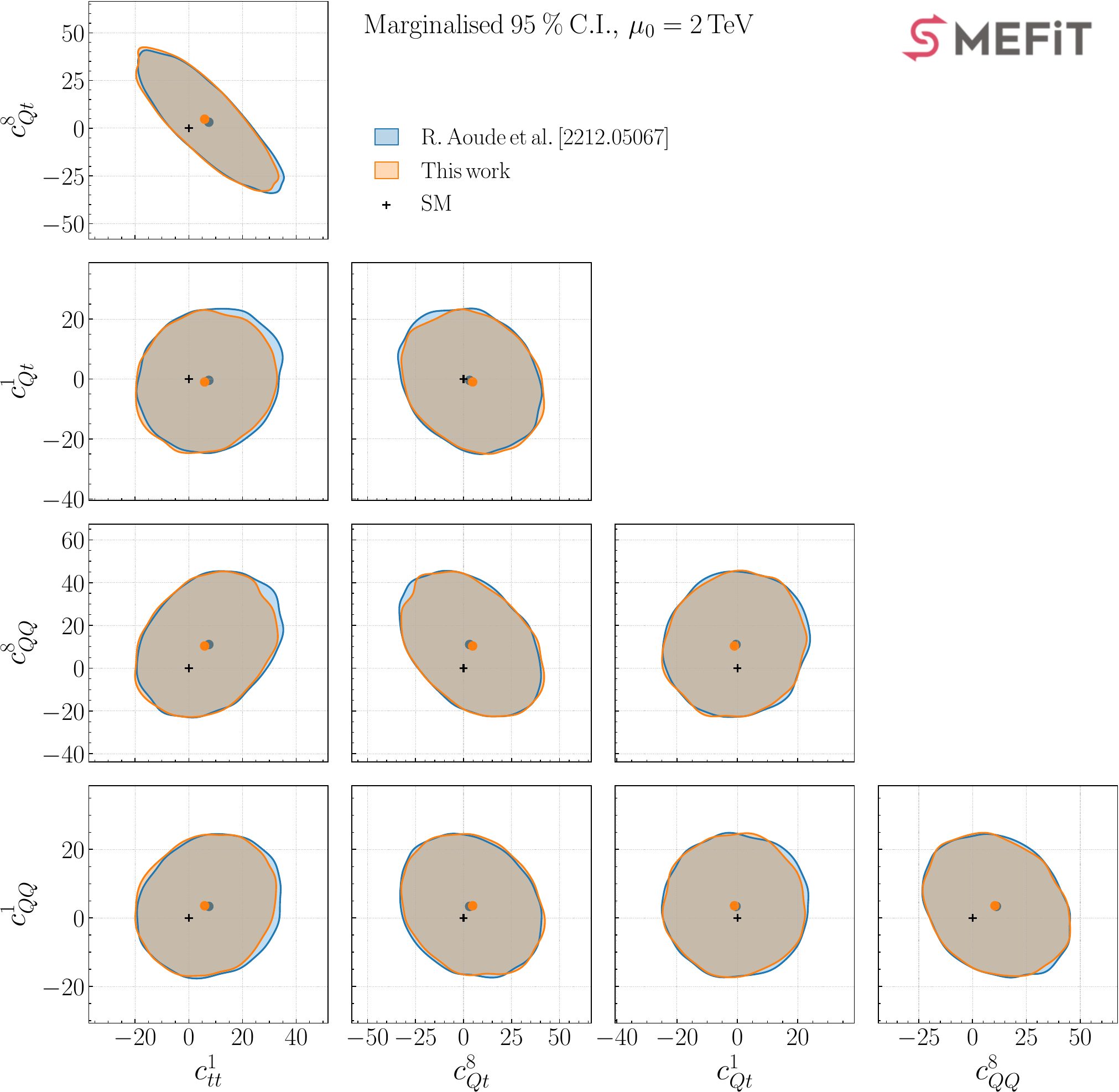}
\caption{Comparison at $\mu_0=2$ TeV of a quadratic EFT fit to $t\bar{t}t\bar{t}$ and $t\bar{t}b\bar{b}$ cross-sections in the presence of RGE with the independent results from~\cite{Aoude:2022aro} based on the same dataset.
}
\label{fig:benchmark-4Q-Claudio}
\end{figure}

Second, we compare our results for the implementation of RGEs in \smefit with related studies in the literature which consider RGE effects to a subset of our dataset and/or our operator basis. 
Specifically, we compare with the outcome of the analysis in~\cite{Aoude:2022aro} in a fit which constraints four-heavy-quark operators, and where the same input dataset is used. 
The benchmark comparison shown in Fig.~\ref{fig:benchmark-4Q-Claudio}, corresponding to a linear fit to the $t\bar{t}t\bar{t}$ and $t\bar{t}b\bar{b}$ production cross-sections in the presence of RGE effects, reveals excellent agreement with the independent calculation of~\cite{Aoude:2022aro} when applied to the same dataset. 
We note that at the linear EFT level one cannot constrain four-heavy-quark operators, and that 
the bounds obtained in Fig.~\ref{fig:benchmark-4Q-Claudio} arise entirely from the information provided by the RGEs.

%--------------------------
%--------------------------

Additionally, we checked our results against~\cite{Allwicher:2024sso}. We find very good agreement after setting our input uncertainties to the same values, since ~\cite{Allwicher:2024sso} originally used uncertainties from~\cite{DeBlas:2019qco} instead of the updated ones in Snowmass~\cite{deBlas:2022ofj}.
\section{Implementation and performance}
\label{app:implementation}

In this appendix we provide technical details about the implementation of RGE effects in \smefit\ and the associated fit performance.

As mentioned in Sect.~\ref{sec:settings}, RGE is implemented in \smefit~ through an interface with the {\sc\small wilson} package~\cite{Aebischer:2018bkb}. Specifically, using a custom-defined dictionary, we map our operator definitions to those of the generic Warsaw basis as defined in {\sc\small wilson} and evaluate the RG evolution matrices $\Gamma$ in Eq.~\eqref{eq:theory_EFT_3} in the first step of the fitting procedure. 
Since, in the approximation used, $\Gamma$  does not depend on the Wilson coefficients, a single evaluation at the beginning of each fit suffices.

To ensure a consistent implementation, we have modified certain functionalities of {\sc\small wilson}. 
As discussed in Sect.~\ref{sec:settings}, we linearize the RG equations by switching off the dependence of Wilson coefficients on the running of the SM parameters and by neglecting dimension-six input parameter shifts in the dimension-six anomalous dimension matrix.
Additionally, we disable the flavour rotation that {\sc\small wilson} applies at the end of the running, which diagonalizes the down-quark Yukawa matrix. 
While this step is not directly related to the RGEs themselves, it concerns the definition of the flavour basis at different scales and introduces additional non-linearities that we don't consider here. 

Here we adopt a scheme where the only non-zero Yukawa coupling is that of the top quark, $y_t$. 
In this framework, the flavour symmetry is preserved during the evolution, rendering the flavour rotation trivial in the SM. However, dimension-six Yukawa operators can induce non-zero corrections to the Yukawa couplings of light quarks.
In such cases, the flavour rotation should in principle be applied, thereby moving us out of our operator basis. Since these Wilson coefficients contribute to our observables and are subject to experimental constraints, we aim to constrain them in the fit but choose to neglect RGE-related effects which activate operators outside our operator basis, such
as the light quark Yukawas.

RGE running effects in \smefit\ can be enabled by specifying a new entry in the run card together with some associated parameters: 

\begin{python}
rge:
  init_scale: 5000.0 # The reference scale $\mu_0$ for the Wilson coefficients
  obs_scale: dynamic # "float" or "dynamic"
  scale_variation: 0.5  # Apply a scale factor to the dynamic observable scale
  smeft_accuracy: integrate # Options: integrate, leadinglog
  yukawa: top # Options: top, full, or none
  adm_QCD: False # If true, the EW couplings are set to zero
\end{python}
The different options that are available
to the user to configure before performing the fit are the following:

\begin{itemize}

\item \texttt{init\_scale}: this parameter specifies (in GeV)
the value of $\mu_0$ at which Wilson coefficients $\boldsymbol{c}(\mu_0)$ are fitted, with Eq.~(\ref{eq:theory_EFT_3}) being used to relate them to the physical observables.
When matching to UV-complete models, $\mu_0$ coincides with the matching scale.

\item \texttt{obs\_scale}: choose between a fixed common scale $\mu$ for the fitted observables and a dynamical, bin-dependent scale.
In the former case, a single evolution matrix 
$\Gamma$  needs to be computed for all the data points entering the fit.
In the latter case, a different matrix $\Gamma$ 
 is computed for each datapoint at the beginning of the fit.

\item \texttt{scale\_variations}: this option allows the user to apply a constant rescaling factor to the dynamic observable scales $\mu$, enabling a direct assessment of the impact of scale dependence on the fit as shown in Sect.~\ref{sec:results}.

\item \texttt{smeft\_accuracy}: this option, inherited from {\sc\small wilson}, allows the user to determine the level of precision in solving the RGEs. Specifically, one can choose between performing a leading-log (LL) approximation or a full all-order resummation.

\item \texttt{yukawa}: this option allows for further customization of the Yukawa matrices, enabling the user to include the full Yukawa matrices in the solution to the RGE equations, only the top quark Yukawa, or to exclude them entirely from the calculation.

\item \texttt{adm\_QCD}: if set to true, this option disables the inclusion of EW couplings in the solution of the RGE equations.

\end{itemize}

\paragraph{Scale setting procedure.}
In the dynamic scale setting approach, a characteristic energy scale, $\mu_m$, must be specified for each data point in the associated theory file. 
Assuming that $n_{\rm dat}$ cross-sections are included in the fit, \smefit evaluates the RGE matrices:
\begin{equation}
\label{eq:RGEmatrices}
\Gamma_{ij}^{(m)}\big(\mu_m, \mu_0, \bar{g}\big), 
\quad m = 1, \ldots, n_{\rm dat}, 
\quad i, j = 1, \ldots, n_{\rm op},
\end{equation}
using a caching system.
If the same scale $\mu_m$ appears multiple times in the theory files (as happens frequently, see Fig.~\ref{fig:scales}), the calculation is performed only once. These matrices are then used to build Eq.~(\ref{eq:theory_EFT_3}) from Eq.~(\ref{eq:theory_EFT_1}), thereby incorporating the RGE effects into the fitted cross-sections.
This operation is performed only during the initialization phase, so the runtime for the actual fits based on Eq.~(\ref{eq:theory_EFT_3}) is effectively identical to that for fits using Eq.~(\ref{eq:theory_EFT_1}), without incurring additional overhead.

Within its current implementation, the actual overhead for computing the RGE matrices in \smefit~can be substantial but still affordable.
For instance, in our fit with the broadest dataset (including FCC-ee projections), the calculation of the $\Gamma$ matrices at the beginning of the fit takes about one hour. 
While optimisation is certainly possible, for example using fast grid techniques as in {\sc\small PineAPPL}~\cite{Carrazza:2020gss}, the current level of affordable overhead does not make this a priority.

\paragraph{High-statistics EFT calculation of top-quark pair production.}
While not strictly related to the RGE implementation, another technical improvement in the present work that is worth mentioning concerns the calculation of EFT cross-sections for a subset of the top-quark pair production datasets. 
 These are now calculated with much higher Monte Carlo statistics to address numerical fluctuations observed in some datasets, specially in the high energy tails of differential distributions where statistics are heavily suppressed. 
These fluctuations had non-negligible effects on the bounds for certain two-light-two-heavy four-quark operators in the linear $\mathcal{O}\lp 1/\Lambda^2\rp$ fit, where constraints are generally very loose to begin with. 
However, these changes are entirely irrelevant for the quadratic, $\mathcal{O}\lp 1/ \Lambda^4\rp$, fit.
As customary, all the theoretical predictions used in our fits are publicly available from the {\sc\small SMEFiT} database repository:
\begin{center}
\url{https://github.com/LHCfitNikhef/smefit_database} 
\end{center}
\section{Details on the UV models}
\label{app:uv_models}
In this appendix we provide some additional information about the UV models considered in Sect.~\ref{sec:uv_fcc} and on the assumptions made on the structure of their couplings to the SM fields.
In general, we follow the notation and conventions from~\cite{deBlas:2017xtg}.
We show the SM gauge representations of the different fields considered in our models in the second column of Table~\ref{tab:uv_gauge_couplings}.
In the third column of Table~\ref{tab:uv_gauge_couplings}, we show the couplings between the heavy UV particles and the SM that we consider besides the fixed gauge couplings.

%%%%%%%%%%%%%%%%%%%%%%%%%%%%%%%%%%
\begin{table}[htbp]
\begin{center}
  \renewcommand{\arraystretch}{1.35}
\resizebox{0.85\textwidth}{!}{
\begin{tabular}{|c|c|c|}
\hline
{\bf Model Label} & $\qquad${\bf SM irreps}$\qquad$ & {\bf UV couplings} \\
\hline
\multicolumn{3}{|c|}{\bf Heavy Scalar Models} \\
\hline
$S$ & $(1,1)_{0\phantom{/-3}}$ & $\kappa_{\mathcal{S}}$, $\kappa_{\mathcal{S}^3}$, $\lambda_{\mathcal{S}}$\\
$S_1$ & $(1,1)_{1\phantom{/-}}$ & $\left(y_{S_1}\right)_{12}\left(y_{S_1}\right)_{21}$\\
$\phi$ & $(1,2)_{1/2\phantom{-}}$ & $\lambda_{\phi},\,\, (y_{\phi}^u)_{33}$    \\
$\Xi$  & $(1,3)_{0\phantom{/-3}}$ & $\kappa_{\Xi}$, $\lambda_{\Xi}$ \\
$\Xi$  & $(1,3)_{1\phantom{/-3}}$ & $\kappa_{\Xi_1}$, $\lambda_{\Xi_1}$, $\lambda'_{\Xi_1}$ \\
$\Theta_1$  & $(1,4)_{1/2}$ & $\lambda_{\Theta_1}$ \\
$\Theta_3$  & $(1,4)_{3/2}$ & $\lambda_{\Theta_3}$ \\
$\omega_1$  & $(3,1)_{-1/3}$ & $\left(y_{\omega_1}^{qq}\right)_{33}$ \\
$\omega_4$  & $(3,1)_{-4/3}$ & $\left(y_{\omega_4}^{uu}\right)_{33}$ \\
$\zeta$  & $(3,3)_{-1/3}$ & $\left(y_{\zeta}^{qq}\right)_{33}$ \\
$\Omega_1$  & $(6,1)_{1/3\phantom{-}}$ & $\left(y_{\Omega_1}^{qq}\right)_{33}$ \\
$\Omega_4$  & $(6,1)_{4/3\phantom{-}}$ & $\left(y_{\omega_4}\right)_{33}$ \\
$\Upsilon$  & $(6,3)_{1/3\phantom{-}}$ & $\left(y_{\Upsilon} \right)_{33}$\\
$\Phi$  & $(8,2)_{1/2\phantom{-}}$ &  $\left(y_{\Phi}^{qu}\right)_{33}$ \\[1mm]
\hline
\multicolumn{3}{|c|}{\bf Heavy Vector Models} \\
\hline
$\mathcal{B}$ & $(1,1)_{0\phantom{/-3}}$ & $\left(g_{\mathcal{B}}^{u}\right)_{33},\,\,\left(g_{\mathcal{B}}^{q}\right)_{33},\,\,
g_{\mathcal{B}}^{\varphi},\,\,\left(g_{\mathcal{B}}^{e}\right)_{11},\,\, \left(g_{\mathcal{B}}^{e}\right)_{22},
\left(g_{\mathcal{B}}^{e}\right)_{33},\,\,
\left(g_{\mathcal{B}}^{\ell}\right)_{11}, \,\,
\left(g_{\mathcal{B}}^{\ell}\right)_{22}, \,\,
\left(g_{\mathcal{B}}^{\ell}\right)_{33}, \,\,
$ \\
$\mathcal{B}_1$ & $(1,1)_{1\phantom{/-3}}$ & $g_{B_1}^{\varphi}$ \\
$\mathcal{W}$ & $(1,3)_{0\phantom{/-3}}$ & $\left(g_{\mathcal{W}}^\ell\right)_{11},\,\left(g_{\mathcal{W}}^\ell\right)_{22}$, $\left(g_{\mathcal{W}}^\ell\right)_{33}$, $g_{\mathcal{W}}^\varphi$, $\left(g_{\mathcal{W}}^q\right)_{33}$ \\
$\mathcal{W}_1$ & $(1,3)_{1\phantom{/-3}}$ & $g_{\mathcal{W}_1}^{\varphi}$ \\
$\mathcal{G}$ & $(8,1)_{0\phantom{/-3}}$ & $\left(g_{\mathcal{G}}^{q}\right)_{33},\,\,\left(g_{\mathcal{G}}^{u}\right)_{33}$ \\
$\mathcal{H}$ & $(8,3)_{0\phantom{/-3}}$ & $\left(g_{\mathcal{H}}\right)_{33}$ \\
$\mathcal{Q}_5$ & $(3,2)_{-5/6}$ & $\left( g_{\mathcal{Q}_5}^{uq} \right)_{33}$ \\
$\mathcal{Y}_5$ & $(\bar{6},2)_{-5/6}$ & $\left( g_{\mathcal{Y}_5} \right)_{33}$ \\
\hline
\multicolumn{3}{|c|}{\bf Heavy Fermion Models} \\
\hline
 $N$ & $(1,1)_{0\phantom{/-3}}$ & $\left(\lambda_N^e\right)_3$ \\
$E$ & $(1,1)_{-1\phantom{/3}}$ & $\left(\lambda_E\right)_3$ \\
$\Delta_1$ & $(1,2)_{-1/2}$ & $\left(\lambda_{\Delta_1}\right)_3$\\
$\Delta_3$ & $(1,2)_{-3/2}$ & $\left(\lambda_{\Delta_3}\right)_3$ \\
$\Sigma$ & $(1,3)_{0\phantom{/-3}}$ & $\left(\lambda_{\Sigma}\right)_3$ \\
$\Sigma_1$ & $(1,3)_{-1\phantom{/3}}$ & $\left(\lambda_{\Sigma_1}\right)_3$ \\
$U$ & $(3,1)_{2/3\phantom{-}}$ & $\left(\lambda_{U}\right)_3$ \\
$D$ & $(3,1)_{-1/3}$ & $\left(\lambda_{D}\right)_3$\\
$Q_1$ & $(3,2)_{1/6\phantom{-}}$ & $\left(\lambda_{Q_1}^{u} \right)_3$ \\
$Q_7$ & $(3,2)_{7/6\phantom{-}}$ & $\left(\lambda_{Q_7}\right)_3$ \\
$T_1$ & $(3,3)_{-1/3}$ & $\left(\lambda_{T_1}\right)_3$  \\
$T_2$ & $(3,3)_{2/3\phantom{-}}$ & $\left(\lambda_{T_2}\right)_3$  \\
\hline
\end{tabular}
}
\caption{\label{tab:uv_gauge_couplings} 
SM gauge representations and couplings of the heavy particles included in the single-particle UV-complete models that are considered in this work.
The gauge group representation is in the notation $(\textrm{SU}(3),\textrm{SU}(2))_{\textrm{U}(1)}$.
The model couplings are enforced to respect the {\sc\small SMEFiT} flavour assumption after tree-level matching and follow the notation used in~\cite{deBlas:2017xtg}, except for the couplings of $\Theta_{1,3}$ which follow the normalization in~\cite{Durieux:2022hbu}.
}
\label{tab:UVmodels_list}
\end{center}
\end{table}
%%%%%%%%%%%%%%%%%%%%%%

Compared to our previous work~\cite{terHoeve:2023pvs,Celada:2024mcf}, we have made slight changes to the set of probed models.
We added the electroweak quadruplets $\Theta_{1}$ and $\Theta_{3}$ since our extended dataset allows us to constrain them even at tree-level via $\mathcal{O}_\varphi$. Additionally, we matched them onto SMEFT at 1 loop and the result was cross-checked with the literature~\cite{Durieux:2022cvf}, from where we adopted the normalization of the UV couplings.
The models $S_1$, $\mathcal{B}$, $\mathcal{W}$ and $\mathcal{G}$ had been studied in~\cite{terHoeve:2023pvs} but then partially or fully excluded from our future-collider prospects~\cite{Celada:2024mcf} due to having more than one UV coupling. 
Our new methodology to define the mass reach allowed us to include them again in our study.

\section{Fine-grained impact of the different FCC-ee runs}
\label{app:additional_FCC_results}

\begin{figure}[t]
    \centering
\includegraphics[width=0.5\linewidth]{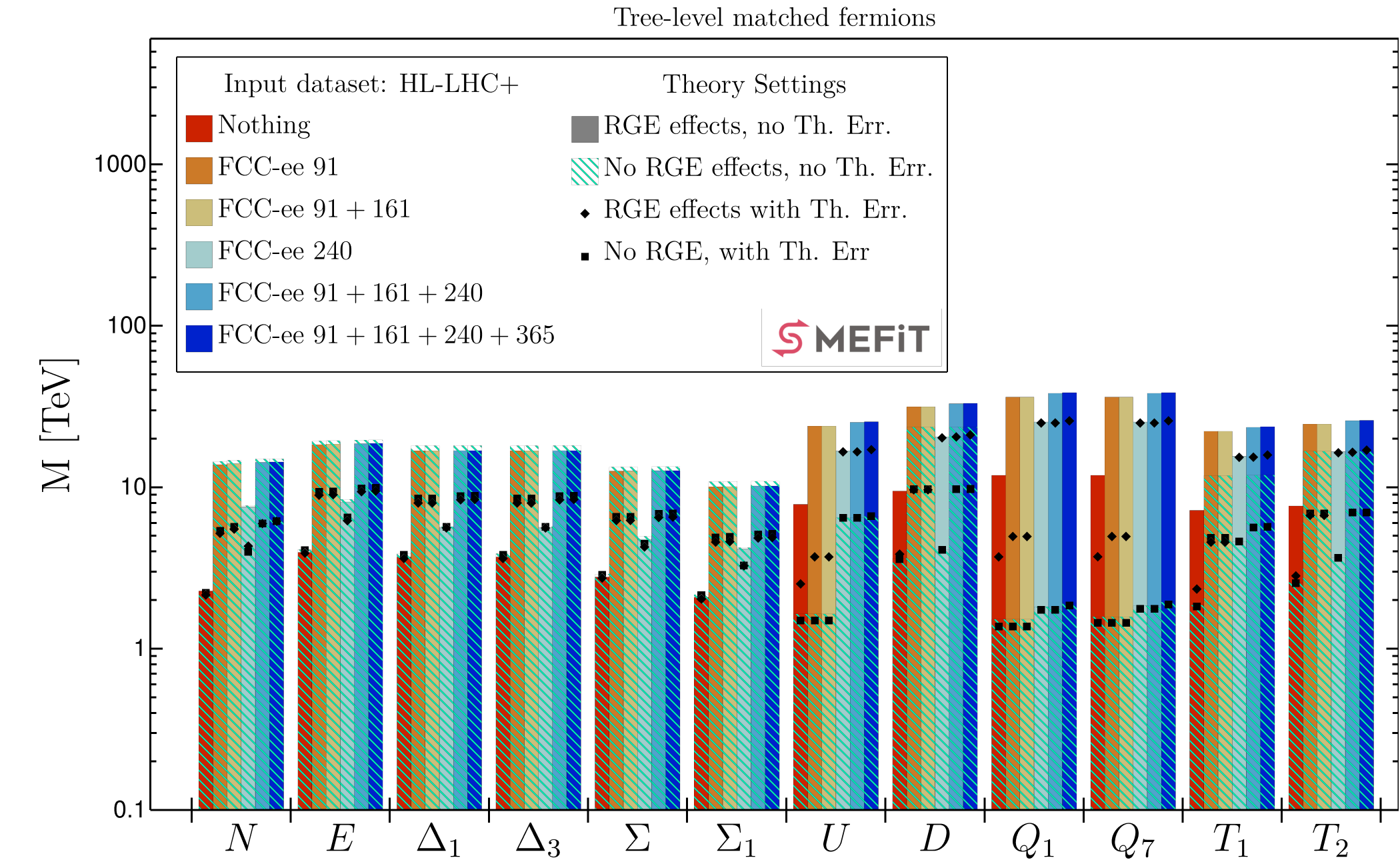}\hfill
\includegraphics[width=0.5\linewidth]{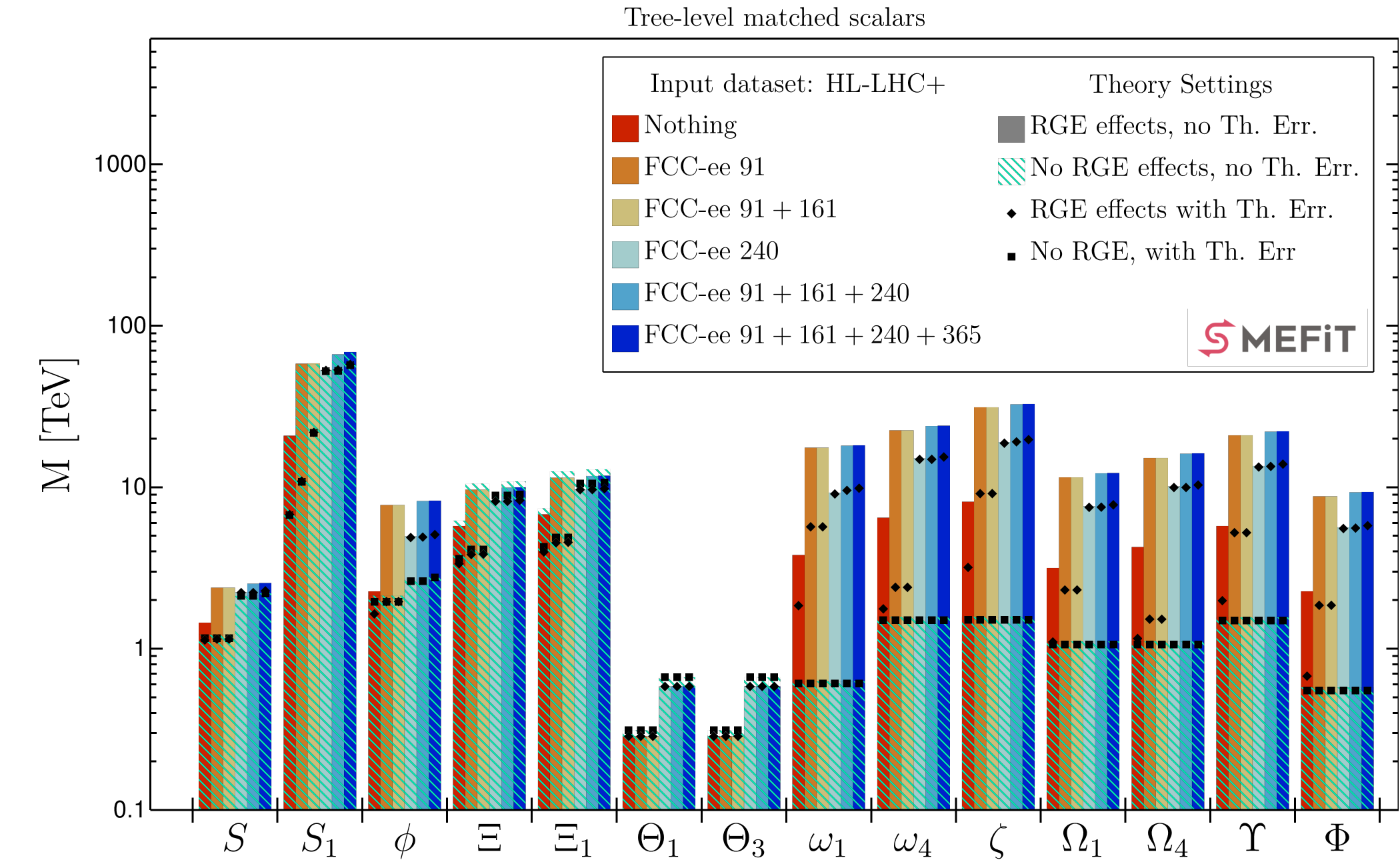}
\includegraphics[width=0.5\linewidth]{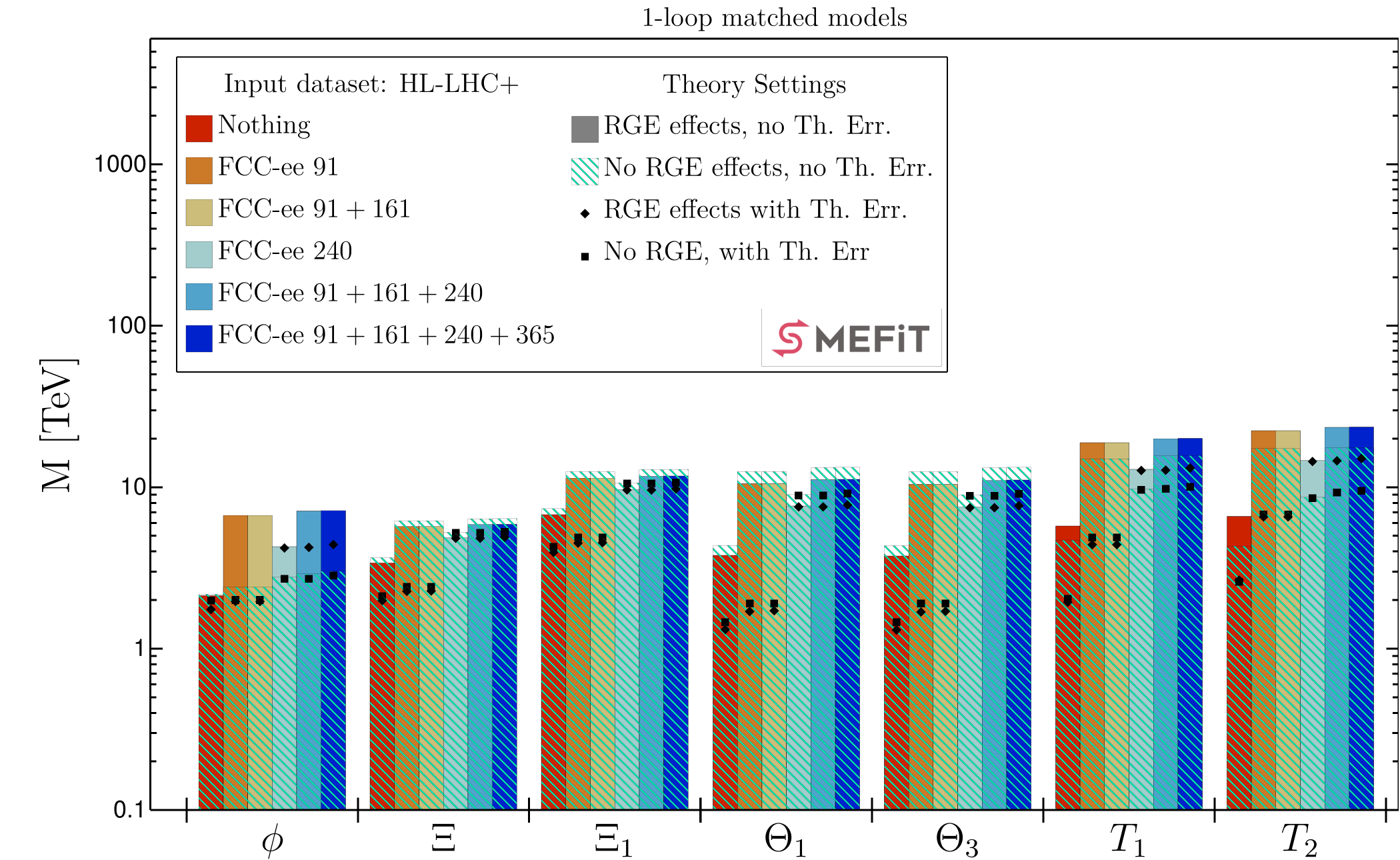}\hfill
\includegraphics[width=0.5\linewidth]{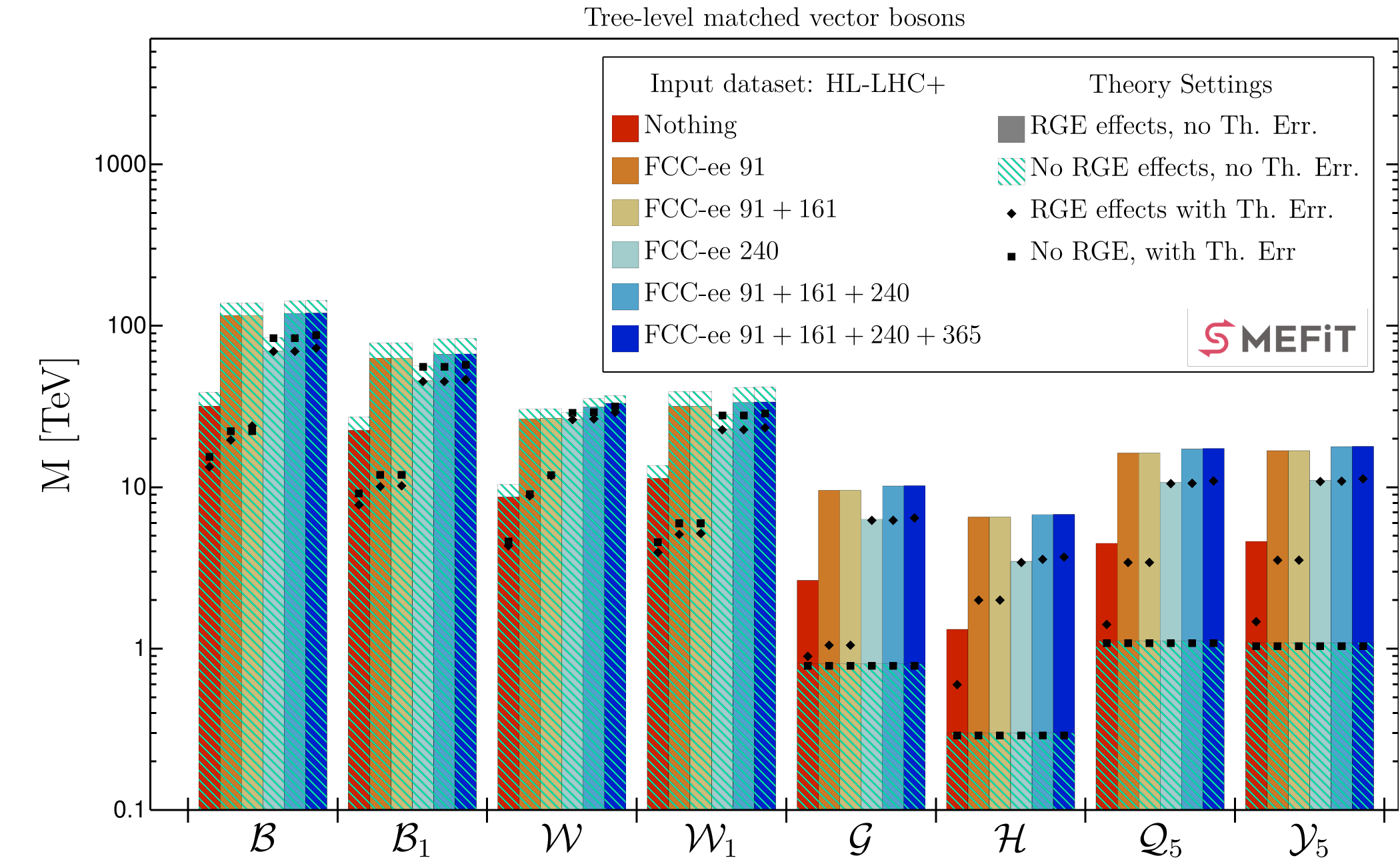}  
    \caption{Similar as Fig.~\ref{fig:mass_reach_FCC} with more granularity in the FCC-ee runs and fixed-scale RGE from the UV mass down to $m_Z=91.19$~GeV in all runs.}
    \label{fig:mass_reach_FCC_detail}
\end{figure}

Fig.~\ref{fig:mass_reach_FCC_detail} provides an extended version of Fig.~\ref{fig:mass_reach_FCC} with the $M_{\rm UV}$ mass reach evaluated for additional combinations of the FCC-ee datasets associated to different $\sqrt{s}$ values.
In particular, compared to Fig.~\ref{fig:mass_reach_FCC}, we add the results of the combined 91 GeV + 161 GeV runs and of the combined 91 GeV + 161 GeV + 240 GeV runs.
Unlike Fig.~\ref{fig:mass_reach_FCC}, in Fig.~\ref{fig:mass_reach_FCC_detail} we have used a fixed-scale RGE running between the mass of the UV particle and $m_{Z}=91.19$~GeV.
This change does not have any qualitative or substantial quantitative effect on the results.

From this fine-grained comparison, it can be seen that the $Z$-pole run is the one that leads the sensitivity to these models in the great majority of cases. Only some models see benefits from the Higgs and top measurements at the $240$~GeV and $365$~GeV runs, while the $161$~GeV run shows little relevance to constrain these simple models.
Nevertheless, as discussed in Sect.~\ref{sec:uv_fcc}, the dominance of the 91 GeV run only holds under the assumption that theory errors on the SM predictions of the EWPOs can be neglected.

%%%%%%%%%%%%%%%%%%%%%%%%%%%%%%%%%%
%%%%%%%%%%%%%%%%%%%%%%%%%%
\section{Numerical bounds}
\label{app:numerical_bounds}
We present in Table \ref{tab:rg_vs_no_rge_all_bounds} the numerical values of the 95\% C.I. bounds on the Wilson coefficients that are obtained from the linear and quadratic EFT fits to real LEP+LHC data presented in Sect.~\ref{sec:results}. In each case, we present bounds both in the presence and absence of RGE effects. 

%%%%%%%%%%%%%%%%%%%%%%%%%%%%%%%%%
\begin{table}[htbp]
    \centering
    \tiny
    \renewcommand{\arraystretch}{1.24}
    \begin{tabular}{l|C{0.8cm}|C{2.3cm}|C{2.3cm}|C{3.1cm}|C{3.1cm}}
        \multirow{2}{*}{Class} & \multirow{2}{*}{DoF}
        & \multicolumn{2}{c|}{ 95\% C.I. bounds, $\mathcal{O}\lp \Lambda^{-2}\rp$} &
        \multicolumn{2}{c}{95\% C.I. bounds, $\mathcal{O}\lp \Lambda^{-4}\rp$,} \\
        &                         & w/o RGE          & w/ RGE      & w/o RGE                            & w/ RGE      \\ \toprule
\multirow{5}{*}{4H}
 & $c_{QQ}^{1}$ & \textemdash & [-88.979, 272.224]& [-3.863, 4.579]& [-3.369, 3.907] \\ \cline{2-6}
 & $c_{QQ}^{8}$ & \textemdash & [-95.053, 312.849]& [-13.587, 10.946]& [-12.282, 8.644] \\ \cline{2-6}
 & $c_{Qt}^{1}$ & \textemdash & [-122.627, 214.164]& [-1.547, 1.432]& [-1.651, 1.523] \\ \cline{2-6}
 & $c_{Qt}^{8}$ & \textemdash & [-253.201, 93.221]& [-3.486, 2.631]& [-3.353, 2.366] \\ \cline{2-6}
 & $c_{tt}^{1}$ & \textemdash & [-70.225, 65.447]& [-0.926, 0.841]& [-1.056, 0.939] \\ \cline{2-6}
\hline
\multirow{14}{*}{2L2H}
 & $c_{Qq}^{1,8}$& [-10.294, 11.688]& [-2.878, 18.133]& [-0.283, 0.187]& [-0.308, 0.135] \\ \cline{2-6}
 & $c_{Qq}^{1,1}$& [-11.900, 10.541]& [-11.967, 9.147]& [-0.135, 0.101]& [-0.134, 0.096] \\ \cline{2-6}
 & $c_{Qq}^{3,8}$& [-7.341, 4.216]& [-7.009, 3.309]& [-0.328, 0.255]& [-0.265, 0.268] \\ \cline{2-6}
 & $c_{Qq}^{3,1}$& [-0.292, 0.150]& [-0.640, 1.357]& [-0.108, 0.102]& [-0.112, 0.088] \\ \cline{2-6}
 & $c_{tq}^{8}$& [-11.496, 10.878]& [-15.948, 1.779]& [-0.471, 0.276]& [-0.420, 0.184] \\ \cline{2-6}
 & $c_{tq}^{1}$& [-4.345, 16.145]& [-3.522, 16.363]& [-0.092, 0.145]& [-0.095, 0.136] \\ \cline{2-6}
 & $c_{tu}^{8}$& [-29.446, 6.278]& [-31.822, 0.617]& [-0.425, 0.185]& [-0.441, 0.161] \\ \cline{2-6}
 & $c_{tu}^{1}$& [-11.330, 30.860]& [-14.434, 15.240]& [-0.151, 0.133]& [-0.149, 0.131] \\ \cline{2-6}
 & $c_{Qu}^{8}$& [-9.941, 27.198]& [-0.558, 20.231]& [-0.750, 0.233]& [-0.698, 0.234] \\ \cline{2-6}
 & $c_{Qu}^{1}$& [-30.497, -0.377]& [-22.609, 2.902]& [-0.145, 0.148]& [-0.125, 0.171] \\ \cline{2-6}
 & $c_{td}^{8}$& [-0.811, 53.138]& [-9.573, 33.092]& [-0.610, 0.276]& [-0.612, 0.245] \\ \cline{2-6}
 & $c_{td}^{1}$& [-56.635, 17.656]& [-25.987, 31.137]& [-0.187, 0.178]& [-0.205, 0.174] \\ \cline{2-6}
 & $c_{Qd}^{8}$& [-48.266, 6.517]& [-14.883, 10.112]& [-1.017, 0.532]& [-1.028, 0.459] \\ \cline{2-6}
 & $c_{Qd}^{1}$& [-12.157, 49.776]& [-15.929, 31.753]& [-0.202, 0.217]& [-0.208, 0.215] \\ \cline{2-6}
\hline
\multirow{23}{*}{2FB}
 & $c_{c \varphi}$& [-0.252, 0.146]& [-0.205, 0.185]& [-0.082, 0.465]& [-0.086, 0.423] \\ \cline{2-6}
 & $c_{b \varphi}$& [-0.039, 0.094]& [-0.053, 0.076]& [-0.044, 0.048]& [-0.041, 0.044] \\ \cline{2-6}
 & $c_{t \varphi}$& [-4.412, 2.839]& [-15.177, 3.453]& [-3.223, 3.160]& [-4.150, 4.382] \\ \cline{2-6}
 & $c_{\tau \varphi}$& [-0.027, 0.041]& [-0.032, 0.046]& [-0.024, 0.046]& [-0.028, 0.049] \\ \cline{2-6}
 & $c_{tG}$& [-0.098, 0.226]& [-0.018, 0.269]& [0.028, 0.185]& [0.002, 0.184] \\ \cline{2-6}
 & $c_{tW}$& [-0.172, 0.149]& [-0.209, 0.169]& [-0.180, 0.142]& [-0.208, 0.161] \\ \cline{2-6}
 & $c_{tZ}$& [-3.146, 26.611]& [-0.377, 0.317]& [-0.692, 1.065]& [-0.402, 0.718] \\ \cline{2-6}
 & $c_{\varphi q}^{(3)}$& [-0.155, -0.009]& [-0.205, 0.050]& [-0.165, -0.006]& [-0.153, 0.028] \\ \cline{2-6}
 & $c_{\varphi Q}^{(3)}$& [-1.065, 0.899]& [-7.685, 2.043]& [-0.685, 0.319]& [-0.723, 0.406] \\ \cline{2-6}
 & $c_{\varphi q}^{(-)}$& [-0.165, 0.302]& [-2.718, 0.939]& [-0.064, 0.237]& [-0.102, 0.230] \\ \cline{2-6}
 & $c_{\varphi Q}^{(-)}$& [-1.849, 2.036]& [-6.145, 30.310]& [-0.707, 1.264]& [-0.775, 1.545] \\ \cline{2-6}
 & $c_{\varphi u}$& [-0.315, 0.546]& [-3.210, 0.923]& [-0.178, 0.174]& [-0.200, 0.199] \\ \cline{2-6}
 & $c_{\varphi d}$& [-1.061, 0.002]& [-3.556, 3.641]& [-0.306, 0.136]& [-0.336, 0.143] \\ \cline{2-6}
 & $c_{\varphi t}$& [-9.445, 7.525]& [-24.391, 33.889]& [-15.233, 1.296]& [-2.241, 0.104] \\ \cline{2-6}
 & $c_{\varphi l_1}$& [-0.357, 0.222]& [-0.362, 0.295]& [-0.127, 0.152]& [-0.131, 0.197] \\ \cline{2-6}
 & $c_{\varphi l_2}$& [-0.414, 0.250]& [-0.279, 0.488]& [-0.232, 0.171]& [-0.144, 0.348] \\ \cline{2-6}
 & $c_{\varphi l_3}$& [-0.411, 0.260]& [-0.562, 0.228]& [-0.151, 0.233]& [-0.251, 0.184] \\ \cline{2-6}
 & $c_{\varphi l_1}^{(3)}$& [-0.151, 0.053]& [-0.082, 0.153]& [-0.169, 0.033]& [-0.145, 0.079] \\ \cline{2-6}
 & $c_{\varphi l_2}^{(3)}$& [-0.155, 0.082]& [-0.248, 0.037]& [-0.138, 0.086]& [-0.221, 0.029] \\ \cline{2-6}
 & $c_{\varphi l_3}^{(3)}$& [-0.234, 0.124]& [-0.075, 0.368]& [-0.297, 0.079]& [-0.198, 0.233] \\ \cline{2-6}
 & $c_{\varphi e}$& [-0.756, 0.425]& [-0.758, 0.580]& [-0.248, 0.251]& [-0.243, 0.345] \\ \cline{2-6}
 & $c_{\varphi \mu}$& [-0.745, 0.430]& [-0.760, 0.581]& [-0.233, 0.274]& [-0.231, 0.373] \\ \cline{2-6}
 & $c_{\varphi \tau}$& [-0.767, 0.405]& [-0.780, 0.558]& [-0.256, 0.247]& [-0.256, 0.338] \\ \cline{2-6}
\hline
\multirow{1}{*}{4l}
 & $c_{ll}$& [-0.112, 0.099]& [-0.320, -0.026]& [-0.069, 0.150]& [-0.190, 0.063] \\ \cline{2-6}
\hline
\multirow{8}{*}{B}
 & $c_{\varphi}$& [-6.866, 5.170]& [-9.209, 8.612]& [-8.871, 2.671]& [-12.517, 3.782] \\ \cline{2-6}
 & $c_{\varphi G}$& [-0.025, 0.012]& [-0.034, 0.014]& [-0.019, 0.006]& [-0.023, 0.003] \\ \cline{2-6}
 & $c_{\varphi B}$& [-0.216, 1.191]& [-0.353, 0.194]& [-0.095, 0.157]& [-0.122, 0.228] \\ \cline{2-6}
 & $c_{\varphi W}$& [-0.238, 0.775]& [-0.470, 0.598]& [-0.072, 0.184]& [-0.065, 0.074] \\ \cline{2-6}
 & $c_{\varphi WB}$& [-0.677, 0.411]& [-0.746, 0.527]& [-0.179, 0.262]& [-0.216, 0.333] \\ \cline{2-6}
 & $c_{WWW}$& [-0.616, 0.566]& [-0.724, 0.610]& [-0.158, 0.232]& [-0.163, 0.245] \\ \cline{2-6}
 & $c_{\varphi \Box}$& [-1.724, 1.865]& [-2.950, 2.197]& [-1.701, 1.149]& [-2.153, 1.498] \\ \cline{2-6}
 & $c_{\varphi D}$& [-0.853, 1.492]& [-3.012, 6.046]& [-0.521, 0.467]& [-0.239, 1.152] \\ \cline{2-6}
\hline
    \end{tabular}
    \caption{\small The 95\% C.I. bounds on the
    EFT coefficients
    from the fit to real LEP and LHC data presented in Fig.~\ref{fig:posterior_rg_vs_no_rg_linear}.
    The reported bounds correspond to $\Lambda=1$ TeV and can thus be rescaled for any other value of $\Lambda$.
    We present results both for linear EFT fits, $\mathcal{O}\lp \Lambda^{-2}\rp$, and for quadratic EFT fits, $\mathcal{O}\lp \Lambda^{-4}\rp$.
    In each case, we indicate both bounds with and without taking RGE effects into account.
    }
    \label{tab:rg_vs_no_rge_all_bounds}
\end{table}
%%%%%%%%%%%%%%%%%%%%%%%%%%%%%%%%

\clearpage

\bibliographystyle{JHEP}
\bibliography{main}

%\clearpage

\end{document}